\documentstyle[aps,12pt]{revtex}
\renewcommand{\baselinestretch}{1.6}
\begin{document}

\begin{center}
{\Large \bf{The well-defined phase of simplicial quantum gravity in four
dimensions}}
\\[1cm]
Wolfgang Beirl, Erwin Gerstenmayer, Harald Markum, and J\"urgen Riedler\\
\it{Institut f\"ur Kernphysik, Technische Universit\"at Wien, A-1040}
Vienna, Austria \\[3cm]
\end{center}

\begin{abstract}
We analyze simplicial quantum gravity in four dimensions using
the Regge approach.
The existence of an entropy dominated phase with small negative
curvature is investigated in detail.
It turns out that observables of the system possess finite expectation values
although the Einstein-Hilbert action is unbounded.
This well-defined phase is found to be stable for a one-parameter family
of measures.
A preliminary study indicates that the influence of the lattice size on
the average curvature is small.
We compare our results with those obtained by dynamical triangulation
and find qualitative correspondence.
\end{abstract}
\begin{quote}
{\small
PACS number(s): 04.60.+n, 12.25.+e}
\end{quote}
\newpage

\section{Introduction}

General relativity relates classical gravitation to the curvature of
space-time and therefore quantum gravity is essentially
a quantum theory of space-time geometry \cite{MTW}.
In the search for such a theory the sum-over-histories approach has
proved to be an elegant and powerful tool \cite{Misner,Hawk,HaHa}.
It leads
directly from the classical action to the quantization of space-time
using the Feynman path integral.
The problems of this approach are well known.
Its perturbation theory is not renormalizable and a unique prescription
for the summation over all 4-geometries does not exist.

Moreover, in the case of pure gravity the action is unbounded from below
due to rapid conformal fluctuations.
At first sight this means that the corresponding Euclidean path-integral is
ill-defined and the quantum theory has no ground state.
However, there are exceptions such as the Hydrogen atom whose
quantum mechanics is well-defined although the Euclidean action
is unbounded from below.
In the case of quantum gravity numerical simulations within
the Regge calculus indicate that the entropy of the system
can suppress geometries with large curvature
leading to a phase with finite expectation values 
\cite{Hamber1,Hamber2,Berg}. 
It is the aim of this article to further investigate the
conditions for the existence of the well-defined phase. We do not give a 
complete picture of the phase structure and the details of the phase
transition but refer the reader to appropriate literature 
\cite{Hamber1,Hamber2,Hamber3}.

We use the path integral
\begin{equation} \label{PI}
 Z = \int D{\bf g} e^{ -{I_E}({\bf g})}
\end{equation}
as starting point for the non-perturbative investigation of
quantum gravity. The functional integral extends over a class of closed
4-geometries $\bf g$ in the Euclidean sector with pure gravitational action
\begin{equation} \label{RA}
-I_E({\bf g}) = {L_P}^{-2} \int d^4x \sqrt{g} ( R - 2 \Lambda ) ,
\end{equation}
where $L_P$ is the Planck length, $R$ the curvature scalar, and
$g$ the determinant of the metric. In general the action contains the
cosmological constant $\Lambda$.

Additional prescriptions are necessary to define the path integral
as a limit of systematic approximations. A direct route to such
approximations is the Regge calculus that uses simplicial lattices
to discretize general relativity in a coordinate independent and
thus very elegant way \cite{Regge,HartJM}.
One considers a simplicial net being the triangulation of a given
4-topology and introduces a metric
by assigning a length to each link in the net.
Increasing systematically the number
of simplices and decreasing the link lengths one reaches the classical
continuum limit if certain conditions are fulfilled
\cite{FriLee1,FriLee2,Chee}.

The Regge approach is a valuable tool to approximate the path integral on a
simplicial net with fixed incidence matrix and varying link lengths
\cite{Hamber2,Berg,Chee,BGM}. A complementary
method using fixed link lengths and varying incidence matrices is
known as dynamical triangulation \cite{DT2,DT3,DT4}.
However, the severe difficulties of the functional integral (\ref{PI}) are
also present on the simplicial lattice.

(i) In general the integration should include a summation over all
possible 4-topologies, but topologies are not classifiable in 4 dimensions
\cite{GeHa,HartTOP,Mark}.

(ii) In the case of pure gravity the action is unbounded due to rapid
conformal fluctuations \cite{Hawk,Mazur,Giddi}. This divergence is also
present in the Regge
action if one does not restrict or even fix all link lengths.
The introduction of a cosmological constant term guarantees a finite
volume of the simplicial lattice, but this is only a necessary and not a
sufficient condition for finite link lengths and a bounded action.
In Sect. 2 we give an explicit example of a configuration with finite
volume but infinite link lengths and action. One has to explain why such
configurations do not contribute to the path integral in the
well-defined phase and this
is the aim of the entropy argument of Berg:
The integral $Z = \int dI n(I) e^{-I}$
is well defined if the density of states $n(I)$
vanishes rapidly enough for $I \to -\infty$. Previous computations
indicate that the entropy of the system can indeed compensate
the unbounded action leading to a phase with small and negative average
curvature \cite{Hamber1,Berg} and one has assumed
that a finite volume will always lead to finite link lengths \cite{Hamber2}.

To give, at least numerically, further evidence for this assumption
we investigate in Sect. 3 the well-defined phase of the Regge approach
with three different methods. Furthermore, we compare our
results with those obtained by dynamical triangulation and find
qualitative agreement.

(iii) A unique definition of the measure $D{\bf g}$ does not
exist, since different physical arguments prefer different classes of measures
to be outlined below \cite{Misner,Leut,deWitt,FaPo,Menotti}.
It is an open question in what manner various
choices of the measure in the path integral affect the well-defined phase
although one expects universality for a certain class \cite{Hamber3,BGM}.
Hence, we use a one-parameter family of
measures and investigate in Sect. 4 the stability of the
entropy dominated region against variations of this parameter.
Again, the comparison with dynamical triangulation shows a remarkable
coincidence of both approaches.

(iv) The discretization of general relativity destroys the
diffeomorphism group and it is not known if it can be recovered in the
continuum limit \cite{HartJM,RoWi}. Until now the existence of a 
reasonable continuum limit
has not been demonstrated for simplicial quantum gravity. Therefore,
the meaning of our numerical results for the continuum remains unclear.
However, it is possible to study the influence of finite-volume effects on
the computed expectation values and it seems that the entropy dominated
phase survives increasing lattice volumes as demonstrated in Sect. 4. 
This might be an important
result for computations within the 'fundamental lenght scale' scenario
proposed recently by Berg et al. \cite{BeKK}.

The article is organized in the following way: The concept of the Regge 
calculus and the simplicial path integral are discussed in Sect. 2.
In Sect. 3 we investigate the entropy dominated phase by checking
various observables for
well-defined expectation values. The stability of this phase against
a variation of the measure and finite-volume effects are studied
in Sect. 4. The conclusion follows in Sect. 5.

\section{Simplicial quantum gravity}

The {\it triangulation} of a manifold $\cal M$ is determined if a
simplicial manifold $\cal K$ homeomorphic to $\cal M$ is given
\cite{HartJM,NaSe}.
A regular triangulation of the 4-torus $T^4$ used in the following
is provided by an algorithm that
divides $T^4$ into $n^4$ hypercubes and then each
hypercube into $24$ 4-simplices \cite{RoWi}.
For the total numbers $N_d$ of d-simplices one has $N_0 = n^4$, $n \geq 3$,
and $N_1 = 15N_0$, $N_2 = 50N_0$, $N_3 = 60N_0$ and $N_4 = 24N_0$.
A triangulation of $T^4$ with a smaller number of simplices is described
in \cite{Kuhn}.
In case of the hypercubic triangulation of $T^4$ the coordination numbers
$N_n/N_m$ are close to those of a {\it random} triangulation as
proposed by Christ et al. \cite{CFL}.

A simplicial net obtains a
metric structure if one assigns a length to each link of the lattice.
In the following we denote the squared length of link $l$
by $q_l \in R^1$.
Since we consider the Euclidean sector, we
require $q_l > 0$ and have to check the Euclidean
triangle inequalities for all simplices \cite{Hamber2,Regge,HartJM}.
In other words, for every 4-simplex $s$ in the lattice we
demand that it can be constructed in $R^4$.
After fulfilling the Euclidean triangle inequalities for every
4-simplex one has a consistent configuration $\{ q_l \}$
that determines the {\it simplicial lattice}.

One can calculate several quantities of a simplicial lattice such as
the area $A_t$ of the triangle $t = 1, ... , N_2$ and the
4-volume $V_s$ of the 4-simplex $s = 1, ... , N_4$.
Another important quantity is the deficit angle $\delta_t$
associated with each triangle $t$. It is defined as
\begin{equation}
\delta_t = 2\pi - \sum_{s \supset t} \theta_{s,t},
\end{equation}
where the sum runs over all 4-simplices $s$ sharing the triangle $t$.
The interior angle $\theta_{s,t}$ is the dihedral angle of the
4-simplex $s$ with the basis triangle $t$ \cite{Hamber2,Regge,HartJM}.

From the viewpoint of differential geometry a simplicial lattice
is a singular 4-geometry with the curvature
concentrated on the triangles and we consider the simplicial lattice as
an approximation to a smooth Riemannian manifold
\cite{Hamber1,Hamber2,Regge,HartJM,FriLee1,FriLee2,Chee}.
The connection of the deficit angle with the curvature
tensor is exhibited by parallel transporting
a vector along a closed loop around a triangle $t$
\cite{Hamber2,Regge,Casel}.

In consideration of the proportionality between the curvature tensor and
the deficit angle and guided by the dimension of the action Regge
proposed \cite{Regge}
\begin{equation} \label{RACT}
  \int d^4x \sqrt{g} R \; \rightarrow \; 2 \sum_t A_t \delta_t .
\end{equation}
An alternative derivation and an action with the same continuum limit can be
found in \cite{Casel}.

One can show that the Einstein-Hilbert action of a smooth
4-geometry converging towards a simplicial lattice approaches
the Regge action \cite{FriLee1}, but
for our purposes it is more important to study the opposite limit,
approximating a given smooth 4-geometry
by increasing the number of simplices
of the lattice and decreasing the link lengths systematically.
A necessary prerequisite in this discussion is the definition of
the {\it fatness} $\phi_s$ of a 4-simplex $s$.
It is a relation between the maximal link length and the 4-volume given by
\begin{equation}
\phi_s = C^2 \frac{{V_s}^2}{\max_{l \in s}({q_l}^4)},
\end{equation}
where the constant is set to $C=24$ in the following.
(Cheeger et al. use a slightly different but equivalent definition
\cite{Chee}.)
If the lattice skeleton approaches a smooth geometry such that
$\phi_s\geq f = const > 0$ one recovers
general relativity in the continuum limit \cite{FriLee2,Chee}.

Notice that the Regge calculus is the only known approximation scheme
of general relativity incorporating the Bianchi identities
\cite{Misner,Hamber2,Regge}. Since
these relations connect the dynamical and constraint equations of
general relativity they are essential for any attempt to quantize gravitation.
In the following we try to assign a concrete meaning to the path integral
(\ref{PI}) in the language of the Regge calculus.

(i) The integration symbol $\int$ stands for a summation over all
4-geometries $({\cal M},{\bf g})$ without boundary.
This can be replaced by a sum over distinct simplicial lattices
$({\cal K},\{ q_l \})$.
One usually restricts the class of incidence matrices to include
only simplicial manifolds ${\cal K}$ with a fixed topology.
Further simplifications lead to two somehow complementary methods.

{\it Dynamical triangulation}:
One assigns the same length to each
link, $q_l = 1 \; \forall l$, and performs a
summation over all possible triangulations $\cal K$ of a given topology
\begin{equation}
 Z = \sum_{\cal K} e^{ -I_E(\cal K) } .
\end{equation}
The main motivation for this approach is its success in two dimensions
\cite{DT2,DT3,DT4,BM}.
In four dimensions the action simplifies for equal link lengths to
$-I_E = c_2 N_2 - c_4 N_4$ with coupling
parameters $c_2, c_4$. It is bounded from below (and above) for
a finite number of simplices and, moreover, since the inequality $N_2 < 10N_4$
holds for every simplicial lattice one has $I_E > 0$
for $c_4>10c_2$.  This can be fulfilled by making the
cosmological constant large enough, in contrary to the continuum theory
and the Regge approach.
Since we try to investigate the possible existence of an entropy dominated
phase for unbounded gravitational action, we use the

{\it Regge approach}:
The triangulation is kept fixed and the path
integral $Z$ reduces to a summation over different configurations
$\{ q_l \}$. Only the link lengths are allowed to vary but not the
incidence matrices \cite{Hamber1,Berg,HartJM,BGM},
\begin{equation} \label{quest}
 Z = \int Dq e^{ -I_E( \{ q_l \} ) }.
\end{equation}
Important motivations for this approach are studies of the
weak field limit which show the correspondence with linearized
quantum gravity \cite{RoWi}.
Comparisons of Regge calculus and mini-superspace approximations
have also been performed with rather promising results
\cite{HartJM,Louko}.

(ii) The gravitational action $I_E( {\bf g} )$ is unbounded in the continuum
due to conformal fluctuations and
this unpleasant feature is also present
in the Regge approach yet for a finite number of simplices \cite{HaHa,HartJM}.
If some of the 4-simplices collapse the corresponding areas can grow,
$A_t \rightarrow \infty$, although their 4-volumes stay finite.
To give an example consider a configuration $\{ q_l \}$ where
first all link lengths are equal to $a$,
i.e. $q_l = a^2 \; \forall l$. Then choose one vertex $v_0$
of the lattice arbitrarily and assign to each link of this vertex
a length $L > a$. It is easy to check that the Euclidean
triangle inequalities are all satisfied in this case.
Now consider $a \rightarrow 0$ and $L \rightarrow \infty$ so that
$a^2 L=const.$ In this case the 4-volume of the 4-simplices with $v_0$
as a vertex vanishes like $a^3 L$ and since the 4-volume
of the other simplices disappears as $a^4$ the total
4-volume of the lattice approaches zero, $\sum_sV_s\rightarrow 0$.
In contrast, the area of triangles with $v_0$ as a vertex diverges
as $aL$ and therefore the action grows without limit,
$\sum_tA_t\delta_t\rightarrow\infty$.

As mentioned in the introduction, it seems that such configurations do not
contribute to the path integral \cite{Hamber1,Hamber2,Berg} in the well-defined
phase and we investigate this possibility in the following section.
Of course, the described divergence cannot be present for
dynamical triangulation with all link lengths equal.

(iii) A unique definition of the gravitational measure $D{\bf g}$ does
not exist since it is not clear which quantities have to be
identified with the 'true' physical degrees of freedom
\cite{Misner,BGM,deWitt,FaPo,Menotti}.
Within the Regge approach it seems natural to translate the
gravitational measure
into a product over the links of the simplicial lattice \cite{Hamber1}
\begin{equation}
Dq = \prod_l \mu(l) dq_l {\cal F}(q_1, .. , q_{N_1}) ,
\end{equation}
where the function ${\cal F}$ ensures that the computations are performed in
the Euclidean sector being equal $1$ if the
generalized triangle inequalities are fulfilled and $0$ otherwise. The weight
function $\mu$ is a (perhaps complicated and non-local) function of
$\{ q_l \}$ associated with each link $l$ \cite{Band,JeviNi}.
The simplest choice is the uniform measure $\mu(l) = 1 \; \forall l$ and
one has argued that it corresponds to the DeWitt measure \cite{Hamber3,HartJM}.
However, considering the results of Misner and Faddeev-Popov it
seems natural to choose $\mu(l)$ so that a scale-invariant
measure results \cite{Hamber1,Berg,Chee}.

Relying on this background the presented numerical studies
have been performed with
\begin{equation} \label{SM1}
 \mu(l) = {q_l}^{\sigma - 1} ,
\end{equation}
where the parameter $\sigma \geq 0$ determines the behavior of the measure
under rescaling. The case $\sigma = 1$ corresponds to the uniform
measure while for $\sigma = 0$ a scale-invariant
measure in its simplest form results.

Metropolis simulations within the Regge approach proceed in the
following way \cite{Hamber1,Hamber2,Berg,BGM}.
Starting with a Euclidean configuration $\{ q_l \}_0$ a Markov chain of
configurations $\{ q_l \}_{\tau}$ is generated to approximate
the canonical ensemble. 
Proposing a new configuration one has to check first the Euclidean
triangle inequalities. If the new configuration is not in the
Euclidean sector it is rejected immediately. Otherwise one has to
check the Metropolis condition to fulfill the requirement of detailed
balance. Given a Euclidean
configuration one changes usually only one link length at once
which simplifies the test of the triangle inequalities.
Scanning one time through the whole lattice
is called a {\it sweep}.
After thermalization of the system the averages over the generated
configurations are taken to approximate the expectation values.

A priori one expects that simulations of simplicial quantum gravity with
unbounded action might not find an equilibrium. Indeed, one has to face
this problem for computations on non-regular triangulations
\cite{BGM,wb}.
To investigate systematically if the entropy of the system can stabilize
the unbounded action we use the
following method: Guided by the analysis of Cheeger et al.
\cite{Chee} we set a lower limit for the fatness of each 4-simplex,
\mbox{$\phi_s \geq f > 0 $}, restricting the configuration space.
In this way we introduce a scale-invariant cutoff allowing the system to
reach an equilibrium after a finite number of sweeps. As long as
\mbox{$f > 0$} the gravitational action is bounded for a finite
number of 4-simplices.
Decreasing stepwise the value of $f$ we investigate the limit
$f\rightarrow 0$ and conclude from the {\em convergence of this process}
if a well-defined phase does exist or not.
Since decreasing $f$ enlarges the significant configuration space one has
to increase the number of sweeps to ensure equilibrium. In the actual
computations the number of sweeps varies from 5k $(f=10^{-3})$ up to
20k $(f=10^{-6})$.

\section{Entropy versus action}

In the following we employ the action in the form
\begin{equation} \label{SA1}
-I_E = \beta \sum_t A_t {\delta}_t  - \lambda \sum_s V_s,
\end{equation}
with a bare cosmological constant $\lambda$ fixing the scale and
we restrict our attention to the simplicial measure (\ref{SM1}) with the
parameter $\sigma$.
Considering the behavior of the path integral $Z(\beta,\lambda)$
under rescaling $q_l \to rq_l$ one obtains
\begin{eqnarray}
Z(\beta,\lambda) = \left( \frac{\beta}{\lambda} \right)^{\sigma N_1}
          Z \left( \frac{\beta^2}{\lambda},\frac{\beta^2}{\lambda}
\right)
\end{eqnarray}
after setting $r=\beta / \lambda$.
Differentiating $\ln Z$ with respect to $\beta$ and $\lambda$ gives
\begin{equation}
 - \beta \langle \sum_t A_t \delta_t \rangle + 2 \lambda
\langle \sum_s V_s \rangle = \sigma N_1
\end{equation}
in the well-defined phase \cite{Hamber2}.
We therefore set the parameter $\lambda$ equal to $\sigma$
so that $\langle V_s \rangle$ takes the value $N_1/2N_4$ for
$\sigma > 0$ at $\beta = 0$. For $\beta>0$, $\langle V_s\rangle$ decreases
slightly as well as $\langle q_l\rangle$. To compensate the changes of
the lattice size as a function of $\beta$ and $\sigma$ we use the 
dimensionless quantity
\begin{equation}
\ell^2=\frac{\beta}{2}\langle q_l\rangle
\end{equation}
to compare the results. The factor $\frac{1}{2}$ respects the original
form of the Regge action (\ref{RACT}) and since $\beta$ corresponds to
$2L_P^{-2}$ the dimensionless observable $\ell^2$ expresses the average
squared lattice spacing in units of the bare Planck length.
Different to conventional lattice field-theory the simplicial lattice
itself is the quantum object, i.e. the lattice spacing
is an observable rather than a parameter and different definitions are
possible \cite{Berg}.

Another important observable of simplicial quantum gravity is the
scale invariant average curvature of a lattice configuration
\cite{Hamber1,Hamber2}
\begin{equation}
\tilde{R} =
 \frac{ \sum_t A_t {\delta}_t }{ \sum_s V_s }
     \left( \frac{1}{N_1} \sum_l q_l \right) .
\end{equation}
The expectation value $\langle \tilde{R} \rangle$ can be understood
as effective cosmological
constant measured in lattice units. This can be seen from the classical field
equations deriving $\Lambda = \frac{1}{4} R$ and obtaining
\begin{eqnarray}
\Lambda = \frac{1}{4} \frac{\int d^4x \sqrt{g} R}{\int d^4x \sqrt{g} }
    & \leftrightarrow &
    \frac{1}{2} \tilde{R},
\end{eqnarray}
with $\Lambda$ the classical cosmological constant, $R$ the curvature
scalar and $g$ the determinant of the metric tensor.

In Fig. 1 the behavior of the expectation value $\langle \tilde{R} \rangle$
as a function of $\ell$ is depicted.
The most interesting region $\langle \tilde{R} \rangle < 0$ is displayed
in more detail in the lower plot. The value of the lower limit $f$
for the fatness decreases as follows: $f = 10^{-n} , n = 3,4,5,6,\infty$.
For every $f$ we increase $\beta$ stepwise leading to increasing $\ell$.
Since the significant configuration
space becomes larger with smaller values of $f$ we performed 20k sweeps
for each data point for $f \leq 10^{-5}$ and 5k sweeps otherwise.
Averages are taken over the last 10k and 2.5k sweeps, respectively, so
that the statistical error is smaller than the symbol size.
This computation was performed with the uniform measure
$(\sigma = 1)$
on the regular hypercubic triangulation of the 4-torus with
$4^4$ vertices. We will see later that the well-defined phase
is stable against changes of the parameter $\sigma$ in a certain range.

For all values of the cutoff $f$ we find a well-defined phase with small
and negative curvature. Increasing $\ell$ one suddenly
enters a region of large positive curvature.
For $f \leq 10^{-5}$ it is difficult to reach a stable equilibrium
across the transition point due to a trend towards
non-reproducible states. This coupling regime is therefore called the
ill-defined phase.

In every case it seems that $\langle \tilde{R} \rangle < 0$ as long as
$\ell<\ell_c$, with $\ell_c=0.45(1)$ for sufficiently small $f$.
Furthermore, we observe for $f \to 0$ a convergence of the function
$\langle \tilde{R} \rangle ( \ell )$ at every $\ell < {\ell}_c$:
As long as $\ell < 0.3$ one sees practically no difference in the average
curvature $\langle \tilde{R} \rangle$ for all values of the cutoff $f$.
This is a strong indication that configurations
with small fatness $\phi_s$ give almost no contribution in this
region. Increasing $\ell$ further one begins to realize differences
in $\langle \tilde{R} \rangle$ for larger $f$. However, for $f \leq 10^{-5}$
there are (nearly) no differences to see over the whole well-defined
phase $0 \leq \ell < \ell_c$. This convergence for $f\to 0$ indicates that
the entropy of the system is indeed able to compensate the unbounded
gravitational action.
Configurations with large average curvature occur with small probability
and give (practically) no contribution to the path integral in the
well-defined phase. Therefore, it might be possible to perform computations
even without any restriction $(f = 0)$ and obtain reasonable
expectation values. One should be aware that the scale of the curvature 
radius should be larger than the
average lattice spacing and smaller than the size of the system, 
$\langle q_l\rangle < \langle q_l\rangle |\langle\tilde{R}\rangle|^{-1} <
N_0^{1/2}\langle q_l\rangle$ \cite{Hamber3}.

To further analyze the probability of 'crumpled' configurations with
collapsing simplices we display the behavior of the average fatness
$\langle \phi_s \rangle$. Fig. 2 shows $\langle \phi_s \rangle$
as a function of $\ell$ for decreasing cutoff $f$. The parameters
of the system and the configurations sampled for the statistics are the same
as before.
It turns out that $\langle \phi_s \rangle$ is much larger
than the lower bound $f$ in the well-defined region.
The convergence of the function $\langle \phi_s \rangle(\ell)$ for
decreasing $f \to 0$ is seen clearly in the well-defined phase.
With the transition to large values of $\langle\tilde{R}\rangle$
the average fatness decreases suddenly as expected.
This indicates the collapse of 4-simplices and the transition to
a 'crumpled' lattice.

Returning to the well-defined phase let us investigate the
behavior of the expectation value
$\langle \delta_t^2 \rangle$.
The following argument will supply us with a further indication
for the stability of the well-defined phase.
Imagine that one incorporates an additional term
$- \alpha \sum_t \delta_t^2$ in the action. Differentiation of
$\ln Z$ then gives
\begin{eqnarray}
\frac{\partial \ln Z}{\partial \beta} =
    \langle \sum_t A_t \delta_t \rangle     &,&
\frac{\partial \ln Z}{\partial \alpha} =
    - \langle \sum_t \delta_t^2 \rangle \nonumber\\
    \Rightarrow \;\;
\frac{\partial \langle \delta_t^2 \rangle}{\partial \beta} &=&
  - \frac{\partial \langle A_t \delta_t \rangle}{\partial \alpha}.
\end{eqnarray}
All computations are performed at $\alpha = 0$ but we
assume that pushing $\alpha$ slightly to a positive value
would shift $\langle A_t \delta_t \rangle$ towards zero
since the additional term prefers flat configurations and supresses
the curved.
(Recall the definition of $\tilde{R}$ to see
that $\tilde{R} < 0$ corresponds to $\sum_t A_t \delta_t < 0$).
This means that $\langle A_t \delta_t \rangle$ will increase for
negative values and decrease for positive.
We thus have
$\frac{\partial}{\partial \alpha} \langle A_t \delta_t \rangle > 0$
and therefore
$\frac{\partial}{\partial \beta} \langle \delta_t^2 \rangle < 0$
in the well-defined phase with negative curvature.
Since $\ell$ is proportional to $\beta^{\frac{1}{2}}$ the fact that
$\langle A_t \delta_t \rangle$ is negative induces that
$\langle \delta_t^2 \rangle$ decreases in this phase with $\ell$.
As seen in Fig. 3 $\langle \delta_t^2 \rangle$ indeed decreases as
expected with the minimum just around the transition point.
This is an indication that the average curvature is really negative
in this phase and at least shows the self consistency of the numerical
results.

It is practically impossible to determine a systematic error for Metropolis
simulations. One cannot exclude for example the possibility that the 
well-defined phase is only a numerical artefact due to metastable states
preventing the system from reaching configurations with large positive 
curvature. To check this we performed computations with inhomogeneous start 
configurations from the ill-defined phase having large positive average 
curvature.

Without a restriction of the
fatness ($f = 0$) and a non-zero coupling $\beta$ the well-defined
phase $\ell < \ell_c$ was reached after a large number
of sweeps. The average curvature returned to small and negative values of
$\tilde{R}$ although the unbounded gravitational action prefers a
large positive curvature. The history of such a run 
is depicted in the upper plot of \mbox{Fig. 4} and compared with the
history for a start with a homogeneous configuration.
The independence of $\langle \tilde{R} \rangle$ from the start
configuration is the strongest indication that entropy is
indeed able to compensate the unbounded action in the well-defined
phase.

Near the transition point very long runs
are necessary to reach the equilibrium for very inhomogeneous
start configurations. Limited resources therefore do not
allow to perform the above check for ${\tilde{R}}_{start} \to \infty$ and
$\ell \to \ell_c$. Thus, we cannot exclude the possibility that near
$\ell_c$ metastable states occur if the phase transition is 
first order as reported in \cite{Hamber3}.
However, applying a (small) lower limit for the fatness
$(f = 10^{-5})$ we found no influence of
the start configuration for $\ell \leq 0.45$ with rather
large ${\tilde{R}}_{start}$. Of course, the curvature of the
start configuration is limited after applying a lower limit on the fatness.
The lower plot of Fig. 4 shows such a history reaching the equilibrium after
50k sweeps.

The two complementary approaches to simplicial quantum gravity are compared in
Fig. 5 by plotting where we depict our results within the Regge approach and 
those obtained by Br\"ugmann within dynamical triangulation \cite{BM}.
The lower curve gives $\langle\tilde R\rangle$ as a function of
$\ell$ for $f=10^{-5}$ and uniform measure taken from Fig. 1.
The upper curve shows the results of Metropolis simulations
performed by Br\"ugmann dynamically triangulating the 4-sphere with
$N_4^c = 16$k \cite{BM}. We depict $\langle \tilde{R} \rangle$ as a function of
$\ell$
and since $q_l = 1 \; \forall l$ we have
\begin{eqnarray}
\tilde{R} = \frac{ a( 2 \pi N_2 - 10 \theta N_4 ) }{ v N_4^c } ,
\end{eqnarray}
where $a, v, \theta$ denote the area, 4-volume and interior angle
associated with equilateral simplices of link lengths equal to 1
and $\ell = \sqrt{\frac{1}{2}\beta}$.

For a given number of 4-simplices the number of triangles in the
lattice is limited.
Since $N_2 < 10 N_4$ there is an upper limit for the average
curvature, \mbox{$\tilde{R} < (a/v)( 20 \pi - 10 \theta )$}, different to
the Regge approach. However, it seems
that both dynamical triangulation and the Regge approach
lead to similar results at least qualitatively.
(One has to remember that the underlying topology is different.)
The average curvature begins for small $\ell$ with small positive values 
and undergoes a transition to large positive curvatures.
The transition point seems to be located close to that of the Regge approach,
but it is unclear if this is pure coincidence.

\section{Influence of the measure and finite-volume effects}

In this section we address first the question of the measure in the
well-defined phase of simplicial quantum gravity
and investigate then the influence of finite-volume effects.

Fig. 6 shows the expectation value $\langle \tilde{R} \rangle$
as a function of $\ell$ for measures with different values of $\sigma$
increasing stepwise from $0$ to
$1.5$ including the scale-invariant and the uniform measure.
For the scale-invariant measure $(\sigma=\lambda=0)$ we use directly the
constraint $\sum_s V_s = const$ rescaling the lattice at every sweep as
proposed by Berg \cite{Berg}.
The computations have been performed on the regular hypercubic
triangulation of the 4-torus with $4^4$ vertices.
To facilitate numerical simulations a lower bound
$f = 10^{-5}$ has been applied for the fatness $\phi_s$ of each 4-simplex.
The results of Sect. 3 suggest that such a restriction of
allowed configurations practically does not affect the behavior of the
system in the well-defined phase.
Averages are taken over at least 10k sweeps after thermalization so that the
statistical error is smaller than the symbol size.

For $\sigma \leq 1$ and $0 \leq \ell \leq 0.3$ the
value of $\langle \tilde{R} \rangle$ seems to be practically independent
of $\sigma$.
Such a stability against a variation of the measure was reported first
by Hamber in the context of $R^2$ theory \cite{Hamber1}. A direct
comparison of scale-invariant and uniform measure for pure gravity with
a cutoff for the link lengths is described in Ref. \cite{BGM}.
Near the transition point, $\ell \approx 0.4$, the influence of
$\sigma$ becomes more pronounced.

The result for $\sigma = 1.5$
differs over the whole range of $\ell$ from those obtained
for $\sigma \leq 1$. At $\ell = 0$ the value of
$\langle \tilde{R} \rangle$ is significantly larger than for $\sigma \leq 1$.
With increasing $\ell$ we find a stable
equilibrium even for $\langle \tilde{R} \rangle > 0$.
This suggests that the considered measures fall into
two qualitatively different classes: While we see almost no dependence
of the expectation values
in the interval $0 \leq \sigma \leq 1$ the influence of the measure
becomes visible for $\sigma > 1$.
The situation for $\ell=0$ is further illustrated in the 
lower plot of Fig. 6 showing 
$\langle \tilde{R} \rangle$ as a function of $\sigma$.
Setting $\sigma = 2$ we found no equilibrium even at $\beta = 0$.

To gain more information about the lattice geometry
the behavior of the areas and deficit angles is examined separately.
Fig. 7 displays the scale-invariant quantity
$\langle A_t \rangle / \langle q_l \rangle$
as a function of $\ell$ for the different
values of $\sigma$. For $0 \leq \ell < 0.4$
this ratio stays almost constant somewhat below the value
$\frac{\sqrt{3}}{4} \approx 0.433$ of equilateral triangles.
The value for $\sigma = 1.5$ lies significantly
below the other data points. Across the transition at $\ell \approx 0.45$ the
ratio $\langle A_t \rangle / \langle q_l \rangle$
decreases, indicating a distortion of the triangles.

The expectation value of the average deficit angle
$\langle \delta_t \rangle$ is depicted as a function of
$\ell$ in Fig. 8.
Surprisingly $\langle \delta_t \rangle$ stays negative for
all values of $\ell$ even after the transition to large
positive curvature. In the well-defined phase the absolute
value of $\langle \delta_t \rangle$ is rather small
compared to $\pi$. One can understand this behavior by examining
the hypercubic triangulation of the
4-torus that contains two different types of triangles. The number
of 4-simplices sharing a triangle of the first type is $6$
while it is $4$ for the second type.
As noticed first by Berg the contributions of these two types
almost cancel each other leaving
a small negative average deficit angle \cite{Berg}.

The dependence of the simplicial path integral on the measure has been
studied also within the framework of dynamical triangulation.
To investigate the influence of different types of the
gravitational measure Br\"ugmann considered an additional
term $ n \sum_v ln( o(v) )$ in the action where $o(v)$ is the 'order'
of the vertex $v$, i.e. the number of 4-simplices that contain
this vertex \cite{BM}. Although the correspondence with the continuum is not
entirely clear, it is plausible that this term reproduces a measure
of the form $\prod_x g^{n/2} \prod_{i \leq j} dg_{ij}$ with
$n = -5$ the scale-invariant and $n = 0$ the uniform measure
of DeWitt.  Results of these computations are depicted in Fig. 9
where again the expectation value $\langle\tilde{R}\rangle$ is plotted 
versus $\ell$ for
different values of $n$. Since Br\"ugmann considers negative couplings
$\beta < 0$ we extend $\ell$ to negative values.

Besides a shift in $\ell$ for dynamical
triangulation, a comparison of Figs. 6 and 9
shows a remarkable qualitative coincidence of the results
with the choice $n=0$ for dynamical triangulation
corresponding to the case $\sigma = 1$ of the Regge approach.
Again the influence of the measure seems to be small between the
scale-invariant and the uniform measure, but turns out to be significant 
for $n=5$.

As for every lattice field theory it is important to study
finite-volume effects for simplicial quantum gravity \cite{Hamber3}. 
To investigate the stability of the well-defined phase for larger 
lattices we increased the number of vertices from $3^4$ to $8^4$.
The results are displayed in Fig. 10 and show $\langle \tilde{R} \rangle$
as a function of $\ell$ computed on the hypercubic triangulation of the
4-torus with the uniform measure $(\sigma = 1)$ and a lower limit 
$f = 10^{-5}$ for the fatness. For the smaller lattices with 
$N_0=3^4, 4^4, 6^4$ 
we performed at least 10k sweeps after thermalization.
Statistics is rather poor in the case of $N_0 = 8^4$ where
the number of sweeps in equilibrium is only 2.5k.
Therefore, one has to consider the three data points for the $8^4$-triangulation
as preliminary.

First results indicate that the critical value of the coupling $\beta$ and
therefore $\ell_c$ slightly decreases for increasing $N_0$.
But it seems that increasing the number of vertices $N_0$
has only a weak influence on the computed expectation values 
of one-point functions and the well-defined phase survives on larger lattices.

\section{Conclusion}

The Regge calculus provides a direct route to systematic
approximations of the quantum-gravity path-integral.
It allows numerical studies of non-perturbative quantum gravity
and has led to astonishing results.

The most important finding is the existence of a well-defined phase.
We studied this phase by applying a lower bound on the fatness of each
simplex and decreasing it systematically. The convergence of this process
supports the 'entropy hypothesis'. Although the gravitational
action is unbounded due to conformal fluctuations the entropy
of the system seems to stabilize the expectation values.
In a certain range of the gravitational coupling a phase with small 
negative average curvature occurs within the Regge approach.
The predicted behavior of the squared deficit angle and the
independence of the simulation from the start configuration further
support the stability of the entropy dominated phase.

The well-defined phase turns out to be stable against
variations of the measure and
an increase of the lattice size.
The influence of the measure has been studied using a one-parameter
family of simple local functions.
However, a wider class of different measures should be investigated
including also non-local types and non-regular triangulations.
The dependence of the results on the lattice size has been
studied to some extent. The fact that the entropy dominated phase
survives on larger lattices is encouraging for studies
of a 'fundamental length scale' scenario \cite{BeKK}.

\section*{Acknowledgment}
We are grateful to Bernd Br\"ugmann for transmitting the data of his
simulations enabling the comparison of Regge approach and dynamical
triangulation. This work was supported in part by "Fonds zur F\"orderung
der wissenschaftlichen Forschung" under contract P9522-PHY.


\newpage
\begin{figure}[p]

\setlength{\unitlength}{0.240900pt}
\ifx\plotpoint\undefined\newsavebox{\plotpoint}\fi
\sbox{\plotpoint}{\rule[-0.175pt]{0.350pt}{0.350pt}}%
\begin{picture}(1500,900)(0,0)
\tenrm
\sbox{\plotpoint}{\rule[-0.175pt]{0.350pt}{0.350pt}}%
\put(264,263){\rule[-0.175pt]{282.335pt}{0.350pt}}
\put(264,179){\rule[-0.175pt]{4.818pt}{0.350pt}}
\put(242,179){\makebox(0,0)[r]{-20}}
\put(1416,179){\rule[-0.175pt]{4.818pt}{0.350pt}}
\put(264,263){\rule[-0.175pt]{4.818pt}{0.350pt}}
\put(242,263){\makebox(0,0)[r]{0}}
\put(1416,263){\rule[-0.175pt]{4.818pt}{0.350pt}}
\put(264,347){\rule[-0.175pt]{4.818pt}{0.350pt}}
\put(242,347){\makebox(0,0)[r]{20}}
\put(1416,347){\rule[-0.175pt]{4.818pt}{0.350pt}}
\put(264,431){\rule[-0.175pt]{4.818pt}{0.350pt}}
\put(242,431){\makebox(0,0)[r]{40}}
\put(1416,431){\rule[-0.175pt]{4.818pt}{0.350pt}}
\put(264,514){\rule[-0.175pt]{4.818pt}{0.350pt}}
\put(242,514){\makebox(0,0)[r]{60}}
\put(1416,514){\rule[-0.175pt]{4.818pt}{0.350pt}}
\put(264,598){\rule[-0.175pt]{4.818pt}{0.350pt}}
\put(242,598){\makebox(0,0)[r]{80}}
\put(1416,598){\rule[-0.175pt]{4.818pt}{0.350pt}}
\put(264,682){\rule[-0.175pt]{4.818pt}{0.350pt}}
\put(242,682){\makebox(0,0)[r]{100}}
\put(1416,682){\rule[-0.175pt]{4.818pt}{0.350pt}}
\put(264,766){\rule[-0.175pt]{4.818pt}{0.350pt}}
\put(242,766){\makebox(0,0)[r]{120}}
\put(1416,766){\rule[-0.175pt]{4.818pt}{0.350pt}}
\put(264,158){\rule[-0.175pt]{0.350pt}{4.818pt}}
\put(264,113){\makebox(0,0){-0.2}}
\put(264,767){\rule[-0.175pt]{0.350pt}{4.818pt}}
\put(381,158){\rule[-0.175pt]{0.350pt}{4.818pt}}
\put(381,113){\makebox(0,0){0}}
\put(381,767){\rule[-0.175pt]{0.350pt}{4.818pt}}
\put(498,158){\rule[-0.175pt]{0.350pt}{4.818pt}}
\put(498,113){\makebox(0,0){0.2}}
\put(498,767){\rule[-0.175pt]{0.350pt}{4.818pt}}
\put(616,158){\rule[-0.175pt]{0.350pt}{4.818pt}}
\put(616,113){\makebox(0,0){0.4}}
\put(616,767){\rule[-0.175pt]{0.350pt}{4.818pt}}
\put(733,158){\rule[-0.175pt]{0.350pt}{4.818pt}}
\put(733,113){\makebox(0,0){0.6}}
\put(733,767){\rule[-0.175pt]{0.350pt}{4.818pt}}
\put(850,158){\rule[-0.175pt]{0.350pt}{4.818pt}}
\put(850,113){\makebox(0,0){0.8}}
\put(850,767){\rule[-0.175pt]{0.350pt}{4.818pt}}
\put(967,158){\rule[-0.175pt]{0.350pt}{4.818pt}}
\put(967,113){\makebox(0,0){1}}
\put(967,767){\rule[-0.175pt]{0.350pt}{4.818pt}}
\put(1084,158){\rule[-0.175pt]{0.350pt}{4.818pt}}
\put(1084,113){\makebox(0,0){1.2}}
\put(1084,767){\rule[-0.175pt]{0.350pt}{4.818pt}}
\put(1202,158){\rule[-0.175pt]{0.350pt}{4.818pt}}
\put(1202,113){\makebox(0,0){1.4}}
\put(1202,767){\rule[-0.175pt]{0.350pt}{4.818pt}}
\put(1319,158){\rule[-0.175pt]{0.350pt}{4.818pt}}
\put(1319,113){\makebox(0,0){1.6}}
\put(1319,767){\rule[-0.175pt]{0.350pt}{4.818pt}}
\put(1436,158){\rule[-0.175pt]{0.350pt}{4.818pt}}
\put(1436,113){\makebox(0,0){1.8}}
\put(1436,767){\rule[-0.175pt]{0.350pt}{4.818pt}}
\put(264,158){\rule[-0.175pt]{282.335pt}{0.350pt}}
\put(1436,158){\rule[-0.175pt]{0.350pt}{151.526pt}}
\put(264,787){\rule[-0.175pt]{282.335pt}{0.350pt}}
\put(45,472){\makebox(0,0)[l]{\shortstack{$\langle \tilde{R} \rangle$}}}
\put(850,68){\makebox(0,0){$\ell$}}
\put(264,158){\rule[-0.175pt]{0.350pt}{151.526pt}}
\sbox{\plotpoint}{\rule[-0.250pt]{0.500pt}{0.500pt}}%
\put(400,745){\makebox(0,0)[l]{$f=0$}}
\put(381,220){\usebox{\plotpoint}}
\put(381,220){\usebox{\plotpoint}}
\put(401,220){\usebox{\plotpoint}}
\put(422,220){\usebox{\plotpoint}}
\put(443,221){\usebox{\plotpoint}}
\put(464,221){\usebox{\plotpoint}}
\put(484,222){\usebox{\plotpoint}}
\put(505,224){\usebox{\plotpoint}}
\put(525,226){\usebox{\plotpoint}}
\put(546,230){\usebox{\plotpoint}}
\put(567,232){\usebox{\plotpoint}}
\put(587,235){\usebox{\plotpoint}}
\put(607,241){\usebox{\plotpoint}}
\put(626,248){\usebox{\plotpoint}}
\put(633,248){\usebox{\plotpoint}}
\put(320,745){\raisebox{-1.2pt}{\makebox(0,0){$\Diamond$}}}
\put(381,220){\raisebox{-1.2pt}{\makebox(0,0){$\Diamond$}}}
\put(477,222){\raisebox{-1.2pt}{\makebox(0,0){$\Diamond$}}}
\put(517,225){\raisebox{-1.2pt}{\makebox(0,0){$\Diamond$}}}
\put(549,231){\raisebox{-1.2pt}{\makebox(0,0){$\Diamond$}}}
\put(575,233){\raisebox{-1.2pt}{\makebox(0,0){$\Diamond$}}}
\put(600,239){\raisebox{-1.2pt}{\makebox(0,0){$\Diamond$}}}
\put(624,248){\raisebox{-1.2pt}{\makebox(0,0){$\Diamond$}}}
\put(633,248){\raisebox{-1.2pt}{\makebox(0,0){$\Diamond$}}}
\put(400,700){\makebox(0,0)[l]{$f=10^{-6}$}}
\put(381,221){\usebox{\plotpoint}}
\put(381,221){\usebox{\plotpoint}}
\put(401,221){\usebox{\plotpoint}}
\put(422,222){\usebox{\plotpoint}}
\put(443,222){\usebox{\plotpoint}}
\put(463,223){\usebox{\plotpoint}}
\put(484,224){\usebox{\plotpoint}}
\put(505,225){\usebox{\plotpoint}}
\put(526,227){\usebox{\plotpoint}}
\put(546,230){\usebox{\plotpoint}}
\put(567,232){\usebox{\plotpoint}}
\put(587,235){\usebox{\plotpoint}}
\put(608,239){\usebox{\plotpoint}}
\put(627,246){\usebox{\plotpoint}}
\put(634,250){\usebox{\plotpoint}}
\put(320,700){\raisebox{-1.2pt}{\makebox(0,0){$\Box$}}}
\put(381,221){\raisebox{-1.2pt}{\makebox(0,0){$\Box$}}}
\put(478,224){\raisebox{-1.2pt}{\makebox(0,0){$\Box$}}}
\put(518,226){\raisebox{-1.2pt}{\makebox(0,0){$\Box$}}}
\put(549,231){\raisebox{-1.2pt}{\makebox(0,0){$\Box$}}}
\put(575,233){\raisebox{-1.2pt}{\makebox(0,0){$\Box$}}}
\put(599,237){\raisebox{-1.2pt}{\makebox(0,0){$\Box$}}}
\put(621,243){\raisebox{-1.2pt}{\makebox(0,0){$\Box$}}}
\put(634,250){\raisebox{-1.2pt}{\makebox(0,0){$\Box$}}}
\put(400,655){\makebox(0,0)[l]{$f=10^{-5}$}}
\put(381,220){\usebox{\plotpoint}}
\put(381,220){\usebox{\plotpoint}}
\put(401,220){\usebox{\plotpoint}}
\put(422,221){\usebox{\plotpoint}}
\put(443,222){\usebox{\plotpoint}}
\put(463,223){\usebox{\plotpoint}}
\put(484,224){\usebox{\plotpoint}}
\put(505,225){\usebox{\plotpoint}}
\put(526,226){\usebox{\plotpoint}}
\put(546,227){\usebox{\plotpoint}}
\put(567,231){\usebox{\plotpoint}}
\put(587,235){\usebox{\plotpoint}}
\put(608,239){\usebox{\plotpoint}}
\put(628,244){\usebox{\plotpoint}}
\put(642,250){\usebox{\plotpoint}}
\put(320,655){\makebox(0,0){$\triangle$}}
\put(381,220){\makebox(0,0){$\triangle$}}
\put(478,224){\makebox(0,0){$\triangle$}}
\put(517,226){\makebox(0,0){$\triangle$}}
\put(548,228){\makebox(0,0){$\triangle$}}
\put(600,238){\makebox(0,0){$\triangle$}}
\put(620,241){\makebox(0,0){$\triangle$}}
\put(642,250){\makebox(0,0){$\triangle$}}
\put(400,610){\makebox(0,0)[l]{$f=10^{-4}$}}
\put(381,220){\usebox{\plotpoint}}
\put(381,220){\usebox{\plotpoint}}
\put(401,220){\usebox{\plotpoint}}
\put(422,221){\usebox{\plotpoint}}
\put(443,222){\usebox{\plotpoint}}
\put(463,223){\usebox{\plotpoint}}
\put(484,224){\usebox{\plotpoint}}
\put(505,225){\usebox{\plotpoint}}
\put(526,226){\usebox{\plotpoint}}
\put(546,227){\usebox{\plotpoint}}
\put(567,228){\usebox{\plotpoint}}
\put(588,229){\usebox{\plotpoint}}
\put(608,232){\usebox{\plotpoint}}
\put(629,236){\usebox{\plotpoint}}
\put(649,239){\usebox{\plotpoint}}
\put(669,244){\usebox{\plotpoint}}
\put(690,248){\usebox{\plotpoint}}
\put(708,258){\usebox{\plotpoint}}
\put(720,274){\usebox{\plotpoint}}
\put(732,291){\usebox{\plotpoint}}
\put(741,310){\usebox{\plotpoint}}
\put(750,329){\usebox{\plotpoint}}
\put(758,348){\usebox{\plotpoint}}
\put(767,366){\usebox{\plotpoint}}
\put(775,385){\usebox{\plotpoint}}
\put(784,404){\usebox{\plotpoint}}
\put(792,421){\usebox{\plotpoint}}
\put(320,610){\makebox(0,0){$\bigtriangledown$}}
\put(381,220){\makebox(0,0){$\bigtriangledown$}}
\put(595,230){\makebox(0,0){$\bigtriangledown$}}
\put(645,239){\makebox(0,0){$\bigtriangledown$}}
\put(688,248){\makebox(0,0){$\bigtriangledown$}}
\put(705,254){\makebox(0,0){$\bigtriangledown$}}
\put(733,292){\makebox(0,0){$\bigtriangledown$}}
\put(792,421){\makebox(0,0){$\bigtriangledown$}}
\put(400,565){\makebox(0,0)[l]{$f=10^{-3}$}}
\put(381,221){\usebox{\plotpoint}}
\put(381,221){\usebox{\plotpoint}}
\put(401,221){\usebox{\plotpoint}}
\put(422,222){\usebox{\plotpoint}}
\put(443,222){\usebox{\plotpoint}}
\put(463,223){\usebox{\plotpoint}}
\put(484,224){\usebox{\plotpoint}}
\put(505,224){\usebox{\plotpoint}}
\put(526,225){\usebox{\plotpoint}}
\put(546,225){\usebox{\plotpoint}}
\put(567,226){\usebox{\plotpoint}}
\put(588,227){\usebox{\plotpoint}}
\put(609,228){\usebox{\plotpoint}}
\put(629,230){\usebox{\plotpoint}}
\put(650,231){\usebox{\plotpoint}}
\put(671,233){\usebox{\plotpoint}}
\put(691,234){\usebox{\plotpoint}}
\put(712,236){\usebox{\plotpoint}}
\put(733,238){\usebox{\plotpoint}}
\put(753,240){\usebox{\plotpoint}}
\put(774,242){\usebox{\plotpoint}}
\put(795,245){\usebox{\plotpoint}}
\put(815,249){\usebox{\plotpoint}}
\put(835,254){\usebox{\plotpoint}}
\put(855,259){\usebox{\plotpoint}}
\put(873,270){\usebox{\plotpoint}}
\put(889,283){\usebox{\plotpoint}}
\put(904,297){\usebox{\plotpoint}}
\put(917,313){\usebox{\plotpoint}}
\put(931,329){\usebox{\plotpoint}}
\put(944,345){\usebox{\plotpoint}}
\put(957,361){\usebox{\plotpoint}}
\put(971,376){\usebox{\plotpoint}}
\put(985,392){\usebox{\plotpoint}}
\put(999,407){\usebox{\plotpoint}}
\put(1012,423){\usebox{\plotpoint}}
\put(1026,438){\usebox{\plotpoint}}
\put(1040,454){\usebox{\plotpoint}}
\put(1053,470){\usebox{\plotpoint}}
\put(1067,485){\usebox{\plotpoint}}
\put(1081,501){\usebox{\plotpoint}}
\put(1096,515){\usebox{\plotpoint}}
\put(1112,528){\usebox{\plotpoint}}
\put(1128,541){\usebox{\plotpoint}}
\put(1145,554){\usebox{\plotpoint}}
\put(1161,567){\usebox{\plotpoint}}
\put(1177,580){\usebox{\plotpoint}}
\put(1193,593){\usebox{\plotpoint}}
\put(1209,606){\usebox{\plotpoint}}
\put(1225,619){\usebox{\plotpoint}}
\put(1241,632){\usebox{\plotpoint}}
\put(1257,645){\usebox{\plotpoint}}
\put(1274,658){\usebox{\plotpoint}}
\put(1290,671){\usebox{\plotpoint}}
\put(1306,684){\usebox{\plotpoint}}
\put(1315,692){\usebox{\plotpoint}}
\put(320,565){\circle{24}}
\put(381,221){\circle{24}}
\put(587,227){\circle{24}}
\put(671,233){\circle{24}}
\put(738,239){\circle{24}}
\put(793,245){\circle{24}}
\put(849,257){\circle{24}}
\put(860,261){\circle{24}}
\put(878,274){\circle{24}}
\put(899,291){\circle{24}}
\put(954,357){\circle{24}}
\put(989,396){\circle{24}}
\put(1044,459){\circle{24}}
\put(1084,505){\circle{24}}
\put(1315,692){\circle{24}}
\end{picture}

\setlength{\unitlength}{0.240900pt}
\ifx\plotpoint\undefined\newsavebox{\plotpoint}\fi
\sbox{\plotpoint}{\rule[-0.175pt]{0.350pt}{0.350pt}}%
\begin{picture}(1500,900)(0,0)
\tenrm
\sbox{\plotpoint}{\rule[-0.175pt]{0.350pt}{0.350pt}}%
\put(264,158){\rule[-0.175pt]{4.818pt}{0.350pt}}
\put(242,158){\makebox(0,0)[r]{-12}}
\put(1416,158){\rule[-0.175pt]{4.818pt}{0.350pt}}
\put(264,248){\rule[-0.175pt]{4.818pt}{0.350pt}}
\put(242,248){\makebox(0,0)[r]{-10}}
\put(1416,248){\rule[-0.175pt]{4.818pt}{0.350pt}}
\put(264,338){\rule[-0.175pt]{4.818pt}{0.350pt}}
\put(242,338){\makebox(0,0)[r]{-8}}
\put(1416,338){\rule[-0.175pt]{4.818pt}{0.350pt}}
\put(264,428){\rule[-0.175pt]{4.818pt}{0.350pt}}
\put(242,428){\makebox(0,0)[r]{-6}}
\put(1416,428){\rule[-0.175pt]{4.818pt}{0.350pt}}
\put(264,517){\rule[-0.175pt]{4.818pt}{0.350pt}}
\put(242,517){\makebox(0,0)[r]{-4}}
\put(1416,517){\rule[-0.175pt]{4.818pt}{0.350pt}}
\put(264,607){\rule[-0.175pt]{4.818pt}{0.350pt}}
\put(242,607){\makebox(0,0)[r]{-2}}
\put(1416,607){\rule[-0.175pt]{4.818pt}{0.350pt}}
\put(264,697){\rule[-0.175pt]{4.818pt}{0.350pt}}
\put(242,697){\makebox(0,0)[r]{0}}
\put(1416,697){\rule[-0.175pt]{4.818pt}{0.350pt}}
\put(264,787){\rule[-0.175pt]{4.818pt}{0.350pt}}
\put(242,787){\makebox(0,0)[r]{2}}
\put(1416,787){\rule[-0.175pt]{4.818pt}{0.350pt}}
\put(381,158){\rule[-0.175pt]{0.350pt}{4.818pt}}
\put(381,113){\makebox(0,0){0}}
\put(381,767){\rule[-0.175pt]{0.350pt}{4.818pt}}
\put(616,158){\rule[-0.175pt]{0.350pt}{4.818pt}}
\put(616,113){\makebox(0,0){0.2}}
\put(616,767){\rule[-0.175pt]{0.350pt}{4.818pt}}
\put(850,158){\rule[-0.175pt]{0.350pt}{4.818pt}}
\put(850,113){\makebox(0,0){0.4}}
\put(850,767){\rule[-0.175pt]{0.350pt}{4.818pt}}
\put(1084,158){\rule[-0.175pt]{0.350pt}{4.818pt}}
\put(1084,113){\makebox(0,0){0.6}}
\put(1084,767){\rule[-0.175pt]{0.350pt}{4.818pt}}
\put(1319,158){\rule[-0.175pt]{0.350pt}{4.818pt}}
\put(1319,113){\makebox(0,0){0.8}}
\put(1319,767){\rule[-0.175pt]{0.350pt}{4.818pt}}
\put(264,158){\rule[-0.175pt]{282.335pt}{0.350pt}}
\put(1436,158){\rule[-0.175pt]{0.350pt}{151.526pt}}
\put(264,787){\rule[-0.175pt]{282.335pt}{0.350pt}}
\put(45,472){\makebox(0,0)[l]{\shortstack{$\langle \tilde{R} \rangle$}}}
\put(850,68){\makebox(0,0){$\ell$}}
\put(264,158){\rule[-0.175pt]{0.350pt}{151.526pt}}
\sbox{\plotpoint}{\rule[-0.250pt]{0.500pt}{0.500pt}}%
\put(400,745){\makebox(0,0)[l]{$f=0$}}
\put(381,242){\usebox{\plotpoint}}
\put(381,242){\usebox{\plotpoint}}
\put(401,243){\usebox{\plotpoint}}
\put(422,245){\usebox{\plotpoint}}
\put(443,246){\usebox{\plotpoint}}
\put(463,248){\usebox{\plotpoint}}
\put(484,249){\usebox{\plotpoint}}
\put(505,251){\usebox{\plotpoint}}
\put(525,252){\usebox{\plotpoint}}
\put(546,254){\usebox{\plotpoint}}
\put(567,255){\usebox{\plotpoint}}
\put(586,262){\usebox{\plotpoint}}
\put(605,272){\usebox{\plotpoint}}
\put(623,281){\usebox{\plotpoint}}
\put(642,290){\usebox{\plotpoint}}
\put(659,302){\usebox{\plotpoint}}
\put(674,316){\usebox{\plotpoint}}
\put(689,330){\usebox{\plotpoint}}
\put(703,345){\usebox{\plotpoint}}
\put(719,358){\usebox{\plotpoint}}
\put(739,364){\usebox{\plotpoint}}
\put(759,370){\usebox{\plotpoint}}
\put(774,382){\usebox{\plotpoint}}
\put(787,399){\usebox{\plotpoint}}
\put(799,416){\usebox{\plotpoint}}
\put(811,432){\usebox{\plotpoint}}
\put(822,450){\usebox{\plotpoint}}
\put(831,469){\usebox{\plotpoint}}
\put(840,487){\usebox{\plotpoint}}
\put(849,506){\usebox{\plotpoint}}
\put(859,524){\usebox{\plotpoint}}
\put(869,540){\usebox{\plotpoint}}
\put(884,540){\usebox{\plotpoint}}
\put(320,745){\raisebox{-1.2pt}{\makebox(0,0){$\Diamond$}}}
\put(381,242){\raisebox{-1.2pt}{\makebox(0,0){$\Diamond$}}}
\put(573,256){\raisebox{-1.2pt}{\makebox(0,0){$\Diamond$}}}
\put(653,296){\raisebox{-1.2pt}{\makebox(0,0){$\Diamond$}}}
\put(717,358){\raisebox{-1.2pt}{\makebox(0,0){$\Diamond$}}}
\put(768,373){\raisebox{-1.2pt}{\makebox(0,0){$\Diamond$}}}
\put(818,442){\raisebox{-1.2pt}{\makebox(0,0){$\Diamond$}}}
\put(867,541){\raisebox{-1.2pt}{\makebox(0,0){$\Diamond$}}}
\put(884,540){\raisebox{-1.2pt}{\makebox(0,0){$\Diamond$}}}
\put(400,700){\makebox(0,0)[l]{$f=10^{-6}$}}
\put(381,247){\usebox{\plotpoint}}
\put(381,247){\usebox{\plotpoint}}
\put(401,250){\usebox{\plotpoint}}
\put(421,254){\usebox{\plotpoint}}
\put(442,258){\usebox{\plotpoint}}
\put(462,262){\usebox{\plotpoint}}
\put(483,266){\usebox{\plotpoint}}
\put(503,269){\usebox{\plotpoint}}
\put(523,273){\usebox{\plotpoint}}
\put(544,277){\usebox{\plotpoint}}
\put(564,281){\usebox{\plotpoint}}
\put(584,285){\usebox{\plotpoint}}
\put(604,291){\usebox{\plotpoint}}
\put(625,296){\usebox{\plotpoint}}
\put(645,301){\usebox{\plotpoint}}
\put(663,311){\usebox{\plotpoint}}
\put(679,323){\usebox{\plotpoint}}
\put(695,336){\usebox{\plotpoint}}
\put(712,349){\usebox{\plotpoint}}
\put(730,359){\usebox{\plotpoint}}
\put(749,366){\usebox{\plotpoint}}
\put(768,374){\usebox{\plotpoint}}
\put(783,389){\usebox{\plotpoint}}
\put(798,404){\usebox{\plotpoint}}
\put(812,418){\usebox{\plotpoint}}
\put(825,435){\usebox{\plotpoint}}
\put(837,452){\usebox{\plotpoint}}
\put(849,468){\usebox{\plotpoint}}
\put(861,486){\usebox{\plotpoint}}
\put(868,505){\usebox{\plotpoint}}
\put(874,525){\usebox{\plotpoint}}
\put(881,544){\usebox{\plotpoint}}
\put(887,560){\usebox{\plotpoint}}
\put(320,700){\raisebox{-1.2pt}{\makebox(0,0){$\Box$}}}
\put(381,247){\raisebox{-1.2pt}{\makebox(0,0){$\Box$}}}
\put(574,283){\raisebox{-1.2pt}{\makebox(0,0){$\Box$}}}
\put(654,304){\raisebox{-1.2pt}{\makebox(0,0){$\Box$}}}
\put(718,354){\raisebox{-1.2pt}{\makebox(0,0){$\Box$}}}
\put(769,375){\raisebox{-1.2pt}{\makebox(0,0){$\Box$}}}
\put(817,423){\raisebox{-1.2pt}{\makebox(0,0){$\Box$}}}
\put(861,485){\raisebox{-1.2pt}{\makebox(0,0){$\Box$}}}
\put(887,560){\raisebox{-1.2pt}{\makebox(0,0){$\Box$}}}
\put(400,655){\makebox(0,0)[l]{$f=10^{-5}$}}
\put(381,237){\usebox{\plotpoint}}
\put(381,237){\usebox{\plotpoint}}
\put(401,241){\usebox{\plotpoint}}
\put(421,246){\usebox{\plotpoint}}
\put(441,251){\usebox{\plotpoint}}
\put(461,255){\usebox{\plotpoint}}
\put(482,260){\usebox{\plotpoint}}
\put(502,265){\usebox{\plotpoint}}
\put(522,269){\usebox{\plotpoint}}
\put(542,274){\usebox{\plotpoint}}
\put(562,279){\usebox{\plotpoint}}
\put(583,284){\usebox{\plotpoint}}
\put(603,289){\usebox{\plotpoint}}
\put(623,294){\usebox{\plotpoint}}
\put(643,300){\usebox{\plotpoint}}
\put(663,306){\usebox{\plotpoint}}
\put(682,313){\usebox{\plotpoint}}
\put(701,321){\usebox{\plotpoint}}
\put(719,330){\usebox{\plotpoint}}
\put(734,346){\usebox{\plotpoint}}
\put(748,361){\usebox{\plotpoint}}
\put(762,376){\usebox{\plotpoint}}
\put(776,391){\usebox{\plotpoint}}
\put(791,406){\usebox{\plotpoint}}
\put(805,421){\usebox{\plotpoint}}
\put(819,436){\usebox{\plotpoint}}
\put(836,448){\usebox{\plotpoint}}
\put(853,460){\usebox{\plotpoint}}
\put(865,476){\usebox{\plotpoint}}
\put(874,495){\usebox{\plotpoint}}
\put(882,514){\usebox{\plotpoint}}
\put(891,533){\usebox{\plotpoint}}
\put(900,552){\usebox{\plotpoint}}
\put(903,558){\usebox{\plotpoint}}
\put(320,655){\makebox(0,0){$\triangle$}}
\put(381,237){\makebox(0,0){$\triangle$}}
\put(574,282){\makebox(0,0){$\triangle$}}
\put(654,303){\makebox(0,0){$\triangle$}}
\put(716,327){\makebox(0,0){$\triangle$}}
\put(818,435){\makebox(0,0){$\triangle$}}
\put(860,465){\makebox(0,0){$\triangle$}}
\put(903,558){\makebox(0,0){$\triangle$}}
\put(400,610){\makebox(0,0)[l]{$f=10^{-4}$}}
\put(381,240){\usebox{\plotpoint}}
\put(381,240){\usebox{\plotpoint}}
\put(401,245){\usebox{\plotpoint}}
\put(421,250){\usebox{\plotpoint}}
\put(441,255){\usebox{\plotpoint}}
\put(461,260){\usebox{\plotpoint}}
\put(481,265){\usebox{\plotpoint}}
\put(501,271){\usebox{\plotpoint}}
\put(521,276){\usebox{\plotpoint}}
\put(541,281){\usebox{\plotpoint}}
\put(561,286){\usebox{\plotpoint}}
\put(581,291){\usebox{\plotpoint}}
\put(602,296){\usebox{\plotpoint}}
\put(622,302){\usebox{\plotpoint}}
\put(642,307){\usebox{\plotpoint}}
\put(662,312){\usebox{\plotpoint}}
\put(682,317){\usebox{\plotpoint}}
\put(702,322){\usebox{\plotpoint}}
\put(722,328){\usebox{\plotpoint}}
\put(742,333){\usebox{\plotpoint}}
\put(762,338){\usebox{\plotpoint}}
\put(782,343){\usebox{\plotpoint}}
\put(803,348){\usebox{\plotpoint}}
\put(819,360){\usebox{\plotpoint}}
\put(835,374){\usebox{\plotpoint}}
\put(850,388){\usebox{\plotpoint}}
\put(866,401){\usebox{\plotpoint}}
\put(881,415){\usebox{\plotpoint}}
\put(897,429){\usebox{\plotpoint}}
\put(912,443){\usebox{\plotpoint}}
\put(925,459){\usebox{\plotpoint}}
\put(938,475){\usebox{\plotpoint}}
\put(952,491){\usebox{\plotpoint}}
\put(965,507){\usebox{\plotpoint}}
\put(978,523){\usebox{\plotpoint}}
\put(991,539){\usebox{\plotpoint}}
\put(1003,556){\usebox{\plotpoint}}
\put(1013,574){\usebox{\plotpoint}}
\put(1024,592){\usebox{\plotpoint}}
\put(1028,599){\usebox{\plotpoint}}
\put(320,610){\makebox(0,0){$\bigtriangledown$}}
\put(381,240){\makebox(0,0){$\bigtriangledown$}}
\put(808,350){\makebox(0,0){$\bigtriangledown$}}
\put(908,439){\makebox(0,0){$\bigtriangledown$}}
\put(995,543){\makebox(0,0){$\bigtriangledown$}}
\put(1028,599){\makebox(0,0){$\bigtriangledown$}}
\put(400,565){\makebox(0,0)[l]{$f=10^{-3}$}}
\put(381,245){\usebox{\plotpoint}}
\put(381,245){\usebox{\plotpoint}}
\put(401,248){\usebox{\plotpoint}}
\put(421,252){\usebox{\plotpoint}}
\put(442,255){\usebox{\plotpoint}}
\put(462,259){\usebox{\plotpoint}}
\put(483,262){\usebox{\plotpoint}}
\put(503,266){\usebox{\plotpoint}}
\put(524,269){\usebox{\plotpoint}}
\put(544,273){\usebox{\plotpoint}}
\put(565,276){\usebox{\plotpoint}}
\put(585,280){\usebox{\plotpoint}}
\put(605,283){\usebox{\plotpoint}}
\put(626,287){\usebox{\plotpoint}}
\put(646,290){\usebox{\plotpoint}}
\put(667,294){\usebox{\plotpoint}}
\put(687,297){\usebox{\plotpoint}}
\put(708,301){\usebox{\plotpoint}}
\put(728,304){\usebox{\plotpoint}}
\put(749,308){\usebox{\plotpoint}}
\put(769,311){\usebox{\plotpoint}}
\put(790,315){\usebox{\plotpoint}}
\put(809,321){\usebox{\plotpoint}}
\put(829,328){\usebox{\plotpoint}}
\put(848,335){\usebox{\plotpoint}}
\put(868,342){\usebox{\plotpoint}}
\put(888,349){\usebox{\plotpoint}}
\put(907,356){\usebox{\plotpoint}}
\put(927,363){\usebox{\plotpoint}}
\put(946,370){\usebox{\plotpoint}}
\put(966,377){\usebox{\plotpoint}}
\put(985,386){\usebox{\plotpoint}}
\put(1003,394){\usebox{\plotpoint}}
\put(1022,403){\usebox{\plotpoint}}
\put(1041,412){\usebox{\plotpoint}}
\put(1060,420){\usebox{\plotpoint}}
\put(1079,429){\usebox{\plotpoint}}
\put(1098,438){\usebox{\plotpoint}}
\put(1115,449){\usebox{\plotpoint}}
\put(1133,459){\usebox{\plotpoint}}
\put(1151,470){\usebox{\plotpoint}}
\put(1169,481){\usebox{\plotpoint}}
\put(1187,491){\usebox{\plotpoint}}
\put(1204,502){\usebox{\plotpoint}}
\put(1218,518){\usebox{\plotpoint}}
\put(1231,534){\usebox{\plotpoint}}
\put(1245,549){\usebox{\plotpoint}}
\put(1258,565){\usebox{\plotpoint}}
\put(1272,581){\usebox{\plotpoint}}
\put(1285,597){\usebox{\plotpoint}}
\put(1299,612){\usebox{\plotpoint}}
\put(1313,628){\usebox{\plotpoint}}
\put(1323,646){\usebox{\plotpoint}}
\put(1334,664){\usebox{\plotpoint}}
\put(1339,673){\usebox{\plotpoint}}
\put(320,565){\circle{24}}
\put(381,245){\circle{24}}
\put(793,316){\circle{24}}
\put(960,375){\circle{24}}
\put(1094,436){\circle{24}}
\put(1204,502){\circle{24}}
\put(1316,632){\circle{24}}
\put(1339,673){\circle{24}}
\end{picture}

  { \small
  Fig. 1. Expectation value $\langle \tilde{R} \rangle$ within the Regge 
  approach as a function of $\ell$ for
  decreasing lower limit $f$ of the fatness of each 4-simplex.
  Computations were performed on a regular triangulation
  of the 4-torus with $4^4$ vertices using the uniform measure ($\sigma = 1$).
  The upper plot gives an overview of all results available whereas the lower
  plot magnifies the well-defined phase. The existence of a well-defined
  phase and the convergence of $\langle \tilde{R} \rangle$ for $f \to 0$ are
  clearly seen. The transition point to positive curvature
  is shifted from $\ell_c \approx 0.8$ for $f = 10^{-3}$ down to
  $\ell_c \approx 0.45$ for $f \leq 10^{-5}$.
  }
\end{figure}

\newpage
\begin{figure}[p]

\setlength{\unitlength}{0.240900pt}
\ifx\plotpoint\undefined\newsavebox{\plotpoint}\fi
\sbox{\plotpoint}{\rule[-0.175pt]{0.350pt}{0.350pt}}%
\begin{picture}(1500,900)(0,0)
\tenrm
\sbox{\plotpoint}{\rule[-0.175pt]{0.350pt}{0.350pt}}%
\put(264,158){\rule[-0.175pt]{282.335pt}{0.350pt}}
\put(264,158){\rule[-0.175pt]{4.818pt}{0.350pt}}
\put(1416,158){\rule[-0.175pt]{4.818pt}{0.350pt}}
\put(264,237){\rule[-0.175pt]{4.818pt}{0.350pt}}
\put(242,237){\makebox(0,0)[r]{0.0025}}
\put(1416,237){\rule[-0.175pt]{4.818pt}{0.350pt}}
\put(264,315){\rule[-0.175pt]{4.818pt}{0.350pt}}
\put(242,315){\makebox(0,0)[r]{0.0050}}
\put(1416,315){\rule[-0.175pt]{4.818pt}{0.350pt}}
\put(264,394){\rule[-0.175pt]{4.818pt}{0.350pt}}
\put(242,394){\makebox(0,0)[r]{0.0075}}
\put(1416,394){\rule[-0.175pt]{4.818pt}{0.350pt}}
\put(264,473){\rule[-0.175pt]{4.818pt}{0.350pt}}
\put(242,473){\makebox(0,0)[r]{0.0100}}
\put(1416,473){\rule[-0.175pt]{4.818pt}{0.350pt}}
\put(264,551){\rule[-0.175pt]{4.818pt}{0.350pt}}
\put(242,551){\makebox(0,0)[r]{0.0125}}
\put(1416,551){\rule[-0.175pt]{4.818pt}{0.350pt}}
\put(264,630){\rule[-0.175pt]{4.818pt}{0.350pt}}
\put(242,630){\makebox(0,0)[r]{0.0150}}
\put(1416,630){\rule[-0.175pt]{4.818pt}{0.350pt}}
\put(264,708){\rule[-0.175pt]{4.818pt}{0.350pt}}
\put(242,708){\makebox(0,0)[r]{0.0175}}
\put(1416,708){\rule[-0.175pt]{4.818pt}{0.350pt}}
\put(264,787){\rule[-0.175pt]{4.818pt}{0.350pt}}
\put(1416,787){\rule[-0.175pt]{4.818pt}{0.350pt}}
\put(264,158){\rule[-0.175pt]{0.350pt}{4.818pt}}
\put(264,113){\makebox(0,0){-0.2}}
\put(264,767){\rule[-0.175pt]{0.350pt}{4.818pt}}
\put(381,158){\rule[-0.175pt]{0.350pt}{4.818pt}}
\put(381,113){\makebox(0,0){0}}
\put(381,767){\rule[-0.175pt]{0.350pt}{4.818pt}}
\put(498,158){\rule[-0.175pt]{0.350pt}{4.818pt}}
\put(498,113){\makebox(0,0){0.2}}
\put(498,767){\rule[-0.175pt]{0.350pt}{4.818pt}}
\put(616,158){\rule[-0.175pt]{0.350pt}{4.818pt}}
\put(616,113){\makebox(0,0){0.4}}
\put(616,767){\rule[-0.175pt]{0.350pt}{4.818pt}}
\put(733,158){\rule[-0.175pt]{0.350pt}{4.818pt}}
\put(733,113){\makebox(0,0){0.6}}
\put(733,767){\rule[-0.175pt]{0.350pt}{4.818pt}}
\put(850,158){\rule[-0.175pt]{0.350pt}{4.818pt}}
\put(850,113){\makebox(0,0){0.8}}
\put(850,767){\rule[-0.175pt]{0.350pt}{4.818pt}}
\put(967,158){\rule[-0.175pt]{0.350pt}{4.818pt}}
\put(967,113){\makebox(0,0){1}}
\put(967,767){\rule[-0.175pt]{0.350pt}{4.818pt}}
\put(1084,158){\rule[-0.175pt]{0.350pt}{4.818pt}}
\put(1084,113){\makebox(0,0){1.2}}
\put(1084,767){\rule[-0.175pt]{0.350pt}{4.818pt}}
\put(1202,158){\rule[-0.175pt]{0.350pt}{4.818pt}}
\put(1202,113){\makebox(0,0){1.4}}
\put(1202,767){\rule[-0.175pt]{0.350pt}{4.818pt}}
\put(1319,158){\rule[-0.175pt]{0.350pt}{4.818pt}}
\put(1319,113){\makebox(0,0){1.6}}
\put(1319,767){\rule[-0.175pt]{0.350pt}{4.818pt}}
\put(1436,158){\rule[-0.175pt]{0.350pt}{4.818pt}}
\put(1436,113){\makebox(0,0){1.8}}
\put(1436,767){\rule[-0.175pt]{0.350pt}{4.818pt}}
\put(264,158){\rule[-0.175pt]{282.335pt}{0.350pt}}
\put(1436,158){\rule[-0.175pt]{0.350pt}{151.526pt}}
\put(264,787){\rule[-0.175pt]{282.335pt}{0.350pt}}
\put(15,472){\makebox(0,0)[l]{\shortstack{$\langle \phi_s \rangle $}}}
\put(850,68){\makebox(0,0){$\ell$}}
\put(264,158){\rule[-0.175pt]{0.350pt}{151.526pt}}
\sbox{\plotpoint}{\rule[-0.250pt]{0.500pt}{0.500pt}}%
\put(1106,722){\makebox(0,0)[l]{$f=0$}}
\put(381,507){\usebox{\plotpoint}}
\put(381,507){\usebox{\plotpoint}}
\put(401,508){\usebox{\plotpoint}}
\put(422,510){\usebox{\plotpoint}}
\put(443,511){\usebox{\plotpoint}}
\put(463,513){\usebox{\plotpoint}}
\put(484,511){\usebox{\plotpoint}}
\put(503,504){\usebox{\plotpoint}}
\put(522,496){\usebox{\plotpoint}}
\put(539,484){\usebox{\plotpoint}}
\put(557,477){\usebox{\plotpoint}}
\put(577,475){\usebox{\plotpoint}}
\put(591,460){\usebox{\plotpoint}}
\put(606,446){\usebox{\plotpoint}}
\put(622,432){\usebox{\plotpoint}}
\put(633,427){\usebox{\plotpoint}}
\put(1350,722){\raisebox{-1.2pt}{\makebox(0,0){$\Diamond$}}}
\put(381,507){\raisebox{-1.2pt}{\makebox(0,0){$\Diamond$}}}
\put(477,514){\raisebox{-1.2pt}{\makebox(0,0){$\Diamond$}}}
\put(517,500){\raisebox{-1.2pt}{\makebox(0,0){$\Diamond$}}}
\put(549,477){\raisebox{-1.2pt}{\makebox(0,0){$\Diamond$}}}
\put(575,478){\raisebox{-1.2pt}{\makebox(0,0){$\Diamond$}}}
\put(600,452){\raisebox{-1.2pt}{\makebox(0,0){$\Diamond$}}}
\put(624,431){\raisebox{-1.2pt}{\makebox(0,0){$\Diamond$}}}
\put(633,427){\raisebox{-1.2pt}{\makebox(0,0){$\Diamond$}}}
\put(1106,677){\makebox(0,0)[l]{$f=10^{-6}$}}
\put(381,507){\usebox{\plotpoint}}
\put(381,507){\usebox{\plotpoint}}
\put(401,504){\usebox{\plotpoint}}
\put(422,502){\usebox{\plotpoint}}
\put(442,499){\usebox{\plotpoint}}
\put(463,497){\usebox{\plotpoint}}
\put(484,496){\usebox{\plotpoint}}
\put(504,496){\usebox{\plotpoint}}
\put(524,492){\usebox{\plotpoint}}
\put(541,480){\usebox{\plotpoint}}
\put(560,475){\usebox{\plotpoint}}
\put(580,471){\usebox{\plotpoint}}
\put(597,460){\usebox{\plotpoint}}
\put(613,447){\usebox{\plotpoint}}
\put(626,431){\usebox{\plotpoint}}
\put(634,417){\usebox{\plotpoint}}
\put(1350,677){\raisebox{-1.2pt}{\makebox(0,0){$\Box$}}}
\put(381,507){\raisebox{-1.2pt}{\makebox(0,0){$\Box$}}}
\put(478,496){\raisebox{-1.2pt}{\makebox(0,0){$\Box$}}}
\put(518,497){\raisebox{-1.2pt}{\makebox(0,0){$\Box$}}}
\put(549,475){\raisebox{-1.2pt}{\makebox(0,0){$\Box$}}}
\put(575,475){\raisebox{-1.2pt}{\makebox(0,0){$\Box$}}}
\put(599,460){\raisebox{-1.2pt}{\makebox(0,0){$\Box$}}}
\put(621,441){\raisebox{-1.2pt}{\makebox(0,0){$\Box$}}}
\put(634,417){\raisebox{-1.2pt}{\makebox(0,0){$\Box$}}}
\put(1106,632){\makebox(0,0)[l]{$f=10^{-5}$}}
\put(381,514){\usebox{\plotpoint}}
\put(381,514){\usebox{\plotpoint}}
\put(401,511){\usebox{\plotpoint}}
\put(422,508){\usebox{\plotpoint}}
\put(442,506){\usebox{\plotpoint}}
\put(463,503){\usebox{\plotpoint}}
\put(484,501){\usebox{\plotpoint}}
\put(504,501){\usebox{\plotpoint}}
\put(525,499){\usebox{\plotpoint}}
\put(545,494){\usebox{\plotpoint}}
\put(562,483){\usebox{\plotpoint}}
\put(579,471){\usebox{\plotpoint}}
\put(596,459){\usebox{\plotpoint}}
\put(615,451){\usebox{\plotpoint}}
\put(630,437){\usebox{\plotpoint}}
\put(643,421){\usebox{\plotpoint}}
\put(657,405){\usebox{\plotpoint}}
\put(670,389){\usebox{\plotpoint}}
\put(683,373){\usebox{\plotpoint}}
\put(696,357){\usebox{\plotpoint}}
\put(710,341){\usebox{\plotpoint}}
\put(723,325){\usebox{\plotpoint}}
\put(736,309){\usebox{\plotpoint}}
\put(749,293){\usebox{\plotpoint}}
\put(762,277){\usebox{\plotpoint}}
\put(776,261){\usebox{\plotpoint}}
\put(789,245){\usebox{\plotpoint}}
\put(802,229){\usebox{\plotpoint}}
\put(815,213){\usebox{\plotpoint}}
\put(828,199){\usebox{\plotpoint}}
\put(1350,632){\makebox(0,0){$\triangle$}}
\put(381,514){\makebox(0,0){$\triangle$}}
\put(478,502){\makebox(0,0){$\triangle$}}
\put(517,501){\makebox(0,0){$\triangle$}}
\put(548,494){\makebox(0,0){$\triangle$}}
\put(600,457){\makebox(0,0){$\triangle$}}
\put(620,450){\makebox(0,0){$\triangle$}}
\put(642,424){\makebox(0,0){$\triangle$}}
\put(828,199){\makebox(0,0){$\triangle$}}
\put(1106,587){\makebox(0,0)[l]{$f=10^{-4}$}}
\put(381,536){\usebox{\plotpoint}}
\put(381,536){\usebox{\plotpoint}}
\put(401,534){\usebox{\plotpoint}}
\put(422,533){\usebox{\plotpoint}}
\put(443,531){\usebox{\plotpoint}}
\put(463,530){\usebox{\plotpoint}}
\put(484,529){\usebox{\plotpoint}}
\put(505,527){\usebox{\plotpoint}}
\put(525,526){\usebox{\plotpoint}}
\put(546,525){\usebox{\plotpoint}}
\put(567,523){\usebox{\plotpoint}}
\put(588,522){\usebox{\plotpoint}}
\put(606,514){\usebox{\plotpoint}}
\put(624,503){\usebox{\plotpoint}}
\put(642,492){\usebox{\plotpoint}}
\put(657,478){\usebox{\plotpoint}}
\put(672,464){\usebox{\plotpoint}}
\put(687,449){\usebox{\plotpoint}}
\put(701,434){\usebox{\plotpoint}}
\put(710,416){\usebox{\plotpoint}}
\put(717,397){\usebox{\plotpoint}}
\put(724,377){\usebox{\plotpoint}}
\put(731,358){\usebox{\plotpoint}}
\put(742,340){\usebox{\plotpoint}}
\put(752,322){\usebox{\plotpoint}}
\put(763,304){\usebox{\plotpoint}}
\put(774,286){\usebox{\plotpoint}}
\put(784,268){\usebox{\plotpoint}}
\put(792,257){\usebox{\plotpoint}}
\put(1350,587){\makebox(0,0){$\bigtriangledown$}}
\put(381,536){\makebox(0,0){$\bigtriangledown$}}
\put(595,522){\makebox(0,0){$\bigtriangledown$}}
\put(645,491){\makebox(0,0){$\bigtriangledown$}}
\put(688,449){\makebox(0,0){$\bigtriangledown$}}
\put(705,431){\makebox(0,0){$\bigtriangledown$}}
\put(733,355){\makebox(0,0){$\bigtriangledown$}}
\put(792,257){\makebox(0,0){$\bigtriangledown$}}
\put(1106,542){\makebox(0,0)[l]{$f=10^{-3}$}}
\put(381,689){\usebox{\plotpoint}}
\put(381,689){\usebox{\plotpoint}}
\put(401,688){\usebox{\plotpoint}}
\put(422,688){\usebox{\plotpoint}}
\put(443,688){\usebox{\plotpoint}}
\put(464,688){\usebox{\plotpoint}}
\put(484,687){\usebox{\plotpoint}}
\put(505,687){\usebox{\plotpoint}}
\put(526,687){\usebox{\plotpoint}}
\put(547,687){\usebox{\plotpoint}}
\put(567,687){\usebox{\plotpoint}}
\put(588,686){\usebox{\plotpoint}}
\put(609,684){\usebox{\plotpoint}}
\put(629,682){\usebox{\plotpoint}}
\put(650,680){\usebox{\plotpoint}}
\put(671,677){\usebox{\plotpoint}}
\put(690,670){\usebox{\plotpoint}}
\put(710,663){\usebox{\plotpoint}}
\put(729,656){\usebox{\plotpoint}}
\put(749,649){\usebox{\plotpoint}}
\put(768,642){\usebox{\plotpoint}}
\put(787,634){\usebox{\plotpoint}}
\put(803,621){\usebox{\plotpoint}}
\put(818,607){\usebox{\plotpoint}}
\put(832,592){\usebox{\plotpoint}}
\put(846,577){\usebox{\plotpoint}}
\put(856,558){\usebox{\plotpoint}}
\put(865,540){\usebox{\plotpoint}}
\put(873,521){\usebox{\plotpoint}}
\put(882,502){\usebox{\plotpoint}}
\put(891,483){\usebox{\plotpoint}}
\put(900,464){\usebox{\plotpoint}}
\put(910,446){\usebox{\plotpoint}}
\put(921,428){\usebox{\plotpoint}}
\put(931,410){\usebox{\plotpoint}}
\put(941,392){\usebox{\plotpoint}}
\put(952,375){\usebox{\plotpoint}}
\put(965,359){\usebox{\plotpoint}}
\put(980,344){\usebox{\plotpoint}}
\put(995,330){\usebox{\plotpoint}}
\put(1013,319){\usebox{\plotpoint}}
\put(1030,308){\usebox{\plotpoint}}
\put(1048,298){\usebox{\plotpoint}}
\put(1068,291){\usebox{\plotpoint}}
\put(1087,285){\usebox{\plotpoint}}
\put(1108,281){\usebox{\plotpoint}}
\put(1128,276){\usebox{\plotpoint}}
\put(1148,272){\usebox{\plotpoint}}
\put(1169,268){\usebox{\plotpoint}}
\put(1189,264){\usebox{\plotpoint}}
\put(1209,260){\usebox{\plotpoint}}
\put(1230,256){\usebox{\plotpoint}}
\put(1250,252){\usebox{\plotpoint}}
\put(1271,247){\usebox{\plotpoint}}
\put(1291,243){\usebox{\plotpoint}}
\put(1311,239){\usebox{\plotpoint}}
\put(1315,239){\usebox{\plotpoint}}
\put(1350,542){\circle{24}}
\put(381,689){\circle{24}}
\put(587,687){\circle{24}}
\put(671,678){\circle{24}}
\put(738,654){\circle{24}}
\put(793,633){\circle{24}}
\put(849,575){\circle{24}}
\put(899,467){\circle{24}}
\put(954,372){\circle{24}}
\put(989,335){\circle{24}}
\put(1044,300){\circle{24}}
\put(1084,286){\circle{24}}
\put(1315,239){\circle{24}}
\end{picture}

  { \small
  Fig. 2. Average fatness $\langle \phi_s \rangle$ versus 
  $\ell$ for decreasing cutoff $f$ for the fatness of each 4-simplex.
  The value of $\langle \phi_s \rangle$ stays much larger than
  $f$ in the well-defined phase and convergence for
  $f \to 0$ is seen. With the transition to large curvatures
  $\langle \phi_s \rangle$ decreases suddenly indicating the
  collapse of 4-simplices into degenerate configurations.
  Configurations and statistics are the same as in Fig. 1.
  }
\end{figure}

\newpage
\begin{figure}[p]

\setlength{\unitlength}{0.240900pt}
\ifx\plotpoint\undefined\newsavebox{\plotpoint}\fi
\sbox{\plotpoint}{\rule[-0.175pt]{0.350pt}{0.350pt}}%
\begin{picture}(1500,900)(0,0)
\tenrm
\sbox{\plotpoint}{\rule[-0.175pt]{0.350pt}{0.350pt}}%
\put(264,158){\rule[-0.175pt]{282.335pt}{0.350pt}}
\put(264,158){\rule[-0.175pt]{4.818pt}{0.350pt}}
\put(242,158){\makebox(0,0)[r]{0}}
\put(1416,158){\rule[-0.175pt]{4.818pt}{0.350pt}}
\put(264,242){\rule[-0.175pt]{4.818pt}{0.350pt}}
\put(242,242){\makebox(0,0)[r]{2}}
\put(1416,242){\rule[-0.175pt]{4.818pt}{0.350pt}}
\put(264,326){\rule[-0.175pt]{4.818pt}{0.350pt}}
\put(242,326){\makebox(0,0)[r]{4}}
\put(1416,326){\rule[-0.175pt]{4.818pt}{0.350pt}}
\put(264,410){\rule[-0.175pt]{4.818pt}{0.350pt}}
\put(242,410){\makebox(0,0)[r]{6}}
\put(1416,410){\rule[-0.175pt]{4.818pt}{0.350pt}}
\put(264,493){\rule[-0.175pt]{4.818pt}{0.350pt}}
\put(242,493){\makebox(0,0)[r]{8}}
\put(1416,493){\rule[-0.175pt]{4.818pt}{0.350pt}}
\put(264,577){\rule[-0.175pt]{4.818pt}{0.350pt}}
\put(242,577){\makebox(0,0)[r]{10}}
\put(1416,577){\rule[-0.175pt]{4.818pt}{0.350pt}}
\put(264,661){\rule[-0.175pt]{4.818pt}{0.350pt}}
\put(242,661){\makebox(0,0)[r]{12}}
\put(1416,661){\rule[-0.175pt]{4.818pt}{0.350pt}}
\put(264,745){\rule[-0.175pt]{4.818pt}{0.350pt}}
\put(242,745){\makebox(0,0)[r]{14}}
\put(1416,745){\rule[-0.175pt]{4.818pt}{0.350pt}}
\put(264,158){\rule[-0.175pt]{0.350pt}{4.818pt}}
\put(264,113){\makebox(0,0){-0.2}}
\put(264,767){\rule[-0.175pt]{0.350pt}{4.818pt}}
\put(381,158){\rule[-0.175pt]{0.350pt}{4.818pt}}
\put(381,113){\makebox(0,0){0}}
\put(381,767){\rule[-0.175pt]{0.350pt}{4.818pt}}
\put(498,158){\rule[-0.175pt]{0.350pt}{4.818pt}}
\put(498,113){\makebox(0,0){0.2}}
\put(498,767){\rule[-0.175pt]{0.350pt}{4.818pt}}
\put(616,158){\rule[-0.175pt]{0.350pt}{4.818pt}}
\put(616,113){\makebox(0,0){0.4}}
\put(616,767){\rule[-0.175pt]{0.350pt}{4.818pt}}
\put(733,158){\rule[-0.175pt]{0.350pt}{4.818pt}}
\put(733,113){\makebox(0,0){0.6}}
\put(733,767){\rule[-0.175pt]{0.350pt}{4.818pt}}
\put(850,158){\rule[-0.175pt]{0.350pt}{4.818pt}}
\put(850,113){\makebox(0,0){0.8}}
\put(850,767){\rule[-0.175pt]{0.350pt}{4.818pt}}
\put(967,158){\rule[-0.175pt]{0.350pt}{4.818pt}}
\put(967,113){\makebox(0,0){1}}
\put(967,767){\rule[-0.175pt]{0.350pt}{4.818pt}}
\put(1084,158){\rule[-0.175pt]{0.350pt}{4.818pt}}
\put(1084,113){\makebox(0,0){1.2}}
\put(1084,767){\rule[-0.175pt]{0.350pt}{4.818pt}}
\put(1202,158){\rule[-0.175pt]{0.350pt}{4.818pt}}
\put(1202,113){\makebox(0,0){1.4}}
\put(1202,767){\rule[-0.175pt]{0.350pt}{4.818pt}}
\put(1319,158){\rule[-0.175pt]{0.350pt}{4.818pt}}
\put(1319,113){\makebox(0,0){1.6}}
\put(1319,767){\rule[-0.175pt]{0.350pt}{4.818pt}}
\put(1436,158){\rule[-0.175pt]{0.350pt}{4.818pt}}
\put(1436,113){\makebox(0,0){1.8}}
\put(1436,767){\rule[-0.175pt]{0.350pt}{4.818pt}}
\put(264,158){\rule[-0.175pt]{282.335pt}{0.350pt}}
\put(1436,158){\rule[-0.175pt]{0.350pt}{151.526pt}}
\put(264,787){\rule[-0.175pt]{282.335pt}{0.350pt}}
\put(45,472){\makebox(0,0)[l]{\shortstack{$\langle \delta_t^2 \rangle$}}}
\put(850,68){\makebox(0,0){$\ell$}}
\put(264,158){\rule[-0.175pt]{0.350pt}{151.526pt}}
\sbox{\plotpoint}{\rule[-0.250pt]{0.500pt}{0.500pt}}%
\put(381,295){\usebox{\plotpoint}}
\put(381,295){\usebox{\plotpoint}}
\put(401,294){\usebox{\plotpoint}}
\put(422,293){\usebox{\plotpoint}}
\put(443,293){\usebox{\plotpoint}}
\put(463,292){\usebox{\plotpoint}}
\put(484,291){\usebox{\plotpoint}}
\put(505,289){\usebox{\plotpoint}}
\put(526,288){\usebox{\plotpoint}}
\put(546,288){\usebox{\plotpoint}}
\put(567,285){\usebox{\plotpoint}}
\put(588,285){\usebox{\plotpoint}}
\put(608,285){\usebox{\plotpoint}}
\put(629,284){\usebox{\plotpoint}}
\put(633,284){\usebox{\plotpoint}}
\put(381,295){\raisebox{-1.2pt}{\makebox(0,0){$\Diamond$}}}
\put(477,292){\raisebox{-1.2pt}{\makebox(0,0){$\Diamond$}}}
\put(517,289){\raisebox{-1.2pt}{\makebox(0,0){$\Diamond$}}}
\put(549,288){\raisebox{-1.2pt}{\makebox(0,0){$\Diamond$}}}
\put(575,285){\raisebox{-1.2pt}{\makebox(0,0){$\Diamond$}}}
\put(600,285){\raisebox{-1.2pt}{\makebox(0,0){$\Diamond$}}}
\put(624,285){\raisebox{-1.2pt}{\makebox(0,0){$\Diamond$}}}
\put(633,284){\raisebox{-1.2pt}{\makebox(0,0){$\Diamond$}}}
\put(381,294){\usebox{\plotpoint}}
\put(381,294){\usebox{\plotpoint}}
\put(401,293){\usebox{\plotpoint}}
\put(422,293){\usebox{\plotpoint}}
\put(443,292){\usebox{\plotpoint}}
\put(464,292){\usebox{\plotpoint}}
\put(484,291){\usebox{\plotpoint}}
\put(505,289){\usebox{\plotpoint}}
\put(526,288){\usebox{\plotpoint}}
\put(546,287){\usebox{\plotpoint}}
\put(567,285){\usebox{\plotpoint}}
\put(588,284){\usebox{\plotpoint}}
\put(609,284){\usebox{\plotpoint}}
\put(629,283){\usebox{\plotpoint}}
\put(634,283){\usebox{\plotpoint}}
\put(381,294){\raisebox{-1.2pt}{\makebox(0,0){$\Box$}}}
\put(478,292){\raisebox{-1.2pt}{\makebox(0,0){$\Box$}}}
\put(518,289){\raisebox{-1.2pt}{\makebox(0,0){$\Box$}}}
\put(549,287){\raisebox{-1.2pt}{\makebox(0,0){$\Box$}}}
\put(575,285){\raisebox{-1.2pt}{\makebox(0,0){$\Box$}}}
\put(599,284){\raisebox{-1.2pt}{\makebox(0,0){$\Box$}}}
\put(621,284){\raisebox{-1.2pt}{\makebox(0,0){$\Box$}}}
\put(634,283){\raisebox{-1.2pt}{\makebox(0,0){$\Box$}}}
\put(381,294){\usebox{\plotpoint}}
\put(381,294){\usebox{\plotpoint}}
\put(401,293){\usebox{\plotpoint}}
\put(422,292){\usebox{\plotpoint}}
\put(443,292){\usebox{\plotpoint}}
\put(463,291){\usebox{\plotpoint}}
\put(484,290){\usebox{\plotpoint}}
\put(505,288){\usebox{\plotpoint}}
\put(526,287){\usebox{\plotpoint}}
\put(546,285){\usebox{\plotpoint}}
\put(567,284){\usebox{\plotpoint}}
\put(588,284){\usebox{\plotpoint}}
\put(608,283){\usebox{\plotpoint}}
\put(629,282){\usebox{\plotpoint}}
\put(650,284){\usebox{\plotpoint}}
\put(669,290){\usebox{\plotpoint}}
\put(689,296){\usebox{\plotpoint}}
\put(709,302){\usebox{\plotpoint}}
\put(729,307){\usebox{\plotpoint}}
\put(749,313){\usebox{\plotpoint}}
\put(769,319){\usebox{\plotpoint}}
\put(789,325){\usebox{\plotpoint}}
\put(809,331){\usebox{\plotpoint}}
\put(828,337){\usebox{\plotpoint}}
\put(381,294){\makebox(0,0){$\triangle$}}
\put(478,291){\makebox(0,0){$\triangle$}}
\put(517,288){\makebox(0,0){$\triangle$}}
\put(548,285){\makebox(0,0){$\triangle$}}
\put(600,284){\makebox(0,0){$\triangle$}}
\put(620,282){\makebox(0,0){$\triangle$}}
\put(642,282){\makebox(0,0){$\triangle$}}
\put(828,337){\makebox(0,0){$\triangle$}}
\put(381,291){\usebox{\plotpoint}}
\put(381,291){\usebox{\plotpoint}}
\put(401,289){\usebox{\plotpoint}}
\put(422,288){\usebox{\plotpoint}}
\put(443,287){\usebox{\plotpoint}}
\put(463,286){\usebox{\plotpoint}}
\put(484,285){\usebox{\plotpoint}}
\put(505,284){\usebox{\plotpoint}}
\put(526,282){\usebox{\plotpoint}}
\put(546,281){\usebox{\plotpoint}}
\put(567,280){\usebox{\plotpoint}}
\put(588,279){\usebox{\plotpoint}}
\put(608,277){\usebox{\plotpoint}}
\put(629,276){\usebox{\plotpoint}}
\put(650,274){\usebox{\plotpoint}}
\put(671,273){\usebox{\plotpoint}}
\put(691,272){\usebox{\plotpoint}}
\put(712,275){\usebox{\plotpoint}}
\put(731,282){\usebox{\plotpoint}}
\put(751,289){\usebox{\plotpoint}}
\put(770,296){\usebox{\plotpoint}}
\put(790,303){\usebox{\plotpoint}}
\put(792,304){\usebox{\plotpoint}}
\put(381,291){\makebox(0,0){$\bigtriangledown$}}
\put(595,279){\makebox(0,0){$\bigtriangledown$}}
\put(645,275){\makebox(0,0){$\bigtriangledown$}}
\put(688,272){\makebox(0,0){$\bigtriangledown$}}
\put(705,273){\makebox(0,0){$\bigtriangledown$}}
\put(733,283){\makebox(0,0){$\bigtriangledown$}}
\put(792,304){\makebox(0,0){$\bigtriangledown$}}
\put(381,280){\usebox{\plotpoint}}
\put(381,280){\usebox{\plotpoint}}
\put(401,278){\usebox{\plotpoint}}
\put(422,277){\usebox{\plotpoint}}
\put(443,276){\usebox{\plotpoint}}
\put(463,275){\usebox{\plotpoint}}
\put(484,273){\usebox{\plotpoint}}
\put(505,272){\usebox{\plotpoint}}
\put(526,271){\usebox{\plotpoint}}
\put(546,270){\usebox{\plotpoint}}
\put(567,269){\usebox{\plotpoint}}
\put(588,267){\usebox{\plotpoint}}
\put(608,265){\usebox{\plotpoint}}
\put(629,263){\usebox{\plotpoint}}
\put(650,261){\usebox{\plotpoint}}
\put(670,259){\usebox{\plotpoint}}
\put(691,256){\usebox{\plotpoint}}
\put(712,254){\usebox{\plotpoint}}
\put(732,252){\usebox{\plotpoint}}
\put(753,250){\usebox{\plotpoint}}
\put(774,248){\usebox{\plotpoint}}
\put(794,246){\usebox{\plotpoint}}
\put(815,246){\usebox{\plotpoint}}
\put(836,246){\usebox{\plotpoint}}
\put(856,247){\usebox{\plotpoint}}
\put(876,255){\usebox{\plotpoint}}
\put(894,264){\usebox{\plotpoint}}
\put(908,279){\usebox{\plotpoint}}
\put(921,295){\usebox{\plotpoint}}
\put(934,311){\usebox{\plotpoint}}
\put(947,327){\usebox{\plotpoint}}
\put(961,343){\usebox{\plotpoint}}
\put(975,358){\usebox{\plotpoint}}
\put(989,373){\usebox{\plotpoint}}
\put(1002,390){\usebox{\plotpoint}}
\put(1014,406){\usebox{\plotpoint}}
\put(1027,423){\usebox{\plotpoint}}
\put(1039,439){\usebox{\plotpoint}}
\put(1052,456){\usebox{\plotpoint}}
\put(1064,473){\usebox{\plotpoint}}
\put(1076,490){\usebox{\plotpoint}}
\put(1090,505){\usebox{\plotpoint}}
\put(1105,519){\usebox{\plotpoint}}
\put(1121,533){\usebox{\plotpoint}}
\put(1136,546){\usebox{\plotpoint}}
\put(1152,560){\usebox{\plotpoint}}
\put(1167,574){\usebox{\plotpoint}}
\put(1183,588){\usebox{\plotpoint}}
\put(1198,602){\usebox{\plotpoint}}
\put(1214,615){\usebox{\plotpoint}}
\put(1230,629){\usebox{\plotpoint}}
\put(1245,643){\usebox{\plotpoint}}
\put(1261,657){\usebox{\plotpoint}}
\put(1276,670){\usebox{\plotpoint}}
\put(1292,684){\usebox{\plotpoint}}
\put(1307,698){\usebox{\plotpoint}}
\put(1315,705){\usebox{\plotpoint}}
\put(381,280){\circle{24}}
\put(587,268){\circle{24}}
\put(671,259){\circle{24}}
\put(738,252){\circle{24}}
\put(793,247){\circle{24}}
\put(849,246){\circle{24}}
\put(860,248){\circle{24}}
\put(878,256){\circle{24}}
\put(899,267){\circle{24}}
\put(954,336){\circle{24}}
\put(989,373){\circle{24}}
\put(1044,445){\circle{24}}
\put(1084,500){\circle{24}}
\put(1315,705){\circle{24}}
\put(400,745){\makebox(0,0)[l]{$f=0$}}
\put(400,700){\makebox(0,0)[l]{$f=10^{-6}$}}
\put(400,655){\makebox(0,0)[l]{$f=10^{-5}$}}
\put(400,610){\makebox(0,0)[l]{$f=10^{-4}$}}
\put(400,565){\makebox(0,0)[l]{$f=10^{-3}$}}
\put(320,745){\raisebox{-1.2pt}{\makebox(0,0){$\Diamond$}}}
\put(320,700){\raisebox{-1.2pt}{\makebox(0,0){$\Box$}}}
\put(320,655){\makebox(0,0){$\triangle$}}
\put(320,610){\makebox(0,0){$\bigtriangledown$}}
\put(320,565){\circle{24}}
\end{picture}

  { \small
  Fig. 3. Average squared deficit angle $\langle \delta_t^2 \rangle$ as 
  function of $\ell$ for decreasing cutoff $f$. A non-monotonic behavior of
  $\langle \delta_t^2 \rangle$ is seen for increasing $\ell$.
  The fact that the average squared deficit
  angle decreases with increasing $\ell$ is a strong indication
  for the stability of the well-defined phase. Near the transition point
  the minimum deviation from flat space is reached and a further increase
  of $\beta$ and thus $\ell$ leads to the ill-defined region with
  fast increasing squared deficit angles (see text). Configurations and 
  statistics are the same as in Fig. 1.
  }
\end{figure}

\newpage
\begin{figure}[p]

\setlength{\unitlength}{0.240900pt}
\ifx\plotpoint\undefined\newsavebox{\plotpoint}\fi
\sbox{\plotpoint}{\rule[-0.175pt]{0.350pt}{0.350pt}}%


  { \small
  Fig. 4. (In)dependence of the numerical results in the well-defined phase
  from the start configuration. The upper plot shows the histories
  of $\tilde{R}$ for a start with homogeneous and
  inhomogeneous configurations, respectively.
  Both simulations converge to the same equilibrium in the well-defined
  phase for non-zero $\beta$ $(\ell \approx 0.29)$. The computer time is denoted
  by $\tau$ with one unit representing $10$ sweeps. Although the unbounded
  action ($f = 0$) favors a large positive curvature the entropy moves
  $\tilde{R}$ to negative values.
  With a small bound for the fatness ($f = 10^{-5}$) independence from the
  start is seen even starting with very inhomogeneous configurations as
  depicted in the lower plot.
           }
\end{figure}

\newpage
\begin{figure}[p]

\setlength{\unitlength}{0.240900pt}
\ifx\plotpoint\undefined\newsavebox{\plotpoint}\fi
\sbox{\plotpoint}{\rule[-0.175pt]{0.350pt}{0.350pt}}%
\begin{picture}(1500,900)(0,0)
\tenrm
\sbox{\plotpoint}{\rule[-0.175pt]{0.350pt}{0.350pt}}%
\put(264,315){\rule[-0.175pt]{282.335pt}{0.350pt}}
\put(1165,178){\rule[-0.175pt]{0.200pt}{141.526pt}}
\put(264,210){\rule[-0.175pt]{4.818pt}{0.350pt}}
\put(242,210){\makebox(0,0)[r]{-10}}
\put(1416,210){\rule[-0.175pt]{4.818pt}{0.350pt}}
\put(264,315){\rule[-0.175pt]{4.818pt}{0.350pt}}
\put(242,315){\makebox(0,0)[r]{0}}
\put(1416,315){\rule[-0.175pt]{4.818pt}{0.350pt}}
\put(264,420){\rule[-0.175pt]{4.818pt}{0.350pt}}
\put(242,420){\makebox(0,0)[r]{10}}
\put(1416,420){\rule[-0.175pt]{4.818pt}{0.350pt}}
\put(264,525){\rule[-0.175pt]{4.818pt}{0.350pt}}
\put(242,525){\makebox(0,0)[r]{20}}
\put(1416,525){\rule[-0.175pt]{4.818pt}{0.350pt}}
\put(264,630){\rule[-0.175pt]{4.818pt}{0.350pt}}
\put(242,630){\makebox(0,0)[r]{30}}
\put(1416,630){\rule[-0.175pt]{4.818pt}{0.350pt}}
\put(264,735){\rule[-0.175pt]{4.818pt}{0.350pt}}
\put(242,735){\makebox(0,0)[r]{40}}
\put(1416,735){\rule[-0.175pt]{4.818pt}{0.350pt}}
\put(354,158){\rule[-0.175pt]{0.350pt}{4.818pt}}
\put(354,113){\makebox(0,0){0}}
\put(354,767){\rule[-0.175pt]{0.350pt}{4.818pt}}
\put(534,158){\rule[-0.175pt]{0.350pt}{4.818pt}}
\put(534,113){\makebox(0,0){0.1}}
\put(534,767){\rule[-0.175pt]{0.350pt}{4.818pt}}
\put(715,158){\rule[-0.175pt]{0.350pt}{4.818pt}}
\put(715,113){\makebox(0,0){0.2}}
\put(715,767){\rule[-0.175pt]{0.350pt}{4.818pt}}
\put(895,158){\rule[-0.175pt]{0.350pt}{4.818pt}}
\put(895,113){\makebox(0,0){0.3}}
\put(895,767){\rule[-0.175pt]{0.350pt}{4.818pt}}
\put(1075,158){\rule[-0.175pt]{0.350pt}{4.818pt}}
\put(1075,113){\makebox(0,0){0.4}}
\put(1075,767){\rule[-0.175pt]{0.350pt}{4.818pt}}
\put(1256,158){\rule[-0.175pt]{0.350pt}{4.818pt}}
\put(1256,113){\makebox(0,0){0.5}}
\put(1256,767){\rule[-0.175pt]{0.350pt}{4.818pt}}
\put(1436,158){\rule[-0.175pt]{0.350pt}{4.818pt}}
\put(1436,113){\makebox(0,0){0.6}}
\put(1436,767){\rule[-0.175pt]{0.350pt}{4.818pt}}
\put(264,158){\rule[-0.175pt]{282.335pt}{0.350pt}}
\put(1436,158){\rule[-0.175pt]{0.350pt}{151.526pt}}
\put(264,787){\rule[-0.175pt]{282.335pt}{0.350pt}}
\put(45,472){\makebox(0,0)[l]{\shortstack{$\langle \tilde{R} \rangle$}}}
\put(850,68){\makebox(0,0){$\ell$}}
\put(264,158){\rule[-0.175pt]{0.350pt}{151.526pt}}
\sbox{\plotpoint}{\rule[-0.250pt]{0.500pt}{0.500pt}}%
\put(399,735){\makebox(0,0)[l]{Dyn.~Triangulation}}
\put(354,208){\usebox{\plotpoint}}
\put(354,208){\usebox{\plotpoint}}
\put(374,208){\usebox{\plotpoint}}
\put(395,209){\usebox{\plotpoint}}
\put(416,210){\usebox{\plotpoint}}
\put(436,210){\usebox{\plotpoint}}
\put(457,211){\usebox{\plotpoint}}
\put(478,212){\usebox{\plotpoint}}
\put(499,212){\usebox{\plotpoint}}
\put(519,213){\usebox{\plotpoint}}
\put(540,214){\usebox{\plotpoint}}
\put(561,214){\usebox{\plotpoint}}
\put(582,215){\usebox{\plotpoint}}
\put(602,216){\usebox{\plotpoint}}
\put(623,217){\usebox{\plotpoint}}
\put(644,217){\usebox{\plotpoint}}
\put(665,218){\usebox{\plotpoint}}
\put(685,219){\usebox{\plotpoint}}
\put(706,220){\usebox{\plotpoint}}
\put(727,221){\usebox{\plotpoint}}
\put(748,221){\usebox{\plotpoint}}
\put(768,222){\usebox{\plotpoint}}
\put(789,224){\usebox{\plotpoint}}
\put(810,225){\usebox{\plotpoint}}
\put(830,226){\usebox{\plotpoint}}
\put(851,227){\usebox{\plotpoint}}
\put(872,229){\usebox{\plotpoint}}
\put(892,232){\usebox{\plotpoint}}
\put(913,236){\usebox{\plotpoint}}
\put(933,239){\usebox{\plotpoint}}
\put(954,242){\usebox{\plotpoint}}
\put(974,245){\usebox{\plotpoint}}
\put(995,249){\usebox{\plotpoint}}
\put(1015,252){\usebox{\plotpoint}}
\put(1036,255){\usebox{\plotpoint}}
\put(1057,257){\usebox{\plotpoint}}
\put(1077,259){\usebox{\plotpoint}}
\put(1097,263){\usebox{\plotpoint}}
\put(1117,270){\usebox{\plotpoint}}
\put(1137,276){\usebox{\plotpoint}}
\put(1157,283){\usebox{\plotpoint}}
\put(308,690){\circle{24}}
\put(354,208){\circle{24}}
\put(651,218){\circle{24}}
\put(773,223){\circle{24}}
\put(869,229){\circle{24}}
\put(1026,254){\circle{24}}
\put(1090,261){\circle{24}}
\put(1157,283){\circle*{24}}
\put(399,690){\makebox(0,0)[l]{Regge Approach}}
\put(354,274){\makebox(0,0)[l]{well-defined}}
\put(1220,590){\makebox(0,0)[l]{ill-defined}}
\put(1332,744){\usebox{\plotpoint}}
\put(1332,744){\usebox{\plotpoint}}
\put(1311,738){\usebox{\plotpoint}}
\put(1291,733){\usebox{\plotpoint}}
\put(1272,726){\usebox{\plotpoint}}
\put(1252,721){\usebox{\plotpoint}}
\put(1232,714){\usebox{\plotpoint}}
\put(1213,706){\usebox{\plotpoint}}
\put(1195,696){\usebox{\plotpoint}}
\put(1179,683){\usebox{\plotpoint}}
\put(1164,669){\usebox{\plotpoint}}
\put(1153,651){\usebox{\plotpoint}}
\put(1142,633){\usebox{\plotpoint}}
\put(1132,615){\usebox{\plotpoint}}
\put(1120,599){\usebox{\plotpoint}}
\put(1107,583){\usebox{\plotpoint}}
\put(1093,567){\usebox{\plotpoint}}
\put(1080,551){\usebox{\plotpoint}}
\put(1066,536){\usebox{\plotpoint}}
\put(1052,520){\usebox{\plotpoint}}
\put(1037,506){\usebox{\plotpoint}}
\put(1021,492){\usebox{\plotpoint}}
\put(1006,478){\usebox{\plotpoint}}
\put(989,465){\usebox{\plotpoint}}
\put(973,453){\usebox{\plotpoint}}
\put(956,441){\usebox{\plotpoint}}
\put(939,429){\usebox{\plotpoint}}
\put(921,418){\usebox{\plotpoint}}
\put(904,407){\usebox{\plotpoint}}
\put(886,396){\usebox{\plotpoint}}
\put(868,386){\usebox{\plotpoint}}
\put(850,375){\usebox{\plotpoint}}
\put(832,366){\usebox{\plotpoint}}
\put(812,359){\usebox{\plotpoint}}
\put(793,351){\usebox{\plotpoint}}
\put(773,344){\usebox{\plotpoint}}
\put(753,340){\usebox{\plotpoint}}
\put(733,335){\usebox{\plotpoint}}
\put(713,330){\usebox{\plotpoint}}
\put(700,327){\usebox{\plotpoint}}
\put(308,735){\makebox(0,0){$\triangle$}}
\put(1332,744){\makebox(0,0){$\triangle$}}
\put(1301,736){\makebox(0,0){$\triangle$}}
\put(1269,726){\makebox(0,0){$\triangle$}}
\put(1235,716){\makebox(0,0){$\triangle$}}
\put(1201,701){\makebox(0,0){$\triangle$}}
\put(1165,671){\makebox(0,0){$\triangle$}}
\put(1127,607){\makebox(0,0){$\triangle$}}
\put(1087,559){\makebox(0,0){$\triangle$}}
\put(1046,514){\makebox(0,0){$\triangle$}}
\put(1001,474){\makebox(0,0){$\triangle$}}
\put(953,439){\makebox(0,0){$\triangle$}}
\put(901,405){\makebox(0,0){$\triangle$}}
\put(843,371){\makebox(0,0){$\triangle$}}
\put(778,346){\makebox(0,0){$\triangle$}}
\put(700,327){\makebox(0,0){$\triangle$}}
\end{picture}

  { \small
  Fig. 5. Expectation value $\langle \tilde{R} \rangle$ versus 
  $\ell$ for the two complementary approaches to simplicial
  quantum gravity. The upper curve shows results from dynamical
  triangulation obtained by Br\"ugmann approximating the 4-sphere
  by 16k 4-simplices \cite{BM}.
  The lower curve was obtained within the Regge approach
  on a regular triangulated 4-torus with $4^4$ vertices.
  The behavior of both functions is rather similar although the underlying
  topologies differ. A phase of small $\langle \tilde{R} \rangle$ turns into
  a region with large curvature for increasing $\ell$. The black
  dot indicates a value close to the transition point $\ell_c$ for the
  Regge approach.
  }
\end{figure}

\newpage
\begin{figure}[p]

\setlength{\unitlength}{0.240900pt}
\ifx\plotpoint\undefined\newsavebox{\plotpoint}\fi
\sbox{\plotpoint}{\rule[-0.175pt]{0.350pt}{0.350pt}}%
\begin{picture}(1500,900)(0,0)
\tenrm
\sbox{\plotpoint}{\rule[-0.175pt]{0.350pt}{0.350pt}}%
\put(264,577){\rule[-0.175pt]{282.335pt}{0.350pt}}
\put(264,158){\rule[-0.175pt]{4.818pt}{0.350pt}}
\put(242,158){\makebox(0,0)[r]{-12}}
\put(1416,158){\rule[-0.175pt]{4.818pt}{0.350pt}}
\put(264,263){\rule[-0.175pt]{4.818pt}{0.350pt}}
\put(242,263){\makebox(0,0)[r]{-9}}
\put(1416,263){\rule[-0.175pt]{4.818pt}{0.350pt}}
\put(264,368){\rule[-0.175pt]{4.818pt}{0.350pt}}
\put(242,368){\makebox(0,0)[r]{-6}}
\put(1416,368){\rule[-0.175pt]{4.818pt}{0.350pt}}
\put(264,473){\rule[-0.175pt]{4.818pt}{0.350pt}}
\put(242,473){\makebox(0,0)[r]{-3}}
\put(1416,473){\rule[-0.175pt]{4.818pt}{0.350pt}}
\put(264,577){\rule[-0.175pt]{4.818pt}{0.350pt}}
\put(242,577){\makebox(0,0)[r]{0}}
\put(1416,577){\rule[-0.175pt]{4.818pt}{0.350pt}}
\put(264,682){\rule[-0.175pt]{4.818pt}{0.350pt}}
\put(242,682){\makebox(0,0)[r]{3}}
\put(1416,682){\rule[-0.175pt]{4.818pt}{0.350pt}}
\put(264,787){\rule[-0.175pt]{4.818pt}{0.350pt}}
\put(242,787){\makebox(0,0)[r]{6}}
\put(1416,787){\rule[-0.175pt]{4.818pt}{0.350pt}}
\put(264,158){\rule[-0.175pt]{0.350pt}{4.818pt}}
\put(264,113){\makebox(0,0){-0.1}}
\put(264,767){\rule[-0.175pt]{0.350pt}{4.818pt}}
\put(459,158){\rule[-0.175pt]{0.350pt}{4.818pt}}
\put(459,113){\makebox(0,0){0}}
\put(459,767){\rule[-0.175pt]{0.350pt}{4.818pt}}
\put(655,158){\rule[-0.175pt]{0.350pt}{4.818pt}}
\put(655,113){\makebox(0,0){0.1}}
\put(655,767){\rule[-0.175pt]{0.350pt}{4.818pt}}
\put(850,158){\rule[-0.175pt]{0.350pt}{4.818pt}}
\put(850,113){\makebox(0,0){0.2}}
\put(850,767){\rule[-0.175pt]{0.350pt}{4.818pt}}
\put(1045,158){\rule[-0.175pt]{0.350pt}{4.818pt}}
\put(1045,113){\makebox(0,0){0.3}}
\put(1045,767){\rule[-0.175pt]{0.350pt}{4.818pt}}
\put(1241,158){\rule[-0.175pt]{0.350pt}{4.818pt}}
\put(1241,113){\makebox(0,0){0.4}}
\put(1241,767){\rule[-0.175pt]{0.350pt}{4.818pt}}
\put(1436,158){\rule[-0.175pt]{0.350pt}{4.818pt}}
\put(1436,113){\makebox(0,0){0.5}}
\put(1436,767){\rule[-0.175pt]{0.350pt}{4.818pt}}
\put(264,158){\rule[-0.175pt]{282.335pt}{0.350pt}}
\put(1436,158){\rule[-0.175pt]{0.350pt}{151.526pt}}
\put(264,787){\rule[-0.175pt]{282.335pt}{0.350pt}}
\put(45,472){\makebox(0,0)[l]{\shortstack{$\langle \tilde{R} \rangle$}}}
\put(850,68){\makebox(0,0){$\ell$}}
\put(264,158){\rule[-0.175pt]{0.350pt}{151.526pt}}
\sbox{\plotpoint}{\rule[-0.250pt]{0.500pt}{0.500pt}}%
\put(400,742){\makebox(0,0)[l]{$\sigma=1.50$}}
\put(459,287){\usebox{\plotpoint}}
\put(459,287){\usebox{\plotpoint}}
\put(479,291){\usebox{\plotpoint}}
\put(499,295){\usebox{\plotpoint}}
\put(520,299){\usebox{\plotpoint}}
\put(540,303){\usebox{\plotpoint}}
\put(560,307){\usebox{\plotpoint}}
\put(581,311){\usebox{\plotpoint}}
\put(601,316){\usebox{\plotpoint}}
\put(621,320){\usebox{\plotpoint}}
\put(642,324){\usebox{\plotpoint}}
\put(662,328){\usebox{\plotpoint}}
\put(682,332){\usebox{\plotpoint}}
\put(703,336){\usebox{\plotpoint}}
\put(723,341){\usebox{\plotpoint}}
\put(743,345){\usebox{\plotpoint}}
\put(764,349){\usebox{\plotpoint}}
\put(784,353){\usebox{\plotpoint}}
\put(804,358){\usebox{\plotpoint}}
\put(822,368){\usebox{\plotpoint}}
\put(840,378){\usebox{\plotpoint}}
\put(859,388){\usebox{\plotpoint}}
\put(877,397){\usebox{\plotpoint}}
\put(895,407){\usebox{\plotpoint}}
\put(914,417){\usebox{\plotpoint}}
\put(932,427){\usebox{\plotpoint}}
\put(950,437){\usebox{\plotpoint}}
\put(968,447){\usebox{\plotpoint}}
\put(986,458){\usebox{\plotpoint}}
\put(1004,469){\usebox{\plotpoint}}
\put(1021,479){\usebox{\plotpoint}}
\put(1039,490){\usebox{\plotpoint}}
\put(1057,500){\usebox{\plotpoint}}
\put(1071,515){\usebox{\plotpoint}}
\put(1081,533){\usebox{\plotpoint}}
\put(1092,551){\usebox{\plotpoint}}
\put(1102,569){\usebox{\plotpoint}}
\put(1113,587){\usebox{\plotpoint}}
\put(1123,605){\usebox{\plotpoint}}
\put(1133,623){\usebox{\plotpoint}}
\put(1144,641){\usebox{\plotpoint}}
\put(1154,659){\usebox{\plotpoint}}
\put(1164,677){\usebox{\plotpoint}}
\put(1175,695){\usebox{\plotpoint}}
\put(1178,700){\usebox{\plotpoint}}
\put(317,742){\raisebox{-1.2pt}{\makebox(0,0){$\Diamond$}}}
\put(459,287){\raisebox{-1.2pt}{\makebox(0,0){$\Diamond$}}}
\put(801,357){\raisebox{-1.2pt}{\makebox(0,0){$\Diamond$}}}
\put(947,435){\raisebox{-1.2pt}{\makebox(0,0){$\Diamond$}}}
\put(1066,506){\raisebox{-1.2pt}{\makebox(0,0){$\Diamond$}}}
\put(1178,700){\raisebox{-1.2pt}{\makebox(0,0){$\Diamond$}}}
\put(400,697){\makebox(0,0)[l]{$\sigma=1.00$}}
\put(459,219){\usebox{\plotpoint}}
\put(459,219){\usebox{\plotpoint}}
\put(479,221){\usebox{\plotpoint}}
\put(500,223){\usebox{\plotpoint}}
\put(520,225){\usebox{\plotpoint}}
\put(541,227){\usebox{\plotpoint}}
\put(562,230){\usebox{\plotpoint}}
\put(582,232){\usebox{\plotpoint}}
\put(603,234){\usebox{\plotpoint}}
\put(624,236){\usebox{\plotpoint}}
\put(644,239){\usebox{\plotpoint}}
\put(665,241){\usebox{\plotpoint}}
\put(685,243){\usebox{\plotpoint}}
\put(706,245){\usebox{\plotpoint}}
\put(727,248){\usebox{\plotpoint}}
\put(747,250){\usebox{\plotpoint}}
\put(768,252){\usebox{\plotpoint}}
\put(789,255){\usebox{\plotpoint}}
\put(809,257){\usebox{\plotpoint}}
\put(830,260){\usebox{\plotpoint}}
\put(850,262){\usebox{\plotpoint}}
\put(871,265){\usebox{\plotpoint}}
\put(892,268){\usebox{\plotpoint}}
\put(912,270){\usebox{\plotpoint}}
\put(933,274){\usebox{\plotpoint}}
\put(953,278){\usebox{\plotpoint}}
\put(973,282){\usebox{\plotpoint}}
\put(994,285){\usebox{\plotpoint}}
\put(1014,289){\usebox{\plotpoint}}
\put(1033,298){\usebox{\plotpoint}}
\put(1052,307){\usebox{\plotpoint}}
\put(1070,316){\usebox{\plotpoint}}
\put(1089,325){\usebox{\plotpoint}}
\put(1108,334){\usebox{\plotpoint}}
\put(1126,343){\usebox{\plotpoint}}
\put(1145,352){\usebox{\plotpoint}}
\put(1164,361){\usebox{\plotpoint}}
\put(1182,370){\usebox{\plotpoint}}
\put(1202,377){\usebox{\plotpoint}}
\put(1221,384){\usebox{\plotpoint}}
\put(1241,391){\usebox{\plotpoint}}
\put(1260,400){\usebox{\plotpoint}}
\put(1274,414){\usebox{\plotpoint}}
\put(1289,429){\usebox{\plotpoint}}
\put(1304,444){\usebox{\plotpoint}}
\put(1318,458){\usebox{\plotpoint}}
\put(1329,469){\usebox{\plotpoint}}
\put(317,697){\raisebox{-1.2pt}{\makebox(0,0){$\Box$}}}
\put(459,219){\raisebox{-1.2pt}{\makebox(0,0){$\Box$}}}
\put(781,254){\raisebox{-1.2pt}{\makebox(0,0){$\Box$}}}
\put(913,271){\raisebox{-1.2pt}{\makebox(0,0){$\Box$}}}
\put(1017,290){\raisebox{-1.2pt}{\makebox(0,0){$\Box$}}}
\put(1188,373){\raisebox{-1.2pt}{\makebox(0,0){$\Box$}}}
\put(1257,397){\raisebox{-1.2pt}{\makebox(0,0){$\Box$}}}
\put(1329,469){\raisebox{-1.2pt}{\makebox(0,0){$\Box$}}}
\put(400,652){\makebox(0,0)[l]{$\sigma=0.50$}}
\put(459,216){\usebox{\plotpoint}}
\put(459,216){\usebox{\plotpoint}}
\put(479,218){\usebox{\plotpoint}}
\put(500,221){\usebox{\plotpoint}}
\put(520,223){\usebox{\plotpoint}}
\put(541,226){\usebox{\plotpoint}}
\put(562,228){\usebox{\plotpoint}}
\put(582,231){\usebox{\plotpoint}}
\put(603,233){\usebox{\plotpoint}}
\put(623,236){\usebox{\plotpoint}}
\put(644,238){\usebox{\plotpoint}}
\put(665,241){\usebox{\plotpoint}}
\put(685,243){\usebox{\plotpoint}}
\put(706,246){\usebox{\plotpoint}}
\put(726,248){\usebox{\plotpoint}}
\put(747,251){\usebox{\plotpoint}}
\put(768,253){\usebox{\plotpoint}}
\put(788,256){\usebox{\plotpoint}}
\put(809,258){\usebox{\plotpoint}}
\put(829,261){\usebox{\plotpoint}}
\put(850,263){\usebox{\plotpoint}}
\put(871,266){\usebox{\plotpoint}}
\put(891,268){\usebox{\plotpoint}}
\put(912,271){\usebox{\plotpoint}}
\put(932,273){\usebox{\plotpoint}}
\put(953,276){\usebox{\plotpoint}}
\put(974,278){\usebox{\plotpoint}}
\put(994,281){\usebox{\plotpoint}}
\put(1015,283){\usebox{\plotpoint}}
\put(1035,286){\usebox{\plotpoint}}
\put(1056,288){\usebox{\plotpoint}}
\put(1077,291){\usebox{\plotpoint}}
\put(1097,293){\usebox{\plotpoint}}
\put(1118,296){\usebox{\plotpoint}}
\put(1138,298){\usebox{\plotpoint}}
\put(1159,302){\usebox{\plotpoint}}
\put(1179,307){\usebox{\plotpoint}}
\put(1199,312){\usebox{\plotpoint}}
\put(1219,317){\usebox{\plotpoint}}
\put(1239,324){\usebox{\plotpoint}}
\put(1249,328){\usebox{\plotpoint}}
\put(317,652){\makebox(0,0){$\triangle$}}
\put(459,216){\makebox(0,0){$\triangle$}}
\put(1151,300){\makebox(0,0){$\triangle$}}
\put(1218,317){\makebox(0,0){$\triangle$}}
\put(1249,328){\makebox(0,0){$\triangle$}}
\put(400,607){\makebox(0,0)[l]{$\sigma=0.25$}}
\put(459,219){\usebox{\plotpoint}}
\put(459,219){\usebox{\plotpoint}}
\put(479,221){\usebox{\plotpoint}}
\put(500,223){\usebox{\plotpoint}}
\put(520,225){\usebox{\plotpoint}}
\put(541,227){\usebox{\plotpoint}}
\put(562,230){\usebox{\plotpoint}}
\put(582,232){\usebox{\plotpoint}}
\put(603,234){\usebox{\plotpoint}}
\put(624,236){\usebox{\plotpoint}}
\put(644,239){\usebox{\plotpoint}}
\put(665,241){\usebox{\plotpoint}}
\put(685,243){\usebox{\plotpoint}}
\put(706,245){\usebox{\plotpoint}}
\put(727,247){\usebox{\plotpoint}}
\put(747,250){\usebox{\plotpoint}}
\put(768,252){\usebox{\plotpoint}}
\put(789,254){\usebox{\plotpoint}}
\put(809,256){\usebox{\plotpoint}}
\put(830,259){\usebox{\plotpoint}}
\put(851,261){\usebox{\plotpoint}}
\put(871,263){\usebox{\plotpoint}}
\put(892,265){\usebox{\plotpoint}}
\put(912,268){\usebox{\plotpoint}}
\put(933,270){\usebox{\plotpoint}}
\put(954,272){\usebox{\plotpoint}}
\put(974,274){\usebox{\plotpoint}}
\put(995,276){\usebox{\plotpoint}}
\put(1016,279){\usebox{\plotpoint}}
\put(1036,281){\usebox{\plotpoint}}
\put(1057,283){\usebox{\plotpoint}}
\put(1078,285){\usebox{\plotpoint}}
\put(1098,288){\usebox{\plotpoint}}
\put(1119,290){\usebox{\plotpoint}}
\put(1135,292){\usebox{\plotpoint}}
\put(1159,298){\usebox{\plotpoint}}
\put(1179,304){\usebox{\plotpoint}}
\put(1199,309){\usebox{\plotpoint}}
\put(1219,314){\usebox{\plotpoint}}
\put(1239,320){\usebox{\plotpoint}}
\put(1259,325){\usebox{\plotpoint}}
\put(1279,331){\usebox{\plotpoint}}
\put(1282,332){\usebox{\plotpoint}}
\put(317,607){\circle{24}}
\put(459,219){\circle{24}}
\put(1135,292){\circle{24}}
\put(1282,332){\circle{24}}
\put(400,552){\makebox(0,0)[l]{$\sigma=0.00$}}
\put(747,239){\usebox{\plotpoint}}
\put(747,239){\usebox{\plotpoint}}
\put(767,241){\usebox{\plotpoint}}
\put(788,244){\usebox{\plotpoint}}
\put(808,247){\usebox{\plotpoint}}
\put(829,250){\usebox{\plotpoint}}
\put(849,253){\usebox{\plotpoint}}
\put(870,256){\usebox{\plotpoint}}
\put(890,258){\usebox{\plotpoint}}
\put(911,260){\usebox{\plotpoint}}
\put(932,262){\usebox{\plotpoint}}
\put(952,264){\usebox{\plotpoint}}
\put(973,268){\usebox{\plotpoint}}
\put(993,272){\usebox{\plotpoint}}
\put(1013,276){\usebox{\plotpoint}}
\put(1034,280){\usebox{\plotpoint}}
\put(1054,286){\usebox{\plotpoint}}
\put(1073,293){\usebox{\plotpoint}}
\put(1093,299){\usebox{\plotpoint}}
\put(1113,303){\usebox{\plotpoint}}
\put(1134,307){\usebox{\plotpoint}}
\put(1154,311){\usebox{\plotpoint}}
\put(1175,315){\usebox{\plotpoint}}
\put(1195,320){\usebox{\plotpoint}}
\put(1215,325){\usebox{\plotpoint}}
\put(1235,330){\usebox{\plotpoint}}
\put(1255,335){\usebox{\plotpoint}}
\put(1275,340){\usebox{\plotpoint}}
\put(1293,352){\usebox{\plotpoint}}
\put(1310,363){\usebox{\plotpoint}}
\put(1328,374){\usebox{\plotpoint}}
\put(1329,375){\usebox{\plotpoint}}
\put(317,552){\raisebox{-0.0pt}{\makebox(0,0){$\bigtriangledown$}}}
\put(459,228){\raisebox{-0.0pt}{\makebox(0,0){$\bigtriangledown$}}}
\put(747,239){\raisebox{-0.0pt}{\makebox(0,0){$\bigtriangledown$}}}
\put(866,256){\raisebox{-0.0pt}{\makebox(0,0){$\bigtriangledown$}}}
\put(957,265){\raisebox{-0.0pt}{\makebox(0,0){$\bigtriangledown$}}}
\put(1035,281){\raisebox{-0.0pt}{\makebox(0,0){$\bigtriangledown$}}}
\put(1103,302){\raisebox{-0.0pt}{\makebox(0,0){$\bigtriangledown$}}}
\put(1165,313){\raisebox{-0.0pt}{\makebox(0,0){$\bigtriangledown$}}}
\put(1223,327){\raisebox{-0.0pt}{\makebox(0,0){$\bigtriangledown$}}}
\put(1276,341){\raisebox{-0.0pt}{\makebox(0,0){$\bigtriangledown$}}}
\put(1329,375){\raisebox{-0.0pt}{\makebox(0,0){$\bigtriangledown$}}}
\end{picture}

\setlength{\unitlength}{0.240900pt}
\ifx\plotpoint\undefined\newsavebox{\plotpoint}\fi
\sbox{\plotpoint}{\rule[-0.175pt]{0.350pt}{0.350pt}}%
\begin{picture}(1500,900)(0,0)
\tenrm
\sbox{\plotpoint}{\rule[-0.175pt]{0.350pt}{0.350pt}}%
\put(264,158){\rule[-0.175pt]{4.818pt}{0.350pt}}
\put(242,158){\makebox(0,0)[r]{-11}}
\put(1416,158){\rule[-0.175pt]{4.818pt}{0.350pt}}
\put(264,237){\rule[-0.175pt]{4.818pt}{0.350pt}}
\put(242,237){\makebox(0,0)[r]{-10.5}}
\put(1416,237){\rule[-0.175pt]{4.818pt}{0.350pt}}
\put(264,315){\rule[-0.175pt]{4.818pt}{0.350pt}}
\put(242,315){\makebox(0,0)[r]{-10}}
\put(1416,315){\rule[-0.175pt]{4.818pt}{0.350pt}}
\put(264,394){\rule[-0.175pt]{4.818pt}{0.350pt}}
\put(242,394){\makebox(0,0)[r]{-9.5}}
\put(1416,394){\rule[-0.175pt]{4.818pt}{0.350pt}}
\put(264,473){\rule[-0.175pt]{4.818pt}{0.350pt}}
\put(242,473){\makebox(0,0)[r]{-9}}
\put(1416,473){\rule[-0.175pt]{4.818pt}{0.350pt}}
\put(264,551){\rule[-0.175pt]{4.818pt}{0.350pt}}
\put(242,551){\makebox(0,0)[r]{-8.5}}
\put(1416,551){\rule[-0.175pt]{4.818pt}{0.350pt}}
\put(264,630){\rule[-0.175pt]{4.818pt}{0.350pt}}
\put(242,630){\makebox(0,0)[r]{-8}}
\put(1416,630){\rule[-0.175pt]{4.818pt}{0.350pt}}
\put(264,708){\rule[-0.175pt]{4.818pt}{0.350pt}}
\put(242,708){\makebox(0,0)[r]{-7.5}}
\put(1416,708){\rule[-0.175pt]{4.818pt}{0.350pt}}
\put(264,787){\rule[-0.175pt]{4.818pt}{0.350pt}}
\put(242,787){\makebox(0,0)[r]{-7}}
\put(1416,787){\rule[-0.175pt]{4.818pt}{0.350pt}}
\put(333,158){\rule[-0.175pt]{0.350pt}{4.818pt}}
\put(333,113){\makebox(0,0){0}}
\put(333,767){\rule[-0.175pt]{0.350pt}{4.818pt}}
\put(505,158){\rule[-0.175pt]{0.350pt}{4.818pt}}
\put(505,113){\makebox(0,0){0.25}}
\put(505,767){\rule[-0.175pt]{0.350pt}{4.818pt}}
\put(678,158){\rule[-0.175pt]{0.350pt}{4.818pt}}
\put(678,113){\makebox(0,0){0.5}}
\put(678,767){\rule[-0.175pt]{0.350pt}{4.818pt}}
\put(850,158){\rule[-0.175pt]{0.350pt}{4.818pt}}
\put(850,113){\makebox(0,0){0.75}}
\put(850,767){\rule[-0.175pt]{0.350pt}{4.818pt}}
\put(1022,158){\rule[-0.175pt]{0.350pt}{4.818pt}}
\put(1022,113){\makebox(0,0){1}}
\put(1022,767){\rule[-0.175pt]{0.350pt}{4.818pt}}
\put(1195,158){\rule[-0.175pt]{0.350pt}{4.818pt}}
\put(1195,113){\makebox(0,0){1.25}}
\put(1195,767){\rule[-0.175pt]{0.350pt}{4.818pt}}
\put(1367,158){\rule[-0.175pt]{0.350pt}{4.818pt}}
\put(1367,113){\makebox(0,0){1.5}}
\put(1367,767){\rule[-0.175pt]{0.350pt}{4.818pt}}
\put(264,158){\rule[-0.175pt]{282.335pt}{0.350pt}}
\put(1436,158){\rule[-0.175pt]{0.350pt}{151.526pt}}
\put(264,787){\rule[-0.175pt]{282.335pt}{0.350pt}}
\put(45,472){\makebox(0,0)[l]{\shortstack{$\langle \tilde{R} \rangle$}}}
\put(850,68){\makebox(0,0){$\sigma$}}
\put(264,158){\rule[-0.175pt]{0.350pt}{151.526pt}}
\sbox{\plotpoint}{\rule[-0.250pt]{0.500pt}{0.500pt}}%
\put(400,708){\makebox(0,0)[l]{$\ell=0$}}
\put(333,317){\usebox{\plotpoint}}
\put(333,317){\usebox{\plotpoint}}
\put(353,311){\usebox{\plotpoint}}
\put(373,306){\usebox{\plotpoint}}
\put(393,301){\usebox{\plotpoint}}
\put(413,296){\usebox{\plotpoint}}
\put(433,291){\usebox{\plotpoint}}
\put(453,286){\usebox{\plotpoint}}
\put(473,281){\usebox{\plotpoint}}
\put(494,276){\usebox{\plotpoint}}
\put(514,273){\usebox{\plotpoint}}
\put(535,272){\usebox{\plotpoint}}
\put(555,270){\usebox{\plotpoint}}
\put(576,269){\usebox{\plotpoint}}
\put(597,268){\usebox{\plotpoint}}
\put(618,266){\usebox{\plotpoint}}
\put(638,265){\usebox{\plotpoint}}
\put(659,264){\usebox{\plotpoint}}
\put(680,263){\usebox{\plotpoint}}
\put(700,263){\usebox{\plotpoint}}
\put(721,264){\usebox{\plotpoint}}
\put(742,265){\usebox{\plotpoint}}
\put(763,266){\usebox{\plotpoint}}
\put(783,267){\usebox{\plotpoint}}
\put(804,267){\usebox{\plotpoint}}
\put(825,268){\usebox{\plotpoint}}
\put(846,269){\usebox{\plotpoint}}
\put(866,270){\usebox{\plotpoint}}
\put(887,270){\usebox{\plotpoint}}
\put(908,271){\usebox{\plotpoint}}
\put(929,272){\usebox{\plotpoint}}
\put(949,273){\usebox{\plotpoint}}
\put(970,274){\usebox{\plotpoint}}
\put(991,274){\usebox{\plotpoint}}
\put(1012,275){\usebox{\plotpoint}}
\put(1030,282){\usebox{\plotpoint}}
\put(1047,294){\usebox{\plotpoint}}
\put(1063,307){\usebox{\plotpoint}}
\put(1080,319){\usebox{\plotpoint}}
\put(1097,332){\usebox{\plotpoint}}
\put(1113,344){\usebox{\plotpoint}}
\put(1130,356){\usebox{\plotpoint}}
\put(1147,369){\usebox{\plotpoint}}
\put(1163,381){\usebox{\plotpoint}}
\put(1180,394){\usebox{\plotpoint}}
\put(1196,406){\usebox{\plotpoint}}
\put(1211,421){\usebox{\plotpoint}}
\put(1225,436){\usebox{\plotpoint}}
\put(1240,451){\usebox{\plotpoint}}
\put(1254,466){\usebox{\plotpoint}}
\put(1269,480){\usebox{\plotpoint}}
\put(1284,495){\usebox{\plotpoint}}
\put(1298,510){\usebox{\plotpoint}}
\put(1313,525){\usebox{\plotpoint}}
\put(1327,540){\usebox{\plotpoint}}
\put(1342,554){\usebox{\plotpoint}}
\put(1356,569){\usebox{\plotpoint}}
\put(1367,580){\usebox{\plotpoint}}
\put(333,708){\circle{28}}
\put(333,317){\circle{28}}
\put(505,274){\circle{28}}
\put(678,263){\circle{28}}
\put(1022,276){\circle{28}}
\put(1195,405){\circle{28}}
\put(1367,580){\circle{28}}
\end{picture}

  { \small
  Fig. 6. Expectation value $\langle \tilde{R} \rangle$
  as a function of $\ell$ within the Regge approach for
  different types of the measure parametrized by $\sigma \geq 0$.
  Computations have been performed on a regular tringulated
  4-torus with $4^4$ vertices applying a cutoff $f = 10^{-5}$
  for faster convergence.
  The upper plot shows that the behavior of $\langle \tilde{R} \rangle$
  in the region of small $\ell$ is almost independent of $\sigma$
  as long as $\sigma \leq 1$. This stability ceases for
  $\sigma > 1$ as depicted in some detail in the lower plot where
  $\langle \tilde{R} \rangle$ is given for increasing $\sigma$ in the
  case of pure entropy, $\ell = 0 \leftrightarrow \beta = 0$.
  }
\end{figure}

\newpage
\begin{figure}[p]

\setlength{\unitlength}{0.240900pt}
\ifx\plotpoint\undefined\newsavebox{\plotpoint}\fi
\sbox{\plotpoint}{\rule[-0.175pt]{0.350pt}{0.350pt}}%
\begin{picture}(1500,900)(0,0)
\tenrm
\sbox{\plotpoint}{\rule[-0.175pt]{0.350pt}{0.350pt}}%
\put(264,158){\rule[-0.175pt]{4.818pt}{0.350pt}}
\put(242,158){\makebox(0,0)[r]{0.300}}
\put(1416,158){\rule[-0.175pt]{4.818pt}{0.350pt}}
\put(264,263){\rule[-0.175pt]{4.818pt}{0.350pt}}
\put(242,263){\makebox(0,0)[r]{0.325}}
\put(1416,263){\rule[-0.175pt]{4.818pt}{0.350pt}}
\put(264,368){\rule[-0.175pt]{4.818pt}{0.350pt}}
\put(242,368){\makebox(0,0)[r]{0.350}}
\put(1416,368){\rule[-0.175pt]{4.818pt}{0.350pt}}
\put(264,473){\rule[-0.175pt]{4.818pt}{0.350pt}}
\put(242,473){\makebox(0,0)[r]{0.375}}
\put(1416,473){\rule[-0.175pt]{4.818pt}{0.350pt}}
\put(264,577){\rule[-0.175pt]{4.818pt}{0.350pt}}
\put(242,577){\makebox(0,0)[r]{0.400}}
\put(1416,577){\rule[-0.175pt]{4.818pt}{0.350pt}}
\put(264,682){\rule[-0.175pt]{4.818pt}{0.350pt}}
\put(242,682){\makebox(0,0)[r]{0.425}}
\put(1416,682){\rule[-0.175pt]{4.818pt}{0.350pt}}
\put(264,787){\rule[-0.175pt]{4.818pt}{0.350pt}}
\put(242,787){\makebox(0,0)[r]{0.450}}
\put(1416,787){\rule[-0.175pt]{4.818pt}{0.350pt}}
\put(264,158){\rule[-0.175pt]{0.350pt}{4.818pt}}
\put(264,113){\makebox(0,0){-0.1}}
\put(264,767){\rule[-0.175pt]{0.350pt}{4.818pt}}
\put(459,158){\rule[-0.175pt]{0.350pt}{4.818pt}}
\put(459,113){\makebox(0,0){0}}
\put(459,767){\rule[-0.175pt]{0.350pt}{4.818pt}}
\put(655,158){\rule[-0.175pt]{0.350pt}{4.818pt}}
\put(655,113){\makebox(0,0){0.1}}
\put(655,767){\rule[-0.175pt]{0.350pt}{4.818pt}}
\put(850,158){\rule[-0.175pt]{0.350pt}{4.818pt}}
\put(850,113){\makebox(0,0){0.2}}
\put(850,767){\rule[-0.175pt]{0.350pt}{4.818pt}}
\put(1045,158){\rule[-0.175pt]{0.350pt}{4.818pt}}
\put(1045,113){\makebox(0,0){0.3}}
\put(1045,767){\rule[-0.175pt]{0.350pt}{4.818pt}}
\put(1241,158){\rule[-0.175pt]{0.350pt}{4.818pt}}
\put(1241,113){\makebox(0,0){0.4}}
\put(1241,767){\rule[-0.175pt]{0.350pt}{4.818pt}}
\put(1436,158){\rule[-0.175pt]{0.350pt}{4.818pt}}
\put(1436,113){\makebox(0,0){0.5}}
\put(1436,767){\rule[-0.175pt]{0.350pt}{4.818pt}}
\put(264,158){\rule[-0.175pt]{282.335pt}{0.350pt}}
\put(1436,158){\rule[-0.175pt]{0.350pt}{151.526pt}}
\put(264,787){\rule[-0.175pt]{282.335pt}{0.350pt}}
\put(45,472){\makebox(0,0)[l]{\shortstack{$\langle A_t \rangle$\\$\langle q_\ell \rangle$}}}
\put(45,471){\line(1,0){80}}
\put(850,68){\makebox(0,0){$\ell$}}
\put(264,158){\rule[-0.175pt]{0.350pt}{151.526pt}}
\sbox{\plotpoint}{\rule[-0.250pt]{0.500pt}{0.500pt}}%
\put(459,405){\usebox{\plotpoint}}
\put(459,405){\usebox{\plotpoint}}
\put(479,403){\usebox{\plotpoint}}
\put(500,402){\usebox{\plotpoint}}
\put(521,401){\usebox{\plotpoint}}
\put(541,399){\usebox{\plotpoint}}
\put(562,398){\usebox{\plotpoint}}
\put(583,397){\usebox{\plotpoint}}
\put(603,395){\usebox{\plotpoint}}
\put(624,394){\usebox{\plotpoint}}
\put(645,393){\usebox{\plotpoint}}
\put(666,391){\usebox{\plotpoint}}
\put(686,390){\usebox{\plotpoint}}
\put(707,389){\usebox{\plotpoint}}
\put(728,387){\usebox{\plotpoint}}
\put(748,386){\usebox{\plotpoint}}
\put(769,385){\usebox{\plotpoint}}
\put(790,383){\usebox{\plotpoint}}
\put(811,381){\usebox{\plotpoint}}
\put(831,379){\usebox{\plotpoint}}
\put(852,377){\usebox{\plotpoint}}
\put(872,375){\usebox{\plotpoint}}
\put(893,372){\usebox{\plotpoint}}
\put(914,370){\usebox{\plotpoint}}
\put(934,368){\usebox{\plotpoint}}
\put(955,365){\usebox{\plotpoint}}
\put(975,362){\usebox{\plotpoint}}
\put(996,359){\usebox{\plotpoint}}
\put(1016,355){\usebox{\plotpoint}}
\put(1037,352){\usebox{\plotpoint}}
\put(1057,349){\usebox{\plotpoint}}
\put(1077,343){\usebox{\plotpoint}}
\put(1097,336){\usebox{\plotpoint}}
\put(1116,329){\usebox{\plotpoint}}
\put(1136,322){\usebox{\plotpoint}}
\put(1156,315){\usebox{\plotpoint}}
\put(1175,308){\usebox{\plotpoint}}
\put(1178,308){\usebox{\plotpoint}}
\put(459,405){\raisebox{-1.2pt}{\makebox(0,0){$\Diamond$}}}
\put(801,383){\raisebox{-1.2pt}{\makebox(0,0){$\Diamond$}}}
\put(947,367){\raisebox{-1.2pt}{\makebox(0,0){$\Diamond$}}}
\put(1066,348){\raisebox{-1.2pt}{\makebox(0,0){$\Diamond$}}}
\put(1178,308){\raisebox{-1.2pt}{\makebox(0,0){$\Diamond$}}}
\put(459,471){\usebox{\plotpoint}}
\put(459,471){\usebox{\plotpoint}}
\put(479,470){\usebox{\plotpoint}}
\put(500,469){\usebox{\plotpoint}}
\put(521,469){\usebox{\plotpoint}}
\put(541,468){\usebox{\plotpoint}}
\put(562,468){\usebox{\plotpoint}}
\put(583,467){\usebox{\plotpoint}}
\put(604,467){\usebox{\plotpoint}}
\put(624,466){\usebox{\plotpoint}}
\put(645,466){\usebox{\plotpoint}}
\put(666,465){\usebox{\plotpoint}}
\put(687,465){\usebox{\plotpoint}}
\put(707,464){\usebox{\plotpoint}}
\put(728,464){\usebox{\plotpoint}}
\put(749,463){\usebox{\plotpoint}}
\put(770,463){\usebox{\plotpoint}}
\put(790,463){\usebox{\plotpoint}}
\put(811,463){\usebox{\plotpoint}}
\put(832,463){\usebox{\plotpoint}}
\put(853,463){\usebox{\plotpoint}}
\put(874,463){\usebox{\plotpoint}}
\put(894,463){\usebox{\plotpoint}}
\put(915,462){\usebox{\plotpoint}}
\put(936,461){\usebox{\plotpoint}}
\put(956,460){\usebox{\plotpoint}}
\put(977,459){\usebox{\plotpoint}}
\put(998,458){\usebox{\plotpoint}}
\put(1019,457){\usebox{\plotpoint}}
\put(1039,454){\usebox{\plotpoint}}
\put(1060,452){\usebox{\plotpoint}}
\put(1080,449){\usebox{\plotpoint}}
\put(1101,446){\usebox{\plotpoint}}
\put(1122,443){\usebox{\plotpoint}}
\put(1142,441){\usebox{\plotpoint}}
\put(1163,438){\usebox{\plotpoint}}
\put(1183,435){\usebox{\plotpoint}}
\put(1204,435){\usebox{\plotpoint}}
\put(1225,435){\usebox{\plotpoint}}
\put(1245,435){\usebox{\plotpoint}}
\put(1266,432){\usebox{\plotpoint}}
\put(1286,427){\usebox{\plotpoint}}
\put(1306,421){\usebox{\plotpoint}}
\put(1326,416){\usebox{\plotpoint}}
\put(1329,416){\usebox{\plotpoint}}
\put(459,471){\raisebox{-1.2pt}{\makebox(0,0){$\Box$}}}
\put(781,463){\raisebox{-1.2pt}{\makebox(0,0){$\Box$}}}
\put(913,463){\raisebox{-1.2pt}{\makebox(0,0){$\Box$}}}
\put(1017,458){\raisebox{-1.2pt}{\makebox(0,0){$\Box$}}}
\put(1188,435){\raisebox{-1.2pt}{\makebox(0,0){$\Box$}}}
\put(1257,435){\raisebox{-1.2pt}{\makebox(0,0){$\Box$}}}
\put(1329,416){\raisebox{-1.2pt}{\makebox(0,0){$\Box$}}}
\put(459,505){\usebox{\plotpoint}}
\put(459,505){\usebox{\plotpoint}}
\put(479,504){\usebox{\plotpoint}}
\put(500,504){\usebox{\plotpoint}}
\put(521,503){\usebox{\plotpoint}}
\put(542,503){\usebox{\plotpoint}}
\put(562,502){\usebox{\plotpoint}}
\put(583,502){\usebox{\plotpoint}}
\put(604,501){\usebox{\plotpoint}}
\put(625,501){\usebox{\plotpoint}}
\put(645,500){\usebox{\plotpoint}}
\put(666,500){\usebox{\plotpoint}}
\put(687,500){\usebox{\plotpoint}}
\put(708,499){\usebox{\plotpoint}}
\put(728,499){\usebox{\plotpoint}}
\put(749,498){\usebox{\plotpoint}}
\put(770,498){\usebox{\plotpoint}}
\put(791,497){\usebox{\plotpoint}}
\put(811,497){\usebox{\plotpoint}}
\put(832,496){\usebox{\plotpoint}}
\put(853,496){\usebox{\plotpoint}}
\put(874,496){\usebox{\plotpoint}}
\put(894,495){\usebox{\plotpoint}}
\put(915,495){\usebox{\plotpoint}}
\put(936,494){\usebox{\plotpoint}}
\put(957,494){\usebox{\plotpoint}}
\put(977,493){\usebox{\plotpoint}}
\put(998,493){\usebox{\plotpoint}}
\put(1019,492){\usebox{\plotpoint}}
\put(1040,492){\usebox{\plotpoint}}
\put(1060,491){\usebox{\plotpoint}}
\put(1081,491){\usebox{\plotpoint}}
\put(1102,491){\usebox{\plotpoint}}
\put(1123,490){\usebox{\plotpoint}}
\put(1143,490){\usebox{\plotpoint}}
\put(1164,489){\usebox{\plotpoint}}
\put(1185,488){\usebox{\plotpoint}}
\put(1205,487){\usebox{\plotpoint}}
\put(1226,486){\usebox{\plotpoint}}
\put(1247,484){\usebox{\plotpoint}}
\put(1249,484){\usebox{\plotpoint}}
\put(459,505){\makebox(0,0){$\triangle$}}
\put(1151,490){\makebox(0,0){$\triangle$}}
\put(1218,487){\makebox(0,0){$\triangle$}}
\put(1249,484){\makebox(0,0){$\triangle$}}
\put(459,518){\usebox{\plotpoint}}
\put(459,518){\usebox{\plotpoint}}
\put(479,517){\usebox{\plotpoint}}
\put(500,517){\usebox{\plotpoint}}
\put(521,516){\usebox{\plotpoint}}
\put(542,516){\usebox{\plotpoint}}
\put(562,516){\usebox{\plotpoint}}
\put(583,515){\usebox{\plotpoint}}
\put(604,515){\usebox{\plotpoint}}
\put(625,515){\usebox{\plotpoint}}
\put(645,514){\usebox{\plotpoint}}
\put(666,514){\usebox{\plotpoint}}
\put(687,513){\usebox{\plotpoint}}
\put(708,513){\usebox{\plotpoint}}
\put(728,513){\usebox{\plotpoint}}
\put(749,512){\usebox{\plotpoint}}
\put(770,512){\usebox{\plotpoint}}
\put(791,512){\usebox{\plotpoint}}
\put(811,511){\usebox{\plotpoint}}
\put(832,511){\usebox{\plotpoint}}
\put(853,511){\usebox{\plotpoint}}
\put(874,510){\usebox{\plotpoint}}
\put(894,510){\usebox{\plotpoint}}
\put(915,509){\usebox{\plotpoint}}
\put(936,509){\usebox{\plotpoint}}
\put(957,509){\usebox{\plotpoint}}
\put(977,508){\usebox{\plotpoint}}
\put(998,508){\usebox{\plotpoint}}
\put(1019,508){\usebox{\plotpoint}}
\put(1040,507){\usebox{\plotpoint}}
\put(1060,507){\usebox{\plotpoint}}
\put(1081,506){\usebox{\plotpoint}}
\put(1102,506){\usebox{\plotpoint}}
\put(1123,506){\usebox{\plotpoint}}
\put(1135,506){\usebox{\plotpoint}}
\put(1143,504){\usebox{\plotpoint}}
\put(1164,503){\usebox{\plotpoint}}
\put(1185,501){\usebox{\plotpoint}}
\put(1205,500){\usebox{\plotpoint}}
\put(1226,499){\usebox{\plotpoint}}
\put(1247,498){\usebox{\plotpoint}}
\put(1268,496){\usebox{\plotpoint}}
\put(1282,496){\usebox{\plotpoint}}
\put(459,518){\circle{24}}
\put(1135,506){\circle{24}}
\put(1282,496){\circle{24}}
\put(459,527){\usebox{\plotpoint}}
\put(459,527){\usebox{\plotpoint}}
\put(479,526){\usebox{\plotpoint}}
\put(500,526){\usebox{\plotpoint}}
\put(521,526){\usebox{\plotpoint}}
\put(542,526){\usebox{\plotpoint}}
\put(562,526){\usebox{\plotpoint}}
\put(583,526){\usebox{\plotpoint}}
\put(604,526){\usebox{\plotpoint}}
\put(625,526){\usebox{\plotpoint}}
\put(645,526){\usebox{\plotpoint}}
\put(666,526){\usebox{\plotpoint}}
\put(687,526){\usebox{\plotpoint}}
\put(708,526){\usebox{\plotpoint}}
\put(728,526){\usebox{\plotpoint}}
\put(747,526){\usebox{\plotpoint}}
\put(767,525){\usebox{\plotpoint}}
\put(788,524){\usebox{\plotpoint}}
\put(809,524){\usebox{\plotpoint}}
\put(829,523){\usebox{\plotpoint}}
\put(850,523){\usebox{\plotpoint}}
\put(871,522){\usebox{\plotpoint}}
\put(892,522){\usebox{\plotpoint}}
\put(913,522){\usebox{\plotpoint}}
\put(933,522){\usebox{\plotpoint}}
\put(954,522){\usebox{\plotpoint}}
\put(975,521){\usebox{\plotpoint}}
\put(995,520){\usebox{\plotpoint}}
\put(1016,519){\usebox{\plotpoint}}
\put(1037,518){\usebox{\plotpoint}}
\put(1058,516){\usebox{\plotpoint}}
\put(1078,515){\usebox{\plotpoint}}
\put(1099,513){\usebox{\plotpoint}}
\put(1120,512){\usebox{\plotpoint}}
\put(1140,511){\usebox{\plotpoint}}
\put(1161,511){\usebox{\plotpoint}}
\put(1182,509){\usebox{\plotpoint}}
\put(1203,508){\usebox{\plotpoint}}
\put(1223,506){\usebox{\plotpoint}}
\put(1244,505){\usebox{\plotpoint}}
\put(1265,504){\usebox{\plotpoint}}
\put(1285,501){\usebox{\plotpoint}}
\put(1305,495){\usebox{\plotpoint}}
\put(1325,489){\usebox{\plotpoint}}
\put(1329,489){\usebox{\plotpoint}}
\put(459,527){\raisebox{-0.0pt}{\makebox(0,0){$\bigtriangledown$}}}
\put(747,526){\raisebox{-0.0pt}{\makebox(0,0){$\bigtriangledown$}}}
\put(866,523){\raisebox{-0.0pt}{\makebox(0,0){$\bigtriangledown$}}}
\put(957,522){\raisebox{-0.0pt}{\makebox(0,0){$\bigtriangledown$}}}
\put(1035,519){\raisebox{-0.0pt}{\makebox(0,0){$\bigtriangledown$}}}
\put(1103,513){\raisebox{-0.0pt}{\makebox(0,0){$\bigtriangledown$}}}
\put(1165,511){\raisebox{-0.0pt}{\makebox(0,0){$\bigtriangledown$}}}
\put(1223,507){\raisebox{-0.0pt}{\makebox(0,0){$\bigtriangledown$}}}
\put(1276,504){\raisebox{-0.0pt}{\makebox(0,0){$\bigtriangledown$}}}
\put(1329,489){\raisebox{-0.0pt}{\makebox(0,0){$\bigtriangledown$}}}
\put(400,742){\makebox(0,0)[l]{$\sigma=1.50$}}
\put(400,697){\makebox(0,0)[l]{$\sigma=1.00$}}
\put(400,652){\makebox(0,0)[l]{$\sigma=0.50$}}
\put(400,607){\makebox(0,0)[l]{$\sigma=0.25$}}
\put(738,742){\makebox(0,0)[l]{$\sigma=0.00$}}
\put(317,742){\raisebox{-1.2pt}{\makebox(0,0){$\Diamond$}}}
\put(317,697){\raisebox{-1.2pt}{\makebox(0,0){$\Box$}}}
\put(317,652){\makebox(0,0){$\triangle$}}
\put(317,607){\circle{24}}
\put(655,742){\raisebox{-0.0pt}{\makebox(0,0){$\bigtriangledown$}}}
\end{picture}

  { \small
  Fig. 7. Ratio $\langle A_t \rangle / \langle q_l \rangle$ as a
  function of $\ell$ for different measures parametrized
  by $\sigma \geq 0$. In the well-defined phase this
  scale-invariant quantity stays almost constant and
  somewhat below the maximum value $\frac{\sqrt{3}}{4} \approx 0.433$ for
  equilateral triangles. For the scale-invariant measure ($\sigma = 0$)
  the upper curve results and the expectation values decrease slightly with
  increasing $\sigma$.
  Configurations and statistics are the same as in Fig. 6.
  }
\end{figure}

\newpage
\begin{figure}[p]

\setlength{\unitlength}{0.240900pt}
\ifx\plotpoint\undefined\newsavebox{\plotpoint}\fi
\sbox{\plotpoint}{\rule[-0.175pt]{0.350pt}{0.350pt}}%
\begin{picture}(1500,900)(0,0)
\tenrm
\sbox{\plotpoint}{\rule[-0.175pt]{0.350pt}{0.350pt}}%
\put(264,661){\rule[-0.175pt]{282.335pt}{0.350pt}}
\put(264,158){\rule[-0.175pt]{4.818pt}{0.350pt}}
\put(242,158){\makebox(0,0)[r]{-0.100}}
\put(1416,158){\rule[-0.175pt]{4.818pt}{0.350pt}}
\put(264,284){\rule[-0.175pt]{4.818pt}{0.350pt}}
\put(242,284){\makebox(0,0)[r]{-0.075}}
\put(1416,284){\rule[-0.175pt]{4.818pt}{0.350pt}}
\put(264,410){\rule[-0.175pt]{4.818pt}{0.350pt}}
\put(242,410){\makebox(0,0)[r]{-0.050}}
\put(1416,410){\rule[-0.175pt]{4.818pt}{0.350pt}}
\put(264,535){\rule[-0.175pt]{4.818pt}{0.350pt}}
\put(242,535){\makebox(0,0)[r]{-0.025}}
\put(1416,535){\rule[-0.175pt]{4.818pt}{0.350pt}}
\put(264,661){\rule[-0.175pt]{4.818pt}{0.350pt}}
\put(242,661){\makebox(0,0)[r]{0.000}}
\put(1416,661){\rule[-0.175pt]{4.818pt}{0.350pt}}
\put(264,787){\rule[-0.175pt]{4.818pt}{0.350pt}}
\put(242,787){\makebox(0,0)[r]{0.025}}
\put(1416,787){\rule[-0.175pt]{4.818pt}{0.350pt}}
\put(264,158){\rule[-0.175pt]{0.350pt}{4.818pt}}
\put(264,113){\makebox(0,0){-0.1}}
\put(264,767){\rule[-0.175pt]{0.350pt}{4.818pt}}
\put(459,158){\rule[-0.175pt]{0.350pt}{4.818pt}}
\put(459,113){\makebox(0,0){0}}
\put(459,767){\rule[-0.175pt]{0.350pt}{4.818pt}}
\put(655,158){\rule[-0.175pt]{0.350pt}{4.818pt}}
\put(655,113){\makebox(0,0){0.1}}
\put(655,767){\rule[-0.175pt]{0.350pt}{4.818pt}}
\put(850,158){\rule[-0.175pt]{0.350pt}{4.818pt}}
\put(850,113){\makebox(0,0){0.2}}
\put(850,767){\rule[-0.175pt]{0.350pt}{4.818pt}}
\put(1045,158){\rule[-0.175pt]{0.350pt}{4.818pt}}
\put(1045,113){\makebox(0,0){0.3}}
\put(1045,767){\rule[-0.175pt]{0.350pt}{4.818pt}}
\put(1241,158){\rule[-0.175pt]{0.350pt}{4.818pt}}
\put(1241,113){\makebox(0,0){0.4}}
\put(1241,767){\rule[-0.175pt]{0.350pt}{4.818pt}}
\put(1436,158){\rule[-0.175pt]{0.350pt}{4.818pt}}
\put(1436,113){\makebox(0,0){0.5}}
\put(1436,767){\rule[-0.175pt]{0.350pt}{4.818pt}}
\put(264,158){\rule[-0.175pt]{282.335pt}{0.350pt}}
\put(1436,158){\rule[-0.175pt]{0.350pt}{151.526pt}}
\put(264,787){\rule[-0.175pt]{282.335pt}{0.350pt}}
\put(45,472){\makebox(0,0)[l]{\shortstack{$\langle \delta_t \rangle$}}}
\put(850,68){\makebox(0,0){$\ell$}}
\put(264,158){\rule[-0.175pt]{0.350pt}{151.526pt}}
\sbox{\plotpoint}{\rule[-0.250pt]{0.500pt}{0.500pt}}%
\put(459,585){\usebox{\plotpoint}}
\put(459,585){\usebox{\plotpoint}}
\put(479,581){\usebox{\plotpoint}}
\put(499,578){\usebox{\plotpoint}}
\put(520,575){\usebox{\plotpoint}}
\put(540,571){\usebox{\plotpoint}}
\put(561,568){\usebox{\plotpoint}}
\put(581,565){\usebox{\plotpoint}}
\put(602,561){\usebox{\plotpoint}}
\put(622,558){\usebox{\plotpoint}}
\put(643,555){\usebox{\plotpoint}}
\put(663,552){\usebox{\plotpoint}}
\put(684,548){\usebox{\plotpoint}}
\put(704,545){\usebox{\plotpoint}}
\put(725,542){\usebox{\plotpoint}}
\put(745,538){\usebox{\plotpoint}}
\put(766,535){\usebox{\plotpoint}}
\put(786,532){\usebox{\plotpoint}}
\put(806,527){\usebox{\plotpoint}}
\put(825,519){\usebox{\plotpoint}}
\put(845,510){\usebox{\plotpoint}}
\put(864,502){\usebox{\plotpoint}}
\put(883,494){\usebox{\plotpoint}}
\put(902,486){\usebox{\plotpoint}}
\put(921,478){\usebox{\plotpoint}}
\put(940,469){\usebox{\plotpoint}}
\put(959,461){\usebox{\plotpoint}}
\put(978,452){\usebox{\plotpoint}}
\put(997,444){\usebox{\plotpoint}}
\put(1015,435){\usebox{\plotpoint}}
\put(1034,427){\usebox{\plotpoint}}
\put(1053,418){\usebox{\plotpoint}}
\put(1070,407){\usebox{\plotpoint}}
\put(1083,390){\usebox{\plotpoint}}
\put(1096,374){\usebox{\plotpoint}}
\put(1109,358){\usebox{\plotpoint}}
\put(1121,342){\usebox{\plotpoint}}
\put(1134,325){\usebox{\plotpoint}}
\put(1147,309){\usebox{\plotpoint}}
\put(1160,293){\usebox{\plotpoint}}
\put(1173,276){\usebox{\plotpoint}}
\put(1178,271){\usebox{\plotpoint}}
\put(459,585){\raisebox{-1.2pt}{\makebox(0,0){$\Diamond$}}}
\put(801,530){\raisebox{-1.2pt}{\makebox(0,0){$\Diamond$}}}
\put(947,467){\raisebox{-1.2pt}{\makebox(0,0){$\Diamond$}}}
\put(1066,413){\raisebox{-1.2pt}{\makebox(0,0){$\Diamond$}}}
\put(1178,271){\raisebox{-1.2pt}{\makebox(0,0){$\Diamond$}}}
\put(459,625){\usebox{\plotpoint}}
\put(459,625){\usebox{\plotpoint}}
\put(479,623){\usebox{\plotpoint}}
\put(500,621){\usebox{\plotpoint}}
\put(521,619){\usebox{\plotpoint}}
\put(541,618){\usebox{\plotpoint}}
\put(562,616){\usebox{\plotpoint}}
\put(583,614){\usebox{\plotpoint}}
\put(603,613){\usebox{\plotpoint}}
\put(624,611){\usebox{\plotpoint}}
\put(645,609){\usebox{\plotpoint}}
\put(665,608){\usebox{\plotpoint}}
\put(686,606){\usebox{\plotpoint}}
\put(707,604){\usebox{\plotpoint}}
\put(727,603){\usebox{\plotpoint}}
\put(748,601){\usebox{\plotpoint}}
\put(769,599){\usebox{\plotpoint}}
\put(790,598){\usebox{\plotpoint}}
\put(810,597){\usebox{\plotpoint}}
\put(831,596){\usebox{\plotpoint}}
\put(852,595){\usebox{\plotpoint}}
\put(872,594){\usebox{\plotpoint}}
\put(893,593){\usebox{\plotpoint}}
\put(914,591){\usebox{\plotpoint}}
\put(934,589){\usebox{\plotpoint}}
\put(955,586){\usebox{\plotpoint}}
\put(976,584){\usebox{\plotpoint}}
\put(996,581){\usebox{\plotpoint}}
\put(1017,578){\usebox{\plotpoint}}
\put(1036,571){\usebox{\plotpoint}}
\put(1055,563){\usebox{\plotpoint}}
\put(1075,556){\usebox{\plotpoint}}
\put(1094,548){\usebox{\plotpoint}}
\put(1113,541){\usebox{\plotpoint}}
\put(1133,533){\usebox{\plotpoint}}
\put(1152,525){\usebox{\plotpoint}}
\put(1171,518){\usebox{\plotpoint}}
\put(1191,511){\usebox{\plotpoint}}
\put(1211,505){\usebox{\plotpoint}}
\put(1231,500){\usebox{\plotpoint}}
\put(1251,494){\usebox{\plotpoint}}
\put(1268,483){\usebox{\plotpoint}}
\put(1284,470){\usebox{\plotpoint}}
\put(1301,457){\usebox{\plotpoint}}
\put(1317,445){\usebox{\plotpoint}}
\put(1329,436){\usebox{\plotpoint}}
\put(459,625){\raisebox{-1.2pt}{\makebox(0,0){$\Box$}}}
\put(781,599){\raisebox{-1.2pt}{\makebox(0,0){$\Box$}}}
\put(913,592){\raisebox{-1.2pt}{\makebox(0,0){$\Box$}}}
\put(1017,579){\raisebox{-1.2pt}{\makebox(0,0){$\Box$}}}
\put(1188,512){\raisebox{-1.2pt}{\makebox(0,0){$\Box$}}}
\put(1257,493){\raisebox{-1.2pt}{\makebox(0,0){$\Box$}}}
\put(1329,436){\raisebox{-1.2pt}{\makebox(0,0){$\Box$}}}
\put(459,609){\usebox{\plotpoint}}
\put(459,609){\usebox{\plotpoint}}
\put(479,607){\usebox{\plotpoint}}
\put(500,605){\usebox{\plotpoint}}
\put(521,604){\usebox{\plotpoint}}
\put(541,602){\usebox{\plotpoint}}
\put(562,600){\usebox{\plotpoint}}
\put(583,599){\usebox{\plotpoint}}
\put(603,597){\usebox{\plotpoint}}
\put(624,595){\usebox{\plotpoint}}
\put(645,594){\usebox{\plotpoint}}
\put(665,592){\usebox{\plotpoint}}
\put(686,590){\usebox{\plotpoint}}
\put(707,589){\usebox{\plotpoint}}
\put(727,587){\usebox{\plotpoint}}
\put(748,585){\usebox{\plotpoint}}
\put(769,584){\usebox{\plotpoint}}
\put(790,582){\usebox{\plotpoint}}
\put(810,581){\usebox{\plotpoint}}
\put(831,579){\usebox{\plotpoint}}
\put(852,577){\usebox{\plotpoint}}
\put(872,576){\usebox{\plotpoint}}
\put(893,574){\usebox{\plotpoint}}
\put(914,572){\usebox{\plotpoint}}
\put(934,571){\usebox{\plotpoint}}
\put(955,569){\usebox{\plotpoint}}
\put(976,567){\usebox{\plotpoint}}
\put(996,566){\usebox{\plotpoint}}
\put(1017,564){\usebox{\plotpoint}}
\put(1038,562){\usebox{\plotpoint}}
\put(1059,561){\usebox{\plotpoint}}
\put(1079,559){\usebox{\plotpoint}}
\put(1100,558){\usebox{\plotpoint}}
\put(1121,556){\usebox{\plotpoint}}
\put(1141,554){\usebox{\plotpoint}}
\put(1162,551){\usebox{\plotpoint}}
\put(1182,548){\usebox{\plotpoint}}
\put(1203,544){\usebox{\plotpoint}}
\put(1223,540){\usebox{\plotpoint}}
\put(1243,535){\usebox{\plotpoint}}
\put(1249,534){\usebox{\plotpoint}}
\put(459,609){\makebox(0,0){$\triangle$}}
\put(1151,554){\makebox(0,0){$\triangle$}}
\put(1218,542){\makebox(0,0){$\triangle$}}
\put(1249,534){\makebox(0,0){$\triangle$}}
\put(459,599){\usebox{\plotpoint}}
\put(459,599){\usebox{\plotpoint}}
\put(479,597){\usebox{\plotpoint}}
\put(500,596){\usebox{\plotpoint}}
\put(521,594){\usebox{\plotpoint}}
\put(541,593){\usebox{\plotpoint}}
\put(562,591){\usebox{\plotpoint}}
\put(583,590){\usebox{\plotpoint}}
\put(603,589){\usebox{\plotpoint}}
\put(624,587){\usebox{\plotpoint}}
\put(645,586){\usebox{\plotpoint}}
\put(666,584){\usebox{\plotpoint}}
\put(686,583){\usebox{\plotpoint}}
\put(707,582){\usebox{\plotpoint}}
\put(728,580){\usebox{\plotpoint}}
\put(748,579){\usebox{\plotpoint}}
\put(769,577){\usebox{\plotpoint}}
\put(790,576){\usebox{\plotpoint}}
\put(811,575){\usebox{\plotpoint}}
\put(831,573){\usebox{\plotpoint}}
\put(852,572){\usebox{\plotpoint}}
\put(873,570){\usebox{\plotpoint}}
\put(893,569){\usebox{\plotpoint}}
\put(914,567){\usebox{\plotpoint}}
\put(935,566){\usebox{\plotpoint}}
\put(955,565){\usebox{\plotpoint}}
\put(976,563){\usebox{\plotpoint}}
\put(997,562){\usebox{\plotpoint}}
\put(1018,560){\usebox{\plotpoint}}
\put(1038,559){\usebox{\plotpoint}}
\put(1059,558){\usebox{\plotpoint}}
\put(1080,556){\usebox{\plotpoint}}
\put(1100,555){\usebox{\plotpoint}}
\put(1121,553){\usebox{\plotpoint}}
\put(1135,553){\usebox{\plotpoint}}
\put(1142,550){\usebox{\plotpoint}}
\put(1162,546){\usebox{\plotpoint}}
\put(1182,542){\usebox{\plotpoint}}
\put(1203,538){\usebox{\plotpoint}}
\put(1223,534){\usebox{\plotpoint}}
\put(1243,529){\usebox{\plotpoint}}
\put(1264,525){\usebox{\plotpoint}}
\put(1282,522){\usebox{\plotpoint}}
\put(459,599){\circle{24}}
\put(1135,553){\circle{24}}
\put(1282,522){\circle{24}}
\put(459,584){\usebox{\plotpoint}}
\put(459,584){\usebox{\plotpoint}}
\put(479,583){\usebox{\plotpoint}}
\put(500,582){\usebox{\plotpoint}}
\put(521,582){\usebox{\plotpoint}}
\put(541,581){\usebox{\plotpoint}}
\put(562,581){\usebox{\plotpoint}}
\put(583,580){\usebox{\plotpoint}}
\put(604,580){\usebox{\plotpoint}}
\put(624,579){\usebox{\plotpoint}}
\put(645,579){\usebox{\plotpoint}}
\put(666,578){\usebox{\plotpoint}}
\put(687,578){\usebox{\plotpoint}}
\put(707,577){\usebox{\plotpoint}}
\put(728,577){\usebox{\plotpoint}}
\put(747,577){\usebox{\plotpoint}}
\put(767,575){\usebox{\plotpoint}}
\put(788,573){\usebox{\plotpoint}}
\put(809,571){\usebox{\plotpoint}}
\put(829,569){\usebox{\plotpoint}}
\put(850,567){\usebox{\plotpoint}}
\put(871,565){\usebox{\plotpoint}}
\put(891,564){\usebox{\plotpoint}}
\put(912,563){\usebox{\plotpoint}}
\put(933,562){\usebox{\plotpoint}}
\put(953,561){\usebox{\plotpoint}}
\put(974,558){\usebox{\plotpoint}}
\put(995,556){\usebox{\plotpoint}}
\put(1015,553){\usebox{\plotpoint}}
\put(1036,550){\usebox{\plotpoint}}
\put(1056,545){\usebox{\plotpoint}}
\put(1076,541){\usebox{\plotpoint}}
\put(1096,536){\usebox{\plotpoint}}
\put(1117,533){\usebox{\plotpoint}}
\put(1137,531){\usebox{\plotpoint}}
\put(1158,528){\usebox{\plotpoint}}
\put(1178,524){\usebox{\plotpoint}}
\put(1199,519){\usebox{\plotpoint}}
\put(1219,514){\usebox{\plotpoint}}
\put(1239,510){\usebox{\plotpoint}}
\put(1259,505){\usebox{\plotpoint}}
\put(1279,500){\usebox{\plotpoint}}
\put(1299,493){\usebox{\plotpoint}}
\put(1318,486){\usebox{\plotpoint}}
\put(1329,483){\usebox{\plotpoint}}
\put(459,584){\raisebox{-0.0pt}{\makebox(0,0){$\bigtriangledown$}}}
\put(747,577){\raisebox{-0.0pt}{\makebox(0,0){$\bigtriangledown$}}}
\put(866,566){\raisebox{-0.0pt}{\makebox(0,0){$\bigtriangledown$}}}
\put(957,561){\raisebox{-0.0pt}{\makebox(0,0){$\bigtriangledown$}}}
\put(1035,551){\raisebox{-0.0pt}{\makebox(0,0){$\bigtriangledown$}}}
\put(1103,535){\raisebox{-0.0pt}{\makebox(0,0){$\bigtriangledown$}}}
\put(1165,528){\raisebox{-0.0pt}{\makebox(0,0){$\bigtriangledown$}}}
\put(1223,514){\raisebox{-0.0pt}{\makebox(0,0){$\bigtriangledown$}}}
\put(1276,502){\raisebox{-0.0pt}{\makebox(0,0){$\bigtriangledown$}}}
\put(1329,483){\raisebox{-0.0pt}{\makebox(0,0){$\bigtriangledown$}}}
\put(400,382){\makebox(0,0)[l]{$\sigma=1.50$}}
\put(400,337){\makebox(0,0)[l]{$\sigma=1.00$}}
\put(400,292){\makebox(0,0)[l]{$\sigma=0.50$}}
\put(400,247){\makebox(0,0)[l]{$\sigma=0.25$}}
\put(400,202){\makebox(0,0)[l]{$\sigma=0.00$}}
\put(317,382){\raisebox{-1.2pt}{\makebox(0,0){$\Diamond$}}}
\put(317,337){\raisebox{-1.2pt}{\makebox(0,0){$\Box$}}}
\put(317,292){\makebox(0,0){$\triangle$}}
\put(317,247){\circle{24}}
\put(317,202){\raisebox{-0.0pt}{\makebox(0,0){$\bigtriangledown$}}}
\end{picture}

  { \small
  Fig. 8. Expectation value $\langle \delta_t \rangle$ as a function of
  $\ell$ for different measure parameters $\sigma$. It is remarkable
  that $\langle \delta_t \rangle$ stays always negative even after
  the transition to positive curvature.
  The curves lie close together
  for $0 \leq \sigma \leq 1$ while the behavior differs significantly
  for $\sigma = 1.5$. The configurations are those of Fig. 6.
  \\[2.0cm]
  }
\end{figure}

\newpage
\begin{figure}[p]

\setlength{\unitlength}{0.240900pt}
\ifx\plotpoint\undefined\newsavebox{\plotpoint}\fi
\sbox{\plotpoint}{\rule[-0.175pt]{0.350pt}{0.350pt}}%
\begin{picture}(1500,900)(0,0)
\tenrm
\sbox{\plotpoint}{\rule[-0.175pt]{0.350pt}{0.350pt}}%
\put(264,263){\rule[-0.175pt]{282.335pt}{0.350pt}}
\put(264,158){\rule[-0.175pt]{4.818pt}{0.350pt}}
\put(242,158){\makebox(0,0)[r]{-10}}
\put(1416,158){\rule[-0.175pt]{4.818pt}{0.350pt}}
\put(264,263){\rule[-0.175pt]{4.818pt}{0.350pt}}
\put(242,263){\makebox(0,0)[r]{0}}
\put(1416,263){\rule[-0.175pt]{4.818pt}{0.350pt}}
\put(264,368){\rule[-0.175pt]{4.818pt}{0.350pt}}
\put(242,368){\makebox(0,0)[r]{10}}
\put(1416,368){\rule[-0.175pt]{4.818pt}{0.350pt}}
\put(264,473){\rule[-0.175pt]{4.818pt}{0.350pt}}
\put(242,473){\makebox(0,0)[r]{20}}
\put(1416,473){\rule[-0.175pt]{4.818pt}{0.350pt}}
\put(264,577){\rule[-0.175pt]{4.818pt}{0.350pt}}
\put(242,577){\makebox(0,0)[r]{30}}
\put(1416,577){\rule[-0.175pt]{4.818pt}{0.350pt}}
\put(264,682){\rule[-0.175pt]{4.818pt}{0.350pt}}
\put(242,682){\makebox(0,0)[r]{40}}
\put(1416,682){\rule[-0.175pt]{4.818pt}{0.350pt}}
\put(264,787){\rule[-0.175pt]{4.818pt}{0.350pt}}
\put(242,787){\makebox(0,0)[r]{50}}
\put(1416,787){\rule[-0.175pt]{4.818pt}{0.350pt}}
\put(381,158){\rule[-0.175pt]{0.350pt}{4.818pt}}
\put(381,113){\makebox(0,0){-1}}
\put(381,767){\rule[-0.175pt]{0.350pt}{4.818pt}}
\put(616,158){\rule[-0.175pt]{0.350pt}{4.818pt}}
\put(616,113){\makebox(0,0){-0.5}}
\put(616,767){\rule[-0.175pt]{0.350pt}{4.818pt}}
\put(850,158){\rule[-0.175pt]{0.350pt}{4.818pt}}
\put(850,113){\makebox(0,0){0}}
\put(850,767){\rule[-0.175pt]{0.350pt}{4.818pt}}
\put(1084,158){\rule[-0.175pt]{0.350pt}{4.818pt}}
\put(1084,113){\makebox(0,0){0.5}}
\put(1084,767){\rule[-0.175pt]{0.350pt}{4.818pt}}
\put(1319,158){\rule[-0.175pt]{0.350pt}{4.818pt}}
\put(1319,113){\makebox(0,0){1}}
\put(1319,767){\rule[-0.175pt]{0.350pt}{4.818pt}}
\put(264,158){\rule[-0.175pt]{282.335pt}{0.350pt}}
\put(1436,158){\rule[-0.175pt]{0.350pt}{151.526pt}}
\put(264,787){\rule[-0.175pt]{282.335pt}{0.350pt}}
\put(45,472){\makebox(0,0)[l]{\shortstack{$\langle \tilde{R} \rangle$}}}
\put(850,68){\makebox(0,0){$\ell$}}
\put(264,158){\rule[-0.175pt]{0.350pt}{151.526pt}}
\sbox{\plotpoint}{\rule[-0.250pt]{0.500pt}{0.500pt}}%
\put(381,761){\makebox(0,0)[l]{$n=5$}}
\put(1342,754){\usebox{\plotpoint}}
\put(1342,754){\usebox{\plotpoint}}
\put(1321,752){\usebox{\plotpoint}}
\put(1300,751){\usebox{\plotpoint}}
\put(1280,745){\usebox{\plotpoint}}
\put(1260,738){\usebox{\plotpoint}}
\put(1243,727){\usebox{\plotpoint}}
\put(1229,713){\usebox{\plotpoint}}
\put(1214,698){\usebox{\plotpoint}}
\put(1199,683){\usebox{\plotpoint}}
\put(1188,666){\usebox{\plotpoint}}
\put(1177,649){\usebox{\plotpoint}}
\put(1166,631){\usebox{\plotpoint}}
\put(1155,614){\usebox{\plotpoint}}
\put(1143,596){\usebox{\plotpoint}}
\put(1132,579){\usebox{\plotpoint}}
\put(1123,560){\usebox{\plotpoint}}
\put(1113,542){\usebox{\plotpoint}}
\put(1103,524){\usebox{\plotpoint}}
\put(1093,505){\usebox{\plotpoint}}
\put(1083,487){\usebox{\plotpoint}}
\put(1073,469){\usebox{\plotpoint}}
\put(1064,451){\usebox{\plotpoint}}
\put(1054,432){\usebox{\plotpoint}}
\put(1039,418){\usebox{\plotpoint}}
\put(1023,405){\usebox{\plotpoint}}
\put(1007,392){\usebox{\plotpoint}}
\put(990,379){\usebox{\plotpoint}}
\put(974,367){\usebox{\plotpoint}}
\put(958,354){\usebox{\plotpoint}}
\put(942,341){\usebox{\plotpoint}}
\put(925,328){\usebox{\plotpoint}}
\put(909,315){\usebox{\plotpoint}}
\put(893,302){\usebox{\plotpoint}}
\put(876,290){\usebox{\plotpoint}}
\put(860,277){\usebox{\plotpoint}}
\put(842,266){\usebox{\plotpoint}}
\put(823,260){\usebox{\plotpoint}}
\put(803,254){\usebox{\plotpoint}}
\put(783,248){\usebox{\plotpoint}}
\put(763,241){\usebox{\plotpoint}}
\put(743,235){\usebox{\plotpoint}}
\put(724,229){\usebox{\plotpoint}}
\put(704,223){\usebox{\plotpoint}}
\put(684,217){\usebox{\plotpoint}}
\put(664,210){\usebox{\plotpoint}}
\put(644,205){\usebox{\plotpoint}}
\put(624,202){\usebox{\plotpoint}}
\put(603,199){\usebox{\plotpoint}}
\put(583,196){\usebox{\plotpoint}}
\put(562,193){\usebox{\plotpoint}}
\put(541,193){\usebox{\plotpoint}}
\put(520,192){\usebox{\plotpoint}}
\put(500,191){\usebox{\plotpoint}}
\put(479,191){\usebox{\plotpoint}}
\put(458,191){\usebox{\plotpoint}}
\put(437,191){\usebox{\plotpoint}}
\put(417,191){\usebox{\plotpoint}}
\put(396,191){\usebox{\plotpoint}}
\put(375,191){\usebox{\plotpoint}}
\put(358,191){\usebox{\plotpoint}}
\put(325,761){\raisebox{-1.2pt}{\makebox(0,0){$\Diamond$}}}
\put(1342,754){\makebox(0,0){$\bigtriangledown$}}
\put(1299,751){\makebox(0,0){$\bigtriangledown$}}
\put(1252,736){\makebox(0,0){$\bigtriangledown$}}
\put(1198,682){\makebox(0,0){$\bigtriangledown$}}
\put(1134,581){\makebox(0,0){$\bigtriangledown$}}
\put(1051,427){\makebox(0,0){$\bigtriangledown$}}
\put(850,269){\makebox(0,0){$\bigtriangledown$}}
\put(649,206){\makebox(0,0){$\bigtriangledown$}}
\put(566,194){\makebox(0,0){$\bigtriangledown$}}
\put(502,192){\makebox(0,0){$\bigtriangledown$}}
\put(448,191){\makebox(0,0){$\bigtriangledown$}}
\put(401,192){\makebox(0,0){$\bigtriangledown$}}
\put(358,191){\makebox(0,0){$\bigtriangledown$}}
\put(381,716){\makebox(0,0)[l]{$n=1$}}
\put(1342,756){\usebox{\plotpoint}}
\put(1342,756){\usebox{\plotpoint}}
\put(1321,755){\usebox{\plotpoint}}
\put(1300,754){\usebox{\plotpoint}}
\put(1279,752){\usebox{\plotpoint}}
\put(1259,751){\usebox{\plotpoint}}
\put(1238,748){\usebox{\plotpoint}}
\put(1218,744){\usebox{\plotpoint}}
\put(1197,739){\usebox{\plotpoint}}
\put(1180,728){\usebox{\plotpoint}}
\put(1162,717){\usebox{\plotpoint}}
\put(1145,706){\usebox{\plotpoint}}
\put(1131,692){\usebox{\plotpoint}}
\put(1123,673){\usebox{\plotpoint}}
\put(1116,654){\usebox{\plotpoint}}
\put(1108,634){\usebox{\plotpoint}}
\put(1101,615){\usebox{\plotpoint}}
\put(1093,596){\usebox{\plotpoint}}
\put(1086,576){\usebox{\plotpoint}}
\put(1078,557){\usebox{\plotpoint}}
\put(1071,538){\usebox{\plotpoint}}
\put(1063,518){\usebox{\plotpoint}}
\put(1056,499){\usebox{\plotpoint}}
\put(1046,481){\usebox{\plotpoint}}
\put(1032,466){\usebox{\plotpoint}}
\put(1017,451){\usebox{\plotpoint}}
\put(1002,436){\usebox{\plotpoint}}
\put(988,422){\usebox{\plotpoint}}
\put(973,407){\usebox{\plotpoint}}
\put(959,392){\usebox{\plotpoint}}
\put(944,377){\usebox{\plotpoint}}
\put(930,362){\usebox{\plotpoint}}
\put(915,348){\usebox{\plotpoint}}
\put(901,333){\usebox{\plotpoint}}
\put(886,318){\usebox{\plotpoint}}
\put(872,303){\usebox{\plotpoint}}
\put(857,288){\usebox{\plotpoint}}
\put(840,278){\usebox{\plotpoint}}
\put(820,271){\usebox{\plotpoint}}
\put(801,265){\usebox{\plotpoint}}
\put(781,259){\usebox{\plotpoint}}
\put(761,252){\usebox{\plotpoint}}
\put(741,246){\usebox{\plotpoint}}
\put(721,240){\usebox{\plotpoint}}
\put(702,233){\usebox{\plotpoint}}
\put(682,227){\usebox{\plotpoint}}
\put(662,221){\usebox{\plotpoint}}
\put(642,215){\usebox{\plotpoint}}
\put(622,212){\usebox{\plotpoint}}
\put(601,208){\usebox{\plotpoint}}
\put(581,204){\usebox{\plotpoint}}
\put(560,201){\usebox{\plotpoint}}
\put(540,200){\usebox{\plotpoint}}
\put(519,199){\usebox{\plotpoint}}
\put(498,198){\usebox{\plotpoint}}
\put(477,198){\usebox{\plotpoint}}
\put(457,197){\usebox{\plotpoint}}
\put(436,197){\usebox{\plotpoint}}
\put(415,197){\usebox{\plotpoint}}
\put(394,197){\usebox{\plotpoint}}
\put(374,197){\usebox{\plotpoint}}
\put(358,197){\usebox{\plotpoint}}
\put(325,716){\raisebox{-1.2pt}{\makebox(0,0){$\Box$}}}
\put(1342,756){\circle{24}}
\put(1299,754){\circle{24}}
\put(1252,751){\circle{24}}
\put(1198,740){\circle{24}}
\put(1134,700){\circle{24}}
\put(1051,486){\circle{24}}
\put(850,281){\circle{24}}
\put(649,217){\circle{24}}
\put(566,202){\circle{24}}
\put(502,199){\circle{24}}
\put(448,197){\circle{24}}
\put(401,197){\circle{24}}
\put(358,197){\circle{24}}
\put(381,671){\makebox(0,0)[l]{$n=0$}}
\put(1342,756){\usebox{\plotpoint}}
\put(1342,756){\usebox{\plotpoint}}
\put(1321,755){\usebox{\plotpoint}}
\put(1300,755){\usebox{\plotpoint}}
\put(1279,753){\usebox{\plotpoint}}
\put(1259,751){\usebox{\plotpoint}}
\put(1238,748){\usebox{\plotpoint}}
\put(1218,745){\usebox{\plotpoint}}
\put(1197,741){\usebox{\plotpoint}}
\put(1178,733){\usebox{\plotpoint}}
\put(1159,725){\usebox{\plotpoint}}
\put(1140,717){\usebox{\plotpoint}}
\put(1125,703){\usebox{\plotpoint}}
\put(1113,686){\usebox{\plotpoint}}
\put(1101,669){\usebox{\plotpoint}}
\put(1089,652){\usebox{\plotpoint}}
\put(1077,635){\usebox{\plotpoint}}
\put(1065,619){\usebox{\plotpoint}}
\put(1053,602){\usebox{\plotpoint}}
\put(1042,584){\usebox{\plotpoint}}
\put(1030,567){\usebox{\plotpoint}}
\put(1019,550){\usebox{\plotpoint}}
\put(1007,533){\usebox{\plotpoint}}
\put(996,515){\usebox{\plotpoint}}
\put(984,498){\usebox{\plotpoint}}
\put(972,481){\usebox{\plotpoint}}
\put(961,464){\usebox{\plotpoint}}
\put(949,446){\usebox{\plotpoint}}
\put(938,429){\usebox{\plotpoint}}
\put(926,412){\usebox{\plotpoint}}
\put(915,395){\usebox{\plotpoint}}
\put(903,377){\usebox{\plotpoint}}
\put(892,360){\usebox{\plotpoint}}
\put(880,343){\usebox{\plotpoint}}
\put(868,326){\usebox{\plotpoint}}
\put(857,309){\usebox{\plotpoint}}
\put(842,295){\usebox{\plotpoint}}
\put(823,288){\usebox{\plotpoint}}
\put(803,281){\usebox{\plotpoint}}
\put(784,274){\usebox{\plotpoint}}
\put(764,267){\usebox{\plotpoint}}
\put(745,260){\usebox{\plotpoint}}
\put(725,252){\usebox{\plotpoint}}
\put(706,245){\usebox{\plotpoint}}
\put(686,238){\usebox{\plotpoint}}
\put(667,231){\usebox{\plotpoint}}
\put(647,224){\usebox{\plotpoint}}
\put(627,220){\usebox{\plotpoint}}
\put(607,215){\usebox{\plotpoint}}
\put(586,211){\usebox{\plotpoint}}
\put(566,207){\usebox{\plotpoint}}
\put(546,205){\usebox{\plotpoint}}
\put(525,203){\usebox{\plotpoint}}
\put(504,201){\usebox{\plotpoint}}
\put(483,200){\usebox{\plotpoint}}
\put(463,200){\usebox{\plotpoint}}
\put(442,199){\usebox{\plotpoint}}
\put(421,199){\usebox{\plotpoint}}
\put(400,199){\usebox{\plotpoint}}
\put(380,199){\usebox{\plotpoint}}
\put(359,199){\usebox{\plotpoint}}
\put(358,199){\usebox{\plotpoint}}
\put(325,671){\makebox(0,0){$\triangle$}}
\put(1342,756){\makebox(0,0){$\triangle$}}
\put(1299,755){\makebox(0,0){$\triangle$}}
\put(1252,751){\makebox(0,0){$\triangle$}}
\put(1198,742){\makebox(0,0){$\triangle$}}
\put(1134,715){\makebox(0,0){$\triangle$}}
\put(1051,598){\makebox(0,0){$\triangle$}}
\put(850,298){\makebox(0,0){$\triangle$}}
\put(649,225){\makebox(0,0){$\triangle$}}
\put(566,207){\makebox(0,0){$\triangle$}}
\put(502,201){\makebox(0,0){$\triangle$}}
\put(448,200){\makebox(0,0){$\triangle$}}
\put(401,199){\makebox(0,0){$\triangle$}}
\put(358,199){\makebox(0,0){$\triangle$}}
\put(381,626){\makebox(0,0)[l]{$n=-1$}}
\put(1342,756){\usebox{\plotpoint}}
\put(1342,756){\usebox{\plotpoint}}
\put(1321,755){\usebox{\plotpoint}}
\put(1300,755){\usebox{\plotpoint}}
\put(1279,753){\usebox{\plotpoint}}
\put(1259,751){\usebox{\plotpoint}}
\put(1238,748){\usebox{\plotpoint}}
\put(1218,745){\usebox{\plotpoint}}
\put(1197,741){\usebox{\plotpoint}}
\put(1178,734){\usebox{\plotpoint}}
\put(1158,727){\usebox{\plotpoint}}
\put(1139,720){\usebox{\plotpoint}}
\put(1122,708){\usebox{\plotpoint}}
\put(1106,695){\usebox{\plotpoint}}
\put(1090,682){\usebox{\plotpoint}}
\put(1073,669){\usebox{\plotpoint}}
\put(1057,656){\usebox{\plotpoint}}
\put(1044,640){\usebox{\plotpoint}}
\put(1032,623){\usebox{\plotpoint}}
\put(1021,606){\usebox{\plotpoint}}
\put(1009,589){\usebox{\plotpoint}}
\put(998,571){\usebox{\plotpoint}}
\put(986,554){\usebox{\plotpoint}}
\put(975,537){\usebox{\plotpoint}}
\put(963,520){\usebox{\plotpoint}}
\put(952,502){\usebox{\plotpoint}}
\put(940,485){\usebox{\plotpoint}}
\put(928,468){\usebox{\plotpoint}}
\put(917,451){\usebox{\plotpoint}}
\put(905,433){\usebox{\plotpoint}}
\put(894,416){\usebox{\plotpoint}}
\put(882,399){\usebox{\plotpoint}}
\put(871,381){\usebox{\plotpoint}}
\put(859,364){\usebox{\plotpoint}}
\put(847,348){\usebox{\plotpoint}}
\put(829,337){\usebox{\plotpoint}}
\put(811,327){\usebox{\plotpoint}}
\put(793,316){\usebox{\plotpoint}}
\put(775,306){\usebox{\plotpoint}}
\put(757,295){\usebox{\plotpoint}}
\put(739,285){\usebox{\plotpoint}}
\put(722,274){\usebox{\plotpoint}}
\put(704,264){\usebox{\plotpoint}}
\put(686,253){\usebox{\plotpoint}}
\put(668,243){\usebox{\plotpoint}}
\put(650,232){\usebox{\plotpoint}}
\put(630,227){\usebox{\plotpoint}}
\put(610,222){\usebox{\plotpoint}}
\put(590,217){\usebox{\plotpoint}}
\put(569,212){\usebox{\plotpoint}}
\put(549,210){\usebox{\plotpoint}}
\put(528,207){\usebox{\plotpoint}}
\put(508,205){\usebox{\plotpoint}}
\put(487,204){\usebox{\plotpoint}}
\put(466,203){\usebox{\plotpoint}}
\put(446,201){\usebox{\plotpoint}}
\put(425,201){\usebox{\plotpoint}}
\put(404,201){\usebox{\plotpoint}}
\put(383,201){\usebox{\plotpoint}}
\put(362,201){\usebox{\plotpoint}}
\put(358,201){\usebox{\plotpoint}}
\put(325,626){\circle{24}}
\put(1342,756){\raisebox{-1.2pt}{\makebox(0,0){$\Box$}}}
\put(1299,755){\raisebox{-1.2pt}{\makebox(0,0){$\Box$}}}
\put(1252,751){\raisebox{-1.2pt}{\makebox(0,0){$\Box$}}}
\put(1198,742){\raisebox{-1.2pt}{\makebox(0,0){$\Box$}}}
\put(1134,718){\raisebox{-1.2pt}{\makebox(0,0){$\Box$}}}
\put(1051,651){\raisebox{-1.2pt}{\makebox(0,0){$\Box$}}}
\put(850,350){\raisebox{-1.2pt}{\makebox(0,0){$\Box$}}}
\put(649,232){\raisebox{-1.2pt}{\makebox(0,0){$\Box$}}}
\put(566,212){\raisebox{-1.2pt}{\makebox(0,0){$\Box$}}}
\put(502,205){\raisebox{-1.2pt}{\makebox(0,0){$\Box$}}}
\put(448,202){\raisebox{-1.2pt}{\makebox(0,0){$\Box$}}}
\put(401,201){\raisebox{-1.2pt}{\makebox(0,0){$\Box$}}}
\put(358,201){\raisebox{-1.2pt}{\makebox(0,0){$\Box$}}}
\put(381,581){\makebox(0,0)[l]{$n=-5$}}
\put(1342,756){\usebox{\plotpoint}}
\put(1342,756){\usebox{\plotpoint}}
\put(1321,755){\usebox{\plotpoint}}
\put(1300,755){\usebox{\plotpoint}}
\put(1279,753){\usebox{\plotpoint}}
\put(1259,752){\usebox{\plotpoint}}
\put(1238,750){\usebox{\plotpoint}}
\put(1217,747){\usebox{\plotpoint}}
\put(1197,744){\usebox{\plotpoint}}
\put(1177,739){\usebox{\plotpoint}}
\put(1157,734){\usebox{\plotpoint}}
\put(1137,728){\usebox{\plotpoint}}
\put(1117,721){\usebox{\plotpoint}}
\put(1098,712){\usebox{\plotpoint}}
\put(1079,704){\usebox{\plotpoint}}
\put(1060,696){\usebox{\plotpoint}}
\put(1041,688){\usebox{\plotpoint}}
\put(1021,682){\usebox{\plotpoint}}
\put(1001,675){\usebox{\plotpoint}}
\put(982,669){\usebox{\plotpoint}}
\put(962,662){\usebox{\plotpoint}}
\put(942,656){\usebox{\plotpoint}}
\put(923,650){\usebox{\plotpoint}}
\put(903,643){\usebox{\plotpoint}}
\put(883,637){\usebox{\plotpoint}}
\put(863,630){\usebox{\plotpoint}}
\put(844,623){\usebox{\plotpoint}}
\put(826,613){\usebox{\plotpoint}}
\put(807,604){\usebox{\plotpoint}}
\put(789,594){\usebox{\plotpoint}}
\put(770,585){\usebox{\plotpoint}}
\put(752,575){\usebox{\plotpoint}}
\put(733,565){\usebox{\plotpoint}}
\put(715,556){\usebox{\plotpoint}}
\put(697,546){\usebox{\plotpoint}}
\put(678,537){\usebox{\plotpoint}}
\put(660,527){\usebox{\plotpoint}}
\put(644,515){\usebox{\plotpoint}}
\put(632,498){\usebox{\plotpoint}}
\put(620,481){\usebox{\plotpoint}}
\put(608,464){\usebox{\plotpoint}}
\put(596,447){\usebox{\plotpoint}}
\put(584,430){\usebox{\plotpoint}}
\put(572,413){\usebox{\plotpoint}}
\put(560,396){\usebox{\plotpoint}}
\put(547,380){\usebox{\plotpoint}}
\put(535,363){\usebox{\plotpoint}}
\put(523,346){\usebox{\plotpoint}}
\put(511,329){\usebox{\plotpoint}}
\put(498,313){\usebox{\plotpoint}}
\put(482,300){\usebox{\plotpoint}}
\put(467,286){\usebox{\plotpoint}}
\put(451,272){\usebox{\plotpoint}}
\put(432,264){\usebox{\plotpoint}}
\put(413,256){\usebox{\plotpoint}}
\put(393,250){\usebox{\plotpoint}}
\put(373,245){\usebox{\plotpoint}}
\put(358,242){\usebox{\plotpoint}}
\put(325,581){\makebox(0,0){$\bigtriangledown$}}
\put(1342,756){\raisebox{-1.2pt}{\makebox(0,0){$\Diamond$}}}
\put(1299,755){\raisebox{-1.2pt}{\makebox(0,0){$\Diamond$}}}
\put(1252,752){\raisebox{-1.2pt}{\makebox(0,0){$\Diamond$}}}
\put(1198,745){\raisebox{-1.2pt}{\makebox(0,0){$\Diamond$}}}
\put(1134,728){\raisebox{-1.2pt}{\makebox(0,0){$\Diamond$}}}
\put(1051,692){\raisebox{-1.2pt}{\makebox(0,0){$\Diamond$}}}
\put(850,626){\raisebox{-1.2pt}{\makebox(0,0){$\Diamond$}}}
\put(649,522){\raisebox{-1.2pt}{\makebox(0,0){$\Diamond$}}}
\put(566,405){\raisebox{-1.2pt}{\makebox(0,0){$\Diamond$}}}
\put(502,317){\raisebox{-1.2pt}{\makebox(0,0){$\Diamond$}}}
\put(448,270){\raisebox{-1.2pt}{\makebox(0,0){$\Diamond$}}}
\put(401,252){\raisebox{-1.2pt}{\makebox(0,0){$\Diamond$}}}
\put(358,242){\raisebox{-1.2pt}{\makebox(0,0){$\Diamond$}}}
\end{picture}

  { \small
  Fig. 9. Investigations of different measures with dynamical triangulation
  of the 4-sphere performed by Br\"ugmann \cite{BM}. An additional term
 $n \sum_v \ln (o_v)$ in the action mimics different proposed
  measures. Besides a shift in $\ell$ the picture
  has a striking similarity to Fig. 6 if one identifies $\sigma = 0$
  with $n = -5$ (scale-invariant measure) and
  $\sigma = 1$ with $n = 0$ (uniform measure).
  Notice the small influence of $n$ in the range $-5 \leq n \leq +1$ and the
  exceptional behavior of $n = +5$.
  }
\end{figure}

\newpage
\begin{figure}[p]

\setlength{\unitlength}{0.240900pt}
\ifx\plotpoint\undefined\newsavebox{\plotpoint}\fi
\sbox{\plotpoint}{\rule[-0.175pt]{0.350pt}{0.350pt}}%
\begin{picture}(1500,900)(0,0)
\tenrm
\sbox{\plotpoint}{\rule[-0.175pt]{0.350pt}{0.350pt}}%
\put(264,158){\rule[-0.175pt]{4.818pt}{0.350pt}}
\put(242,158){\makebox(0,0)[r]{-12}}
\put(1416,158){\rule[-0.175pt]{4.818pt}{0.350pt}}
\put(264,263){\rule[-0.175pt]{4.818pt}{0.350pt}}
\put(242,263){\makebox(0,0)[r]{-10}}
\put(1416,263){\rule[-0.175pt]{4.818pt}{0.350pt}}
\put(264,368){\rule[-0.175pt]{4.818pt}{0.350pt}}
\put(242,368){\makebox(0,0)[r]{-8}}
\put(1416,368){\rule[-0.175pt]{4.818pt}{0.350pt}}
\put(264,473){\rule[-0.175pt]{4.818pt}{0.350pt}}
\put(242,473){\makebox(0,0)[r]{-6}}
\put(1416,473){\rule[-0.175pt]{4.818pt}{0.350pt}}
\put(264,577){\rule[-0.175pt]{4.818pt}{0.350pt}}
\put(242,577){\makebox(0,0)[r]{-4}}
\put(1416,577){\rule[-0.175pt]{4.818pt}{0.350pt}}
\put(264,682){\rule[-0.175pt]{4.818pt}{0.350pt}}
\put(242,682){\makebox(0,0)[r]{-2}}
\put(1416,682){\rule[-0.175pt]{4.818pt}{0.350pt}}
\put(264,787){\rule[-0.175pt]{4.818pt}{0.350pt}}
\put(242,787){\makebox(0,0)[r]{0}}
\put(1416,787){\rule[-0.175pt]{4.818pt}{0.350pt}}
\put(264,158){\rule[-0.175pt]{0.350pt}{4.818pt}}
\put(264,113){\makebox(0,0){-0.1}}
\put(264,767){\rule[-0.175pt]{0.350pt}{4.818pt}}
\put(431,158){\rule[-0.175pt]{0.350pt}{4.818pt}}
\put(431,113){\makebox(0,0){0}}
\put(431,767){\rule[-0.175pt]{0.350pt}{4.818pt}}
\put(599,158){\rule[-0.175pt]{0.350pt}{4.818pt}}
\put(599,113){\makebox(0,0){0.1}}
\put(599,767){\rule[-0.175pt]{0.350pt}{4.818pt}}
\put(766,158){\rule[-0.175pt]{0.350pt}{4.818pt}}
\put(766,113){\makebox(0,0){0.2}}
\put(766,767){\rule[-0.175pt]{0.350pt}{4.818pt}}
\put(934,158){\rule[-0.175pt]{0.350pt}{4.818pt}}
\put(934,113){\makebox(0,0){0.3}}
\put(934,767){\rule[-0.175pt]{0.350pt}{4.818pt}}
\put(1101,158){\rule[-0.175pt]{0.350pt}{4.818pt}}
\put(1101,113){\makebox(0,0){0.4}}
\put(1101,767){\rule[-0.175pt]{0.350pt}{4.818pt}}
\put(1269,158){\rule[-0.175pt]{0.350pt}{4.818pt}}
\put(1269,113){\makebox(0,0){0.5}}
\put(1269,767){\rule[-0.175pt]{0.350pt}{4.818pt}}
\put(1436,158){\rule[-0.175pt]{0.350pt}{4.818pt}}
\put(1436,113){\makebox(0,0){0.6}}
\put(1436,767){\rule[-0.175pt]{0.350pt}{4.818pt}}
\put(264,158){\rule[-0.175pt]{282.335pt}{0.350pt}}
\put(1436,158){\rule[-0.175pt]{0.350pt}{151.526pt}}
\put(264,787){\rule[-0.175pt]{282.335pt}{0.350pt}}
\put(45,472){\makebox(0,0)[l]{\shortstack{$\langle\tilde{R}\rangle$}}}
\put(850,68){\makebox(0,0){$\ell$}}
\put(264,158){\rule[-0.175pt]{0.350pt}{151.526pt}}
\sbox{\plotpoint}{\rule[-0.250pt]{0.500pt}{0.500pt}}%
\put(391,741){\makebox(0,0)[l]{$N_0=3^4$}}
\put(431,261){\usebox{\plotpoint}}
\put(431,261){\usebox{\plotpoint}}
\put(451,265){\usebox{\plotpoint}}
\put(471,270){\usebox{\plotpoint}}
\put(491,274){\usebox{\plotpoint}}
\put(512,279){\usebox{\plotpoint}}
\put(532,283){\usebox{\plotpoint}}
\put(552,288){\usebox{\plotpoint}}
\put(572,292){\usebox{\plotpoint}}
\put(593,297){\usebox{\plotpoint}}
\put(613,301){\usebox{\plotpoint}}
\put(633,306){\usebox{\plotpoint}}
\put(653,310){\usebox{\plotpoint}}
\put(674,315){\usebox{\plotpoint}}
\put(694,319){\usebox{\plotpoint}}
\put(714,324){\usebox{\plotpoint}}
\put(734,328){\usebox{\plotpoint}}
\put(755,333){\usebox{\plotpoint}}
\put(775,337){\usebox{\plotpoint}}
\put(795,342){\usebox{\plotpoint}}
\put(815,346){\usebox{\plotpoint}}
\put(836,351){\usebox{\plotpoint}}
\put(856,355){\usebox{\plotpoint}}
\put(876,360){\usebox{\plotpoint}}
\put(896,364){\usebox{\plotpoint}}
\put(916,371){\usebox{\plotpoint}}
\put(934,381){\usebox{\plotpoint}}
\put(952,392){\usebox{\plotpoint}}
\put(969,403){\usebox{\plotpoint}}
\put(987,413){\usebox{\plotpoint}}
\put(1005,424){\usebox{\plotpoint}}
\put(1023,435){\usebox{\plotpoint}}
\put(1041,445){\usebox{\plotpoint}}
\put(1059,456){\usebox{\plotpoint}}
\put(1076,466){\usebox{\plotpoint}}
\put(1094,477){\usebox{\plotpoint}}
\put(1110,489){\usebox{\plotpoint}}
\put(1117,509){\usebox{\plotpoint}}
\put(1125,528){\usebox{\plotpoint}}
\put(1132,548){\usebox{\plotpoint}}
\put(1140,567){\usebox{\plotpoint}}
\put(1147,586){\usebox{\plotpoint}}
\put(1155,606){\usebox{\plotpoint}}
\put(1162,625){\usebox{\plotpoint}}
\put(1170,644){\usebox{\plotpoint}}
\put(1177,664){\usebox{\plotpoint}}
\put(1182,676){\usebox{\plotpoint}}
\put(320,741){\raisebox{-1.2pt}{\makebox(0,0){$\Diamond$}}}
\put(431,261){\raisebox{-1.2pt}{\makebox(0,0){$\Diamond$}}}
\put(911,368){\raisebox{-1.2pt}{\makebox(0,0){$\Diamond$}}}
\put(1109,486){\raisebox{-1.2pt}{\makebox(0,0){$\Diamond$}}}
\put(1182,676){\raisebox{-1.2pt}{\makebox(0,0){$\Diamond$}}}
\put(391,696){\makebox(0,0)[l]{$N_0=4^4$}}
\put(431,250){\usebox{\plotpoint}}
\put(431,250){\usebox{\plotpoint}}
\put(451,253){\usebox{\plotpoint}}
\put(471,257){\usebox{\plotpoint}}
\put(492,261){\usebox{\plotpoint}}
\put(512,265){\usebox{\plotpoint}}
\put(532,269){\usebox{\plotpoint}}
\put(553,273){\usebox{\plotpoint}}
\put(573,277){\usebox{\plotpoint}}
\put(594,281){\usebox{\plotpoint}}
\put(614,285){\usebox{\plotpoint}}
\put(634,289){\usebox{\plotpoint}}
\put(655,293){\usebox{\plotpoint}}
\put(675,296){\usebox{\plotpoint}}
\put(695,300){\usebox{\plotpoint}}
\put(716,304){\usebox{\plotpoint}}
\put(736,309){\usebox{\plotpoint}}
\put(756,313){\usebox{\plotpoint}}
\put(777,317){\usebox{\plotpoint}}
\put(797,322){\usebox{\plotpoint}}
\put(817,326){\usebox{\plotpoint}}
\put(837,332){\usebox{\plotpoint}}
\put(857,339){\usebox{\plotpoint}}
\put(877,345){\usebox{\plotpoint}}
\put(896,352){\usebox{\plotpoint}}
\put(915,361){\usebox{\plotpoint}}
\put(930,374){\usebox{\plotpoint}}
\put(946,388){\usebox{\plotpoint}}
\put(962,401){\usebox{\plotpoint}}
\put(978,414){\usebox{\plotpoint}}
\put(994,428){\usebox{\plotpoint}}
\put(1009,441){\usebox{\plotpoint}}
\put(1025,455){\usebox{\plotpoint}}
\put(1041,468){\usebox{\plotpoint}}
\put(1057,481){\usebox{\plotpoint}}
\put(1075,492){\usebox{\plotpoint}}
\put(1093,503){\usebox{\plotpoint}}
\put(1111,513){\usebox{\plotpoint}}
\put(1123,530){\usebox{\plotpoint}}
\put(1133,548){\usebox{\plotpoint}}
\put(1143,566){\usebox{\plotpoint}}
\put(1153,584){\usebox{\plotpoint}}
\put(1164,602){\usebox{\plotpoint}}
\put(1174,620){\usebox{\plotpoint}}
\put(1177,625){\usebox{\plotpoint}}
\put(320,696){\raisebox{-1.2pt}{\makebox(0,0){$\Box$}}}
\put(431,250){\raisebox{-1.2pt}{\makebox(0,0){$\Box$}}}
\put(707,303){\raisebox{-1.2pt}{\makebox(0,0){$\Box$}}}
\put(820,327){\raisebox{-1.2pt}{\makebox(0,0){$\Box$}}}
\put(909,356){\raisebox{-1.2pt}{\makebox(0,0){$\Box$}}}
\put(1056,481){\raisebox{-1.2pt}{\makebox(0,0){$\Box$}}}
\put(1115,516){\raisebox{-1.2pt}{\makebox(0,0){$\Box$}}}
\put(1177,625){\raisebox{-1.2pt}{\makebox(0,0){$\Box$}}}
\put(391,651){\makebox(0,0)[l]{$N_0=6^4$}}
\put(431,264){\usebox{\plotpoint}}
\put(431,264){\usebox{\plotpoint}}
\put(451,268){\usebox{\plotpoint}}
\put(471,272){\usebox{\plotpoint}}
\put(491,276){\usebox{\plotpoint}}
\put(512,281){\usebox{\plotpoint}}
\put(532,285){\usebox{\plotpoint}}
\put(552,289){\usebox{\plotpoint}}
\put(573,294){\usebox{\plotpoint}}
\put(593,298){\usebox{\plotpoint}}
\put(613,302){\usebox{\plotpoint}}
\put(634,306){\usebox{\plotpoint}}
\put(654,311){\usebox{\plotpoint}}
\put(674,315){\usebox{\plotpoint}}
\put(694,319){\usebox{\plotpoint}}
\put(715,324){\usebox{\plotpoint}}
\put(735,328){\usebox{\plotpoint}}
\put(755,332){\usebox{\plotpoint}}
\put(776,336){\usebox{\plotpoint}}
\put(796,341){\usebox{\plotpoint}}
\put(816,345){\usebox{\plotpoint}}
\put(837,349){\usebox{\plotpoint}}
\put(857,354){\usebox{\plotpoint}}
\put(877,358){\usebox{\plotpoint}}
\put(898,362){\usebox{\plotpoint}}
\put(916,370){\usebox{\plotpoint}}
\put(932,383){\usebox{\plotpoint}}
\put(949,396){\usebox{\plotpoint}}
\put(965,409){\usebox{\plotpoint}}
\put(981,422){\usebox{\plotpoint}}
\put(998,435){\usebox{\plotpoint}}
\put(1014,447){\usebox{\plotpoint}}
\put(1030,460){\usebox{\plotpoint}}
\put(1046,473){\usebox{\plotpoint}}
\put(1063,486){\usebox{\plotpoint}}
\put(1079,499){\usebox{\plotpoint}}
\put(1095,512){\usebox{\plotpoint}}
\put(1112,525){\usebox{\plotpoint}}
\put(1116,528){\usebox{\plotpoint}}
\put(320,651){\makebox(0,0){$\triangle$}}
\put(431,264){\makebox(0,0){$\triangle$}}
\put(909,365){\makebox(0,0){$\triangle$}}
\put(1116,528){\makebox(0,0){$\triangle$}}
\put(391,606){\makebox(0,0)[l]{$N_0=8^4$}}
\put(431,262){\usebox{\plotpoint}}
\put(431,262){\usebox{\plotpoint}}
\put(451,266){\usebox{\plotpoint}}
\put(471,270){\usebox{\plotpoint}}
\put(491,274){\usebox{\plotpoint}}
\put(512,279){\usebox{\plotpoint}}
\put(532,283){\usebox{\plotpoint}}
\put(552,287){\usebox{\plotpoint}}
\put(573,291){\usebox{\plotpoint}}
\put(593,296){\usebox{\plotpoint}}
\put(613,300){\usebox{\plotpoint}}
\put(634,304){\usebox{\plotpoint}}
\put(654,308){\usebox{\plotpoint}}
\put(674,313){\usebox{\plotpoint}}
\put(695,317){\usebox{\plotpoint}}
\put(715,321){\usebox{\plotpoint}}
\put(735,325){\usebox{\plotpoint}}
\put(756,330){\usebox{\plotpoint}}
\put(776,334){\usebox{\plotpoint}}
\put(796,338){\usebox{\plotpoint}}
\put(816,342){\usebox{\plotpoint}}
\put(837,347){\usebox{\plotpoint}}
\put(857,351){\usebox{\plotpoint}}
\put(877,355){\usebox{\plotpoint}}
\put(898,359){\usebox{\plotpoint}}
\put(909,362){\usebox{\plotpoint}}
\put(916,367){\usebox{\plotpoint}}
\put(933,379){\usebox{\plotpoint}}
\put(950,391){\usebox{\plotpoint}}
\put(967,403){\usebox{\plotpoint}}
\put(984,415){\usebox{\plotpoint}}
\put(1001,428){\usebox{\plotpoint}}
\put(1018,440){\usebox{\plotpoint}}
\put(1035,452){\usebox{\plotpoint}}
\put(1052,464){\usebox{\plotpoint}}
\put(1068,476){\usebox{\plotpoint}}
\put(1084,487){\usebox{\plotpoint}}
\put(320,606){\circle*{19}}
\put(431,262){\circle*{19}}
\put(909,362){\circle*{19}}
\put(1084,487){\circle*{19}}
\end{picture}

  { \small
  Fig. 10. Influence of the lattice size on $\langle \tilde{R} \rangle$ 
  for the Regge approach.
  The number $N_0$ of vertices increases from $3^4$ to $8^4$.
  The well-defined phase survives an extension of the triangulated
  manifold which has almost no influence on the behavior of the computed
  expectation values $\langle \tilde{R} \rangle$. 
  }
\end{figure}


\begin{thebibliography}{99}

\bibitem{MTW} Ch. Misner, K. Thorne, and J. Wheeler, {\em Gravitation}
              (Freeman, San Francisco, 1973).

\bibitem{Misner} Ch. Misner, Rev. Mod. Phys. 29, 497 (1957).

\bibitem{Hawk} S. Hawking, in
 {\em General Relativity, An Einstein Centenary Survey},
 edited by S. Hawking and W. Israel (Cambridge University Press, Cambridge,
 1979).

\bibitem{HaHa} J. Hartle and S. Hawking, Phys. Rev. D28, 2960 (1983).

\bibitem{Hamber1} H. Hamber and R. Williams, Nucl. Phys. B248, 392
 (1984); Phys. Lett. 157B, 368 (1985); Nucl. Phys. B269, 712 (1986).

\bibitem{Hamber2} H. Hamber, in {\em Critical Phenomena, Random
 Systems, Gauge Theories}, Proceedings of the Les Houches
 Summer School, Session XLIII, edited by K. Osterwalder and R. Stora,
 Vol.1 (North Holland, Amsterdam, 1986).

\bibitem{Hamber3} H. Hamber, Phys. Rev. D45, 507 (1992); Nucl. Phys. B400,
 347 (1993); Nucl. Phys. B (Proc. Suppl.) 30, 751 (1993); H. Hamber and
 R. Williams, Phys. Rev. D47, 510 (1993).

\bibitem{Berg} B. Berg, Phys. Rev. Lett. 55, 904 (1985);
               Phys. Lett. B176, 39 (1986).

\bibitem{Regge} T. Regge, Nuovo Cimento 19, 558 (1961).

\bibitem{HartJM} J. Hartle, J. Math. Phys. 26, 804 (1985);
                 J. Math. Phys. 27, 287 (1986);
                 J. Math. Phys. 30, 452 (1989).

\bibitem{FriLee1} R. Friedberg and T. D. Lee, Nucl. Phys. B242, 145 (1984).

\bibitem{FriLee2} G. Feinberg, R. Friedberg, T. D. Lee, and H. Ren,
                  Nucl. Phys. B245, 343 (1984).

\bibitem{Chee} J. Cheeger, W. M\"uller, and R. Schrader,
               Comm. Math. Phys. 92, 405 (1984);
               Lect. Notes in Phys. 160 (Springer, Berlin, 1982).

\bibitem{BGM} W. Beirl, E. Gerstenmayer, and H. Markum, Phys. Rev. Lett. 69,
 713 (1992); Nucl. Phys. B (Proc. Suppl.) 26, 575 (1992);
 W. Beirl, E. Gerstenmayer, H. Markum, and J. Riedler, Nucl. Phys. B
 (Proc. Suppl.) 30, 764 (1993);
 W. Beirl, E. Gerstenmayer, and H. Markum, in {\em Quantum Physics and
 the Universe}, Proceedings of the International Symposium, edited by
 M. Namiki, Y. Aizawa, K. Maeda, and I. Ohba (Pergamon, Oxford, 1993).

\bibitem{DT2}  J. Ambj{\o}rn, B. Durhuus, and J. Fr{\"o}hlich,
                Nucl. Phys.  B257, 433 (1985);
    V. Kazakov, I. Kostov, and A. Migdal, Phys. Lett. 157B, 295 (1985);
 M. Agishtein and A. Migdal, Int. J. Mod. Phys. C1, 165 (1990).

\bibitem{DT3} J. Ambj{\o}rn, Nucl. Phys. B (Proc. Suppl.) 25A, 8 (1992);
  J. Ambj{\o}rn and S. Varsted, Nucl. Phys. B (Proc. Suppl.) 25A, 25 (1992).

\bibitem{DT4} J. Ambj{\o}rn and J. Jurkiewicz, Phys. Lett. B278, 42 (1992);
       M. Agishtein and A. Migdal, Mod. Phys. Lett. 7, 1039 (1992).

\bibitem{GeHa} R. Geroch and J. Hartle, Found. Phys. 16, 533 (1985).

\bibitem{HartTOP} J. Hartle, in {\em Proceedings of the Fourth Marcel Grossmann
                  Meeting on General Relativity}, edited by R. Ruffini
                  (Elsevier, Amsterdam, 1986).

\bibitem{Mark} A. Markov, Proc. Int. Congress Math., 300 (1958).

\bibitem{Mazur} P. Mazur and E. Mottola, Nucl. Phys. B341, 187 (1990).

\bibitem{Giddi} S. Giddings, Int. J. Mod. Phys. A5, 3811 (1990).

\bibitem{Leut} H. Leutwyler, Phys. Rev. 134, B1155 (1964);
               E. Fradkin and G. Vilkovisky, Phys. Rev. D8, 4241 (1973).

\bibitem{deWitt} B. DeWitt, Phys. Rev. 160, 1113 (1967).

\bibitem{FaPo} L. Faddeev and V. Popov, Usp. Fiz. Nauk. 109, 427 (1974);
               Sov. Phys. Usp. 16, 777 (1974).

\bibitem{Menotti} P. Menotti, Nucl. Phys. B (Proc. Suppl.) 17, 29 (1990).

\bibitem{RoWi} M. Ro{\^c}ek and R. Williams, Phys. Lett. 104B, 31 (1981);
               Z. Phys. C21, 371 (1984).

\bibitem{BeKK} B. Berg, B. Krishnan, and M. Katoot, Nucl. Phys. B404,
359 (1993).

\bibitem{NaSe} C. Nash and S. Sen, {\em Topology and Geometry for Physicists}
               (Academic Press, London, 1983).

\bibitem{Kuhn} W. K\"uhnel and G. Lassmann,
               Discrete Comput. Geom. 3, 169 (1988).

\bibitem{CFL} N. Christ, R. Friedberg, and T. D. Lee,
                 Nucl. Phys. B202, 89 (1982).

\bibitem{Casel} M. Caselle, A. d'Adda, and L. Magnea,
                Phys. Lett. B232, 457 (1989).

\bibitem{Louko} J. Louko and P. Tuckey, Class. Quantum Grav. 9, 41 (1992).

\bibitem{BM} B. Br\"ugmann, Nucl. Phys. B (Proc. Suppl.) 30, 760 (1992).

\bibitem{Band} M. Bander, Phys. Rev. Lett. 57, 1825 (1986).

\bibitem{JeviNi} A. Jevicki and M. Ninomiya, Phys. Rev. D33, 1634 (1986).

\bibitem{wb} W. Beirl, Ph. D. Thesis, Technical University of Vienna, 
 1992 (unpublished).

\bibitem{Gross} D. Gross, Phys. Lett. 138B, 185 (1984);
 A. Billoire, D. Gross, and E. Marinari, Phys. Lett. 139B, 75 (1984).

\end{thebibliography}
\end{document}